\documentclass{aa}  
\usepackage{graphicx}
\usepackage{txfonts}
\usepackage{units}
\usepackage{xcolor}
\usepackage{arydshln}
\usepackage[version=4]{mhchem}
\usepackage[colorlinks=true,linkcolor=black,citecolor=blue,urlcolor=blue,draft=false]{hyperref}

\def\Rtap{R_{\rm tap}}

\def\amin{{a_{\rm min}}}
\def\amax{{a_{\rm max}}}
\def\apow{{a_{\rm pow}}}
\def\Nsize{{N_{\rm size}}}
\def\Nbin{{N_{\rm bin}}}
\def\NH{N_{\langle\rm H\rangle}}
\def\nH{n_{\langle\rm H\rangle}}
\def\nd{n_{\rm d}}

\begin{document} 

   \title{Size-dependent charging of dust particles in protoplanetary disks} 
   \subtitle{Can turbulence cause charge separation and lightning?}

   \author{T. Balduin\inst{1,2,3},
          P. Woitke\inst{1,2},
          U. G. J{\o}rgensen\inst{3},
          W.-F. Thi \inst{4}
          \and
          Y. Narita \inst{1}
          }

   \institute{Austrian Academy of Science, Space Research Institute, Schmiedlstrasse 6, A-8042 Graz, Austria\\ \email{Thorsten.Balduin@oeaw.ac.at}
         \and 
   %Centre for Exoplanet Science, University of St Andrews, St Andrews, KY16 9SS, UK
         %\and
   %SUPA, School of Physics and Astronomy, University of St Andrews, St Andrews, KY16 9SS, UK
        % \and
   TU Graz,  Faculty of Mathematics, Physics and Geodesy, Petersgasse 16, 8010 Graz, Austria
         \and 
   Centre for ExoLife Sciences (CELS), Niels Bohr Institute, {\O}stervoldgade 5, DK-1350 Copenhagen, Denmark
         \and
   Max Planck Institute for Extraterrestrial Physics, Giessenbachstrasse 1, D-85748 Garching, Germany
   }

   \date{Received 17.03.2023; accepted 01.08.2023}

% \abstract{}{}{}{}{} 
% 5 {} token are mandatory
 
  \abstract
   {Protoplanetary disk are the foundation of planet formation. Lightning can have a profound impact on the chemistry of planetary atmospheres. The emergence of lightning in a similar manner in protoplanetary disks, would substantially alter the chemistry of protoplanetary disks. }
   {We aim to study under which conditions lightning could emerge within protoplanetary disks.}
   {We employ the {\sc ProDiMo} code to make 2D thermo-chemical models of protoplanetary disks. We included a new way of how the code handles dust grains, which allows the consideration of dust grains of different sizes. We investigate the chemical composition, dust charging behaviour and charge balance of these models, to determine which regions could be most sufficient for lightning.}
   {We identify 6 regions within the disks where the charge balance is dominated by different radiation processes and  find that the emergence of lightning is most probable in the lower and warmer regions of the midplane. This is due to the low electron abundance ($n_{\rm e}/n_{\rm\langle H \rangle}<10^{-15}$) in these regions and dust grains being the most abundant negative charge carriers ($ n_{\rm Z}/n_{\rm\langle H \rangle}> 10^{-13}$). We find that $\rm NH4^+$ is the most abundant positive charge carrier in those regions at the same abundances as the dust grains. We then develop a method of inducing electric fields via turbulence within this mix of dust grains and $\rm NH_4^+$. The electric fields generated with this mechanism are however several orders of magnitude weaker than required to overcome the critical electric field.}
   {}

     % need to look up the available A&A keywords
    %\titlerunning{Can turbulence cause Lightning?}
    \authorrunning{T. Balduin}
    \maketitle
%
%-------------------------------------------------------------------

%\tableofcontents

\section{Introduction}\label{sec:intro}

Dust grains in protoplanetary disks play a crucial role in disk evolution and planet formation, see e.g.\ reviews by \citet{Lesur2023} and \citet{Drazkowska2022}. The dust grains in the midplane tend to pick up the free electrons, which changes the character of the chemistry and has been shown to be connected to the degree of ionization \citep{2011ApJ...736..144B, 2016ApJ...833...92I}. Thus, grain charges are important for dust dynamics and instabilities, such as the 
%\paragraph{Dust Grain Charge}

magneto-rotational instability (MRI), see e.g.
\citet{2008MNRAS.388.1223S}, \citet{ 2011ApJ...735....8P}, \citet{2011ApJ...739...50B} and \citet{Thi2019}. Additionally grain charges and MRI can play an important role in, and can be used to, explain gap formations in disk as shown in \citet{2018A&A...609A..50D},  \citet{2021MNRAS.506.2685R} and \citet{2023arXiv230203430R}.

%\paragraph{Effects of grain charge}
Dust grains accumulate towards the midplane via gravitational settling, enhancing in particular the concentrations of the large grains, see e.g.\ \citet{1995Icar..114..237D}, \citet{2004A&A...421.1075D} and \citet{2018A&A...617A.117R}. Dust grains undergo growth as well, where the charge of the grains can play an important role. Dust grains tend to uniformly charge negatively in the shielded midplane regions, which can create an electrostatic repulsion between the grains large enough to prevent collisions and hence growth. This effect is known as the {\sl charge barrier}, see e.g.\  \citet{2009ApJ...698.1122O}, \citet{2011ApJ...731...95O}, and \citep{2020ApJ...889...64A}.

%\paragraph{Lightning on Earth}
In this work, we want to investigate if charged dust grains can trigger  discharge phenomena similar to lightning on Earth. Lightning is a universal phenomenon, that occurs in atmospheres of several different types of astrophysical objects. Most prominently  lightning occurs in Earth's atmosphere and due to the high temperatures \citep{1968JAtS...25..827O} \citep{1968JAtS...25..839O} \citep{1968JAtS...25..852O} associated with it has several implications on the Earth's atmospheric chemistry. Lightning plays a central role in the abiotic fixation of Nitrogen \citep{1976GeoRL...3..463N} \citep{1980JAtS...37..179H}, which is the creation of several Nitrogen bearing species, e.g. $\mathrm{NO_x}$ \citep{1997JGR...102.5929P}, from elemental Nitrogen $\mathrm{N_2}$. The Nitrogen fixation via lightning is also central in the prebiotic fixation of Nitrogen \citep{2001Natur.412...61N}, therefore making Nitrogen available to the first lifeforms on Earth. Further implications of this process are the abundance of Ozone in earth atmospheres, as Ozone is a product of reactions between $\mathrm{NO_x}$ and elemental Oxygen \citep{2007ACP.....7.2643W} \citep{2013JGRD..11811468M}.
%\paragraph{Lightning on other Solar System Objects Exoplanets and Brown Dwarfs}
Lightning is not just a prominent feature of earth but has been observed on other planets and moons in the solar system. The Pioneer Venus Orbiter missions revealed lightning to discharge on Venus from clouds in the ionosphere \citep{RUSSELL20061344} \citep{Russell2007} \citep{Russell2008} with potential chemical implications discussed by \citep{2015P&SS..113..184D}. Cassini revealed that lightning also occurs on Saturn \citep{2009P&SS...57.1650B} \citep{2010GeoRL..37.9205D}.

There also has been a vast amount of work on the theoretical front, discussing the existence of lightning on objects such as brown dwarfs \citep{2016SGeo...37..705H} \citep{2016MNRAS.461.3927H}  and exoplanets as well \citep{2018ApJ...867..123M} \citep{2019RSPTA.37780398H}.
These findings reveal that lightning is a phenomenon that is ubiquitous.
There has also been a significant amount of work contributed to answer the question if lightning is feasible in protoplanetary disks. Lightning, induced by triboelectric or collisional charging, could be a reasonable mechanism in forming chondrules as shown by \cite{2000Icar..143...87D}. The triboelectric charging they implemented in their work, is not considered in this work. Similar work has been done in \citep{2010MNRAS.401.2641M}, who calculated a critical dust number density at which lightning should occur. Compared to our present work, they are missing an explicit turbulence analysis.
In the realm of lightning in disks one should also mention the works of \citep{2005ApJ...628L.155I}, \citep{2015ApJ...800...47O} and \citep{2019ApJ...878..133O}. In particular the later two investigate how dust grains charge in weakly ionized plasma and additionally study the heating of dust grains by MHD induced electric fields. This work differs in the way electric fields are generated. They investigated how MRI could induce electric fields that heat the dust grains, whilst we will investigate kolmogorov turbulence. In addition any kind of MHD effects are not considered in this work.
Additionally one should mention \citep{2017LPICo1963.2012J}, where they also argue that lightning discharge can be caused by positron emission from pebbles that contain radioactive $\rm Al^{26}$. In our work the main source of ionization would be cosmic rays, and we do not consider radioactive decay.
%\paragraph{Lightning Mechanisms} 
In order for lightning to emerge, certain conditions have to be met. Charges need to build up, then have to be separated, and this separation and build-up have to result in a large enough electric field for an electron cascade to occur \citep{williams_large-scale_1985} \citep{gurevich_runaway_2005} \citep{2008SSRv..137..335S}. In clouds on the Earth, the charges are built up by ice crystals getting charged via collisions with each other \citep{1978JAtS...35.1536T} \citep{1983QJRMS.109..609J} \citep{1993JApMe..32..642S} \citep{1998JGR...10313949S}. During these collisions, faster growing and therefore smaller particles tend to charge positively, whilst the larger particles charge negatively \citep{1987QJRMS.113.1193B} \citep{2001JGR...10620395D}. For charge separation in clouds, an updraft is generally required, and in this updraft  a lower positive layer and upper negative layer forms \citep{1998JGR...10314059S} \citep{2007MWRv..135.2525B}. Another way of lightning to emerge on earth is during volcanic eruptions \citep{1974JGR....79..472B} \citep{2018EGUGA..2012793G}. Whilst the fundamentals are similar to lightning in clouds, some, for us important, details are different. In the volcanic plumes, dust particles are the particles that get triboelectric charged \citep{2013PhRvL.111k8501H}. Triboelectric charging leads to the smaller particles getting charged more negatively and larger particles more positively \citep{LACKS2007107} \citep{2009PhRvL.102b8001F}. Lastly, one should also mention the new experimental findings in triboelectric charging by  \citep{2021A&A...650A..77J}, \citep{2022MNRAS.517L..65W}. In the first publication mentioned they state that triboelectric charging could be instrumental in overcoming the bouncing barrier and in the second they investigate how triboelectric charging could play a role in ionizing the gas phase.
%\paragraph{Conditions in Protoplanetary Disks }
The former explanations might lead the reader to ask the question, if the conditions in protoplanetary disks can lead to a similar emergence of lightning. We do think so as a collection of dust in high densities as found in the midplane of protoplanetary disks, due to gravitational settling, resembles conditions, found on earth and other planets, moons and even brown dwarfs, where lightning has been shown to emerge. In addition to this, turbulent motions, which could result in charge separation, are found in protoplanetary disks.

In Section 2 we explain our model, which includes explanations on the chemistry of our code, all of the numerical basics and improvements we have done to our code and how we implemented cosmic rays. In Section 3 we present the results of our simulations, in particular we investigate the role electrons play in our simulations, how the dust grains get charged and investigate if specific regions of the disk can be favorable for the emergence of lightning. We furthermore investigate how turbulent eddies could induce a charge separation in these areas and investigate if electric fields that emerge from this separation could be large enough for lightning to emerge.  In Section 4 we conclude our results.

%--------------------------------------------------------------------
\section{The thermo-chemical disk model}\label{sec:model}

To simulate the size-dependent charging of dust grains in a protoplanetary disk, we use the \underline{Pro}toplanetary \underline{Di}sk \underline{M}odel ({\sc ProDiMo)} developed by \citet{Woitke2009, Woitke2016}. 
{\sc ProDiMo} is a 2D thermo-chemical disk modelling code, which combines detailed continuum and line radiative transfer with the solution of a chemical network, the determination of the non-LTE population of atomic and molecular states, and the heating/cooling balance for both gas and dust.

The model uses a simple parametric density setup chosen to represent a 2\,Myrs old T\,Tauri star.
All further assumptions about the star, the disk shape, the irradiation of the disk with FUV photons, X-rays and cosmic rays, as well as the dust material composition and opacities, and dust settling, are explained in detail in \citet[][see their table~3]{Woitke2016}.
Here, we only explain the changes made to that code, and to that disk setup, which differ from  \citet{Woitke2016}. 
The selected values for the stellar, disk shape and dust parameters are summarized in Table~\ref{tab:parameters}.

\subsection{Chemical setup}\label{subsec:chem}

We choose the so-called 'large DIANA-standard' chemical network with 235 species as our base network for this work, which has been described in detail by \cite{2017A&A...607A..41K}. 
Most UV, cosmic-ray, two-body and three-body reaction rates are taken from the UMIST 2012 database \citep{UMIST1,UMIST2}. 
In addition, there are rates for the H$_2$-formation on grains, reactions for electronically excited molecular hydrogen $\rm \ce{H_2}^\star$, PAH (Polyaromatic Hydrocarbon) charge chemistry \citep{2019A&A...632A..44T}, and a simple freeze-out chemistry for the ice phases of all neutral molecules, see \citep{Woitke2009} and \cite{2017A&A...607A..41K}.
The large DIANA-standard network also includes X-ray processes with doubly ionized species from \citep{2012PhDT........81A}.

We added a number of dust species to this base chemical network to represent the different charging states $q$ of grains of different sizes $a$.
By solving the chemical rate network, we determine the dust charge distribution function for every given size and charge $f(a,q)$, from which we can then compute, for example, the mean charge of grains of different sizes, including the feedback of the charged grains on all other results of the chemistry.

In contrast to \citet{Woitke2016}, we have chosen to omit the PAH charging states in the chemistry in this paper, but PAH heating is still taken into account.
If present, the PAHs can introduce similar effects as charged grains. Since we want to study the chemical feedback of charged grains in disks in particular, we want to avoid any confusion between PAH charging and dust charging. In addition preliminary tests showed that the impact of PAHs in the midplane is negligible, due to the PAHs freezing out. Therefore, we also omit them to lower the numeric effort.

\subsection{Dust Size Distribution And Settling}\label{subsec:settling}

In {\sc ProDiMo}, the dust component enters into the modelling (i) as an opacity source for the continuum and its effect on line radiative transfer, and (ii) as a set of chemical species that can pick up and release charges in the chemical rate network. The dust grains, however, are neither created nor destroyed, nor do they grow in the chemistry module

Our assumptions about dust settling and opacity are explained in \citet{Woitke2016}. The dust size distribution function before settling $\rm[cm^{-1}]$ is assumed to be
\begin{equation}
    f_0(a)\propto a^{-\apow} \ ,
    \label{eqn:dustsizedist}
\end{equation}
which is the same everywhere in the disk, using
$\Nsize$ log-equidistant size grid points between the minimum grain radius $\amin$ and the maximum radius $\amax$. The proportionality constant in Eq.~(\ref{eqn:dustsizedist}) is derived from the local gas density $\rho(r,z)$, the dust material density $\rho_{\rm gr}$, and the global gas to dust mass ratio assumed to be 100 in this paper. 
In each vertical column, at radius $r$, all grains of size $a$ are then vertically re-distributed, with a scale height that is smaller than the gas scale height. 
Here we use the settling description by \citet{1995Icar..114..237D}, with settling parameter $\alpha_{\rm settle}\!=\!0.01$, as outlined in \citet{Woitke2016}.
After applying the settling, the dust size distribution function  $f(a,r,z)$ is numerically available on $\Nsize$ size grid points at each point $(r,z)$ in the model. $f(a,r,z)$ is the basis for the opacity calculations.

\subsection{Dust Size Bins In the Chemistry}\label{subsec:dustbins}

We use a much smaller number of dust size bins $\Nbin$ (a number between 2 and 9) to represent the position-dependent dust size distribution function after settling $f(a,r,z)$ in the chemistry. 
The idea in the following is to introduce two power-law indices, $\kappa$ and $\zeta$, to make that sparse representation exact with regard to two selectable dust size moments.  
We then argue which of the dust size moments should be selected in order to minimize the numerical errors in the chemistry introduced by the sparse resolution of $f(a,r,z)$ in size space.  
We start by computing the following moment of the dust size distribution function
\begin{equation}
    \langle a^\zeta\rangle(r,z) = \int_\amin^\amax\!\!\! f(a,r,z)\,a^{\,\zeta}\,da \ ,
    \label{eq:dustbins1}
\end{equation}
where $f(a,r,z)$ $\rm[cm^{-1}]$ is normalized as $\int_\amin^\amax f(a,r,z)\,da\!=\!1$. 
For example, with $\zeta\!=\!2$, $n_d\,4\pi\,\langle a^2\rangle$ is the total dust surface per cubic centimeter $\rm[cm^2/cm^3]$.  
Next, we divide $\langle a^\zeta\rangle$ by $\Nbin$, and numerically adjust the bin boundaries $a_j$ such that each size bin contains the same portion of $\langle a^\zeta\rangle$.
\begin{equation}
    \langle a^\zeta\rangle_j
    = \int_{a_{j-1}}^{a_j}\!\!\! f(a,r,z)\,a^{\,\zeta}\,da 
    = \frac{\langle a^\zeta\rangle(r,z)}{\Nbin} \ .
    \label{eq:dustbins2}
\end{equation}
Here, $a_j$ are the size bin boundary values, where $a_0\!=\!\amin$ and $a_\Nbin\!=\!\amax$. Once the $a_j$ are fixed this way, we compute similar quantities
\begin{equation}
    \langle a^\kappa\rangle_j
    = \int_{a_{j-1}}^{a_j}\!\!\! f(a,r,z)\,a^{\,\kappa}\,da 
    \label{eq:dustbins3}
\end{equation}
and define the average size of our grains [cm] in dust bin $j$ as
\begin{equation}
    \bar{a}_j = \left(
    \frac{\langle a^\zeta\rangle_j}
         {\langle a^\kappa\rangle_j}
    \right)^{1/(\zeta-\kappa)} \ .
    \label{eq:dustbins4}
\end{equation}
Finally, the dust particle densities $\rm[cm^{-3}]$ in size bin $j$ are calculated as
\begin{equation}
    n_{{\rm d},j} = \nd(r,z) \frac{\langle a^\kappa\rangle_j}
    {\left(\bar{a}_j\right)^\kappa}
    \label{eq:dustbins5}
\end{equation}
From Eqs.~(\ref{eq:dustbins2}) to (\ref{eq:dustbins5}) it follows that
\begin{eqnarray}
    n_{{\rm d},j}\,(\bar{a}_j)^\kappa
    &=& \nd(r,z) \int_{a_{j-1}}^{a_j}\!\!\! f(a,r,z)\,a^{\,\kappa}\,da \nonumber\\
    n_{{\rm d},j}\,(\bar{a}_j)^\zeta
    &=& \nd(r,z) \int_{a_{j-1}}^{a_j}\!\!\! f(a,r,z)\,a^{\,\zeta}\,da\ . \nonumber
\end{eqnarray}
which means that our dust bins $\{(n_{{\rm d},j},\bar{a}_j)\,|\,j\!=\!1,...\,,\Nbin\}$ exactly represent two dust size moments, where 
the dust size distribution function is weighted with $a^\kappa$ and $a^\zeta$, respectively, see Table~\ref{tab:bins}.

\begin{table}[t]
    \caption{Construction of dust size bins for the chemistry$^{(1)}$.}
    \label{tab:bins}
    \vspace*{-2mm}
    \resizebox{90mm}{!}{\begin{tabular}{r|cccc|c}
      \hline
      &&&&&\\[-2.1ex]
        & bin\,1 & bin\,2 & bin\,3 & bin\,4 & total \\
      \hline
      \multicolumn{6}{c}{$\kappa=0$ and $\zeta=2$}\\
      \hline
      &&&&&\\[-2.1ex]
      $\bar{a}_j\rm\,[\mu m]$ &
        0.0636 & 0.122 & 0.316 & 1.72 & \\
      $n_{{\rm d},j}\rm\,[cm^{-3}]$ & 
        1.36(-4) & 3.70(-5) & 5.48(-6) & 1.85(-7) &
        \bf{1.78(-4)} \\
      length$\rm\,[cm^{-2}]$ &
        8.63(-10) & 4.50(-10) & 1.73(-10) & 3.18(-11) &
        1.52(-9) \\
      area$\rm\,[cm^{-1}]$ &
        6.89(-14) & 6.89(-14) & 6.89(-14) & 6.89(-14) &
        \bf{2.76(-13)} \\
      dust volume &
        1.46(-19) & 2.80(-19) & 7.27(-19) & 3.96(-18) &
        5.11(-18) \\
      \hline
      \multicolumn{6}{c}{$\kappa=1$ and $\zeta=2$}\\
      \hline
      &&&&&\\[-2.1ex]
      $\bar{a}_j\rm\,[\mu m]$ &
        0.0645 & 0.125 & 0.337 & 2.29 & \\      
      $n_{{\rm d},j}\rm\,[cm^{-3}]$ &
        1.32(-4) & 3.51(-5) & 4.83(-6) & 1.04(-7) &
        1.72(-4) \\
      length$\rm\,[cm^{-2}]$ &
        8.51(-10) & 4.39(-10) & 1.63(-10) & 2.39(-11) &
        {\bf 1.48(-9)} \\
      area$\rm\,[cm^{-1}]$ &
        6.89(-14) & 6.89(-14) & 6.89(-14) & 6.89(-14) &
        {\bf 2.76(-13)} \\
      dust volume &
        1.48(-19) & 2.87(-19) & 7.75(-19) & 5.27(-18) &
        6.48(-18) \\
      \hline
      \multicolumn{6}{c}{$\kappa=0$ and $\zeta=3$}\\
      \hline
      &&&&&\\[-2.1ex]
      $\bar{a}_j\rm\,[\mu m]$ &
        0.335 & 329 & 1070 & 2180 & \\
      $n_{{\rm d},j}\rm\,[cm^{-3}]$ & 
        1.78(-4) & 1.88(-13) & 5.45(-15) & 6.43(-16) & 
        {\bf 1.78(-4)} \\
      length$\rm\,[cm^{-2}]$ &
        5.97(-9) & 6.19(-15) & 5.84(-16) & 1.40(-16) &
        5.97(-9) \\
      area$\rm\,[cm^{-1}]$ &
        2.51(-12) & 2.56(-15) & 7.86(-16) & 3.85(-16) &
        2.52(-12) \\
      dust volume &
        2.81(-17) & 2.81(-17) & 2.81(-17) & 2.81(-17) &
        {\bf 1.12(-16)} \\
      \hline
    \end{tabular}}\\[1mm]
    $^{(1)}$: \small Results for an unsettled dust size distribution with
    $\amin\!=\!0.05\rm\,\mu$m, 
    $\amax\!=\!3\,$mm,
    $\apow\!=\!3.5$,
    grain material density $\rho_{\rm gr}\!=\!2.094\rm\,g/cm^3$,
    gas to dust ratio $100$, and gas density  $\nH\!=\!10^{10}\rm\,cm^{-3}$, using $\Nbin\!=\!4$.
    The consecutive lengths are calculated as 
    $\bar{a}_j\,n_{{\rm d},j}$, 
    the surface areas as 
    $4\pi\,\bar{a}_j^2\,n_{{\rm d},j}$, 
    and the volumes as 
    $(4\pi/3)\,\bar{a}_j^3\,n_{{\rm d},j}$. 
    The correct total values are marked in bold face:
    $n_d\!=\!1.78\times 10^{-4}\rm\,cm^{-3}$,
    ${\rm length}\!=\!1.48\times10^{-9}\rm\,cm^{-2}$, 
    ${\rm area}\!=\!2.76\times10^{-13}\rm\,cm^{-1}$ and
    ${\rm volume}\!=\!1.12\times 10^{-16}$.
\end{table}

If $\kappa\!=\!0$ is chosen, the quantity $\langle a^\kappa\rangle_j$ becomes the fraction of dust particles having sizes between $a_{j-1}$ and $a_j$, and consequently, the total dust number density $n_{{\rm d},1} + n_{{\rm d},2} + ... + n_{{\rm d},\Nbin} = \nd$ becomes exact.
In a similar way, choosing $\zeta\!=\!2$ causes the representation of the total dust surface to be exact.
All other dust size moments, however, show deviations from their true integral values.  Increasing the number of dust bins $\Nbin$ would reduce these problems.

Thus, with $\kappa$ and $\zeta$, we can choose two dust size moments, which are represented exactly by the bins and which are most relevant for the problem at hand.  
Since the effects of the charged grains in the chemistry, via their collisional and photoionization rates, scale with their total cross-section, a choice of $\zeta\!=\!2$ seems appropriate.  That scaling, however, is not exact, because there are second order dependencies of the rates on grain charge over radius $q/a$, see Sect.~\ref{subsec:moment_rate_coeff}.
In addition, the conservation of the total charge is an important principle. 
The total number of elementary charges on the grains per volume, assuming $q/a\!=\,$const is valid (see Sect.~\ref{subsec:moment_rate_coeff}), is

\begin{equation}
   Q = \nd \int_\amin^\amax \!\!\!\! q(a)\,f(a)\,da
     ~\approx~ \nd \left(\frac{q}{a}\right) 
         \int_\amin^\amax \!\!\!\! a\,f(a)\,da \ .
\end{equation}
Therefore, it seems important to get the first dust size moment correct as well. 
In contrast, there is no direct link between the total dust particle density and the chemistry, nor any dependencies of the chemistry on the total dust volume. Therefore, our default choice is $\kappa\!=\!1$ and $\zeta\!=\!2$.

Table~\ref{tab:bins} shows that in this case, we only need to represent the smaller grains up to a few microns in size.  A larger value of $\zeta$ would also come with an increased level of computational problems and computing time, for example because we would need to extend the maximum number of charging states $q_{\rm max}$ to many 10000 for large grains, see Sect.~\ref{subsec:moment_rate_coeff}.
%VIELLEICHT: In practise, values around $\zeta\!=\!2.2$, which produce bins that contain sub-micron to about 10\,$\mu$m grains, seem to provide stable results, yet with computationally acceptable efforts.}

\subsection{Dust Charging Reactions}\label{subsec:dustchargereac}
The charging of dust grains is part of the chemistry of {\sc ProDiMo}. We will only focus on the most relevant reactions here, a full description of all considered dust grain charging processes can be found in \citet{Thi2019}. 

The three most important reaction types are electron attachment, photoionization, and charge exchange between dust grains and molecular ions. 

\paragraph{(A) Photoionisation:}
Photoionization describes the process of stripping away electrons from the dust grains via photons with sufficiently high energies. 
\begin{equation}
    \ce{Z} + h\nu ~\rightarrow~ \ce{Z}^+ + e^-,
    \label{eqn:dust_photo}
\end{equation}
where $\ce{Z}$ and $\ce{Z}^+$ stand for grains with charges $q$ and $q+1$, respectively. Here q can have any integer value > 0.

For positively charged and neutral grains this process is called photoejection and for negatively charged grains this process is called photodetachment. According to \cite{Weingartner2001} the energy threshold for photoejection is
\begin{equation}
    h\nu_{\rm pe} = {\rm IP} = W_0+W_c\ ,
\end{equation}
where IP is the ionization potential of the dust grain and $W_0$ the work function of silicate grains, assumed to be 8 eV by \cite{2001ApJ...546..496C} after measurements of \citet{1972JAP....43.1563F}. 
$W_c$ is an additional term that is dependent on the charge of the grain and  increases the ionization potential for positive grains, making it harder to ionize the grain, or decreasing the ionization potential, making it easier to ionize the grain. This additional term $W_c$ is defined as,
\begin{equation}
    W_c=\left(q+\frac{1}{2}\right)\frac{e^2}{a},
\end{equation} 
where $q$ is the charge of the grain normalized by the elementary charge, $e$ the elementary charge and $a$ the grain radius. We note that the factor 1/2 is still open to discussion according to \citet{2003PhRvB..67c5406W}, where a factor of 3/8 was proposed. 
For photodetachment the threshold photon energy $h\nu_{pd}$ is equal to the grain electron affinity EA plus the minimum energy $E_{\rm min}$ at which the tunneling probability becomes relevant. 
\begin{align}
    h\nu_{\rm pd}&= {\rm EA}(q+1,a)+E_{\rm min}(q,a)\\
    {\rm EA}(q,a)&= W_0-E_{\rm bg}+\Big(q+\frac{1}{2}\Big)\frac{e^2}{a}\label{eq:EA}\\
    E_{\rm min}(q,a) &=-(q+1)\frac{e^2}{a}\left[1+\left(\frac{27\AA}{a}\right)^{0.75}\right]^{-1}
    \label{eq:Emin}
\end{align}

where $E_{\rm bg}$ is the band gap, assumed to be 5 eV in \cite{Weingartner2001}, $a$ the dust grains radius and $q$ the grain charge in units of the electron charge.  For the total rate coefficient of photoionization, one has to combine both photodetachment and photoejection. The rate coefficient therefore takes the form of 
\begin{equation}
    k_{\mathrm{ph}}(q)=\pi a^2 \!\!\int_{\nu_{\rm pe}(q)}^{\nu_{\rm max}}\!\!\! \eta_\mathrm{eff}\,Q_{\rm abs}\,J_{\nu}\;d\nu 
    ~+~ \pi a^2 \!\!\int_{\nu_{\rm pd}(q)}^{\nu_{\rm max}}\!\!\!
    \eta_{\rm pd}\,Q_{\rm abs}\,J_{\nu}\;d\nu
    \label{eqn:rate_coeff_ph}
\end{equation}
where $J_{\nu}$ is the direction-averaged flux of photons per area and second $[\rm cm^{-2}s^{-1}]$, $\nu_{\rm pe}$ the threshold frequency of photoejection, $\nu_{\rm pd}$ the threshold frequency for photodetachment, $\nu_{\rm max}$ the maximum frequency of photons, being equivalent to an energy of 13.6\,eV in {\sc ProDiMo}, $Q_{\rm abs}$ is the frequency-dependent absorption efficiency, $\eta_{\rm eff}$ and $\eta_{pd}$ are the yields of the photoejection and photodetachment \citep{Weingartner2001} of silicate and $a$ the dust grain radius.

\paragraph{(B) Electron Attachment:}
The rate coefficient for electron attachment
\begin{equation}
    \ce{Z}+e^- ~\rightarrow~ \ce{Z}^-
    \label{eqn:V}
\end{equation}
is derived by averaging over the Maxwellian velocity distribution of the impinging electrons. Here $\ce{Z}$ and $\ce{Z-}$ stand for grains with charges $q$ and $q-1$, respectively, where $q$ can have any integer value $\leq\!0$. This results in a reaction coefficient $k_e(q)$ of the form
\begin{equation}
    k_e(q) = n_e S_e\sqrt{\frac{8k_b T}{\pi m_e}}\,\sigma_Z\,f_q    
    \label{eqn:rate_coeff_elec}
\end{equation}
where $n_e$ is the electron density, $S_e$ is a sticking coefficient assumed to be larger than 0.3 \citep{1980PASJ...32..405U} and set to 0.5 in our simulations, $T$ is the gas temperature, $m_e$ the electron mass and $k_b$ the Boltzmann constant. $\sigma_Z=\pi a^2$ is the dust grain cross-section, and $a$ the dust grain radius. $f_q$ is a charge-dependent factor that is derived from taking the Coulomb interaction between the incoming electron and charged grain into account. For neutral grains we have $f_q\!=\!1$, for positively charged grains $(q\!>\!0)$ there is an additional attraction which enlarges the rate, and for negatively charged grains ($q\!<\!0$), there is Coulomb repulsion

\begin{equation}
    f_q=
    \begin{cases}
    1+\frac{W_c}{kT}&W_c>0\\
    1&W_c=0\\
    \exp{\frac{W_c}{kT}}&W_c<0\\
    \end{cases} \ ,
    \label{eq:fq}
\end{equation}

\paragraph{(C) Charge Exchange:}
Dust grains can get charged via charge exchanges between dust grains and molecular ions. The most prevalent charge exchange for this study is the exchange of a negative charge between a negatively charged dust grain and a positively charged molecular ion,
\begin{equation}
    \ce{Z}^- + \ce{M}^+ ~\rightarrow~ \ce{Z} + \ce{M}\\
    \label{eqn:charge_exchange}
\end{equation}
The general form of the rate coefficient is similar to the electron attachment. Since we only allow for exothermal charge exchange reactions, the rate coefficient takes the form
\begin{equation}
   k^M_{\rm{ex}}(q)=n_M S_{\!M}\sqrt{\frac{8kT}{\pi m_M}}
   \,\sigma_Z\,f_q^M
    \label{eqn:rate_coeff_grain}
\end{equation}
where, analog to Eq.\,(\ref{eq:fq})
\begin{equation}
    f_q^M=
    \begin{cases}\textstyle
    1-\frac{q\,q^{\rm M}e^2}{a\,kT} & q\,q^{\rm M}<0\\
    1 & q\,q^{\rm M}=0\\ \textstyle
    \exp\Big(-\frac{q\,q^{\rm M}e^2}{a\,kT}\Big) & q\,q^{\rm M}>0\\
    \end{cases} \ ,
\end{equation}
where $n_M$ is the density of the molecules, $q^M$ is the charge of the impinging molecule, usually $+1$,
$S_{\!M}$ is a sticking coefficient set to 1 \citep{Thi2019}, $n_M$ the number density of the molecule and $m_M$ the molecular mass.

Note that we exclude passive ion attachment in our network. Meaning, in our network, molecules and dust grains cannot attach to each other via electromagnetic forces only. We will discuss the potential implications of including passive ion attachment in Section \ref{sec:discussion}.

We exclude endotherm charge exchange reactions between protonated molecules and dust grains, where the work function minus the band gap plus the proton affinity of the molecule exceeds 13.6 eV (see App.\,\ref{AppC} and Tab.\,\ref{tab:proton_aff}). 
In order to identify these endotherm charge exchange reactions, we need to know the heat of formation of negatively charged dust grains $H_f^0(Z^q)$, which is given by the work function minus the band gap plus small correction terms of order $e^2/a$, see App.\,\ref{AppC}. In our model, we consider porous silicate grains internally mixed with amorphous carbon (see Tab.\,\ref{tab:parameters}), but these grains can also be ice coated. For pure silicate grains, values for the work function minus the bad gap lie between  about 3\,eV and 8\,eV \citep{Weingartner2001}. Yet, following the explanations of \cite{workfunctions1}, and looking at the compiled values of work functions of different dust components from \cite{2000Icar..143...87D} and \cite{workfunctions3}, a value between 2\,eV and 6\,eV seems appropriate. 
In this paper we use $H_f^0(Z^q) = q\,\times\,5.89\,$eV, where $q\!<\!0$ for negative grains, 0 for neutral grains and $q\!>\!1$ for positive grains in accordance with \cite{Thi2019}. This is supported by \cite{Rosenberger2001}, who state that a mixture of dust grain material tends to lower the work function overall. According to the data collected by \cite{Thi2019}, water ice (coated) grains should have higher work functions. We can assume that our dust grains are water ice coated according to \cite{Phyllosilicates1}, who showed that at temperatures of 300 – 500 K a water layer should be found on dust grains. This gets further supported by \cite{Phyllosilicates2}, who showed,  with a {\sc ProDiMo} model similar to ours, that phyllosilicates can be found in the midplane of a protoplanetary disk.
\subsection{Chemical Network adjustment for charge and proton exchange reactions}
\label{subsec:charge_exchange}

Particular to this work is that the code has the ability to automatically generate all possible charge exchange reactions between molecules and dust grains and proton exchange reactions between molecules and add them to the network if they meet certain criteria. In addition we allow the code to add endothermic reactions to the network if they meet a predetermined requirement. We test the implications of the different networks this creates in App.\,\ref{sec:charge_exchange}.
\paragraph{Proton Exchange Reactions} 
Proton reactions come in the following form:
\begin{equation}
    \rm M_1 + M_2H^+  \longleftrightarrow \rm M_1H^+ + M_2 
\end{equation}
where $\rm M_1$ and $\rm M_2$ are neutral molecules, 
and $\rm M_1H^+$ and $\rm M_2H^+$ are the corresponding protonated molecules. 
We start our procedure by first identifying pairs of molecules and their protonated counterpart ($\rm M_1,M_2H^+$). If a reaction for the pair already exists in the network, we just move on. However, if a reaction between the pair is not found in the network, we add an approximated reaction to the network with certain assumptions.

The aim is to create a reaction rate with the Arrhenius equation. 
\begin{equation}
    k_{\mathrm{ex}}(T)=\alpha \left(\frac{T}{T_0}\right)^{\beta}{\rm e}^{-\gamma/T},
    \label{eqn:rate_coeff_add_CE_I}
\end{equation}
where $T_0$ refers to the reference temperature of 293 K.
In order to achieve this, we have to provide the different factors $\alpha$, $\beta$ and $\gamma$. For $\alpha$ and $\beta$ we apply the same, following, method.
For each unprotonated molecules, $\rm M_1$, we count how many protonation reactions for $\rm M_1$ already exist in the network. We sum up the different $\alpha_i$ and $\beta_i$ parameters of the existing reactions. We average the summed up parameters by simply dividing them by the number of reactions found in the network, $n_{\rm reac}$
\begin{align}
    \alpha_{\rm mean}=\frac{1}{n_{\rm reac}}\sum \alpha_{i}\\
    \beta_{\rm mean}=\frac{1}{n_{\rm reac}}\sum \beta_{i}   
\end{align}
If we only find one reaction, we assume $\alpha_{\rm mean}$ to be $5\times10^{-10}$ and $\beta_{\rm mean}$ to be -0.5.
We then check pairs of unprotonated and protonated molecules ($\rm M_1,M_2H^+$) and create reaction rates with the calculated $\alpha_{\rm mean}$ and $\beta_{\rm mean}$ and the corresponding product ($\rm M_1H^+,M_2$). Lastly, we need to calculate $\gamma$, given by the reaction enthalpy $\Delta H_r$. This is done by taking the difference of the heats of formation of the products and reactants $H_f^0$, with data taken from measurements at 0\,K from \citet{1997A&AS..121..139M}.
\begin{equation}
    \Delta H_r= 
     H_f^0(M_1H^+)
    +H_f^0(M_2)
    -H_f^0(M_1)
    -H_f^0(M_2H^+)
\end{equation}
Here two cases can arise. In case of an exothermic reaction ($\Delta H_r\!<\!0$) we can just add the reaction and will set $\gamma\!=\!0$. In case of an endothermic reaction ($\Delta H_r\!>\!0$), the reaction is just added when $\Delta H_r/k_b$ is smaller than a predetermined barrier. In cases where $\Delta H_r$ is small enough, $\gamma$ is set to $\Delta H_r/k_b$, and the reaction is added anyway. In our standard case we exclude endothermic reactions, i.e.\  we set the barrier to 0. In the appendix we investigate the impact of such endothermic reactions by using a barrier value of 5000\,K.

\paragraph{Charge Exchange Reactions - Molecules}
For charge exchange reaction we assume two types, non-dissociative and dissociative. We follow the same approach as with the proton exchanges, meaning we have to supply the different parameters of the Arrhenius equation. For $\gamma$ our approach is the same as with the proton exchange reactions. We calculate the difference of the formation enthalpies $dH_f$ and add it as $\gamma$ if the reaction is exothermic or below the predetermined barrier. For $\alpha$ and $\beta$ we simply assume the values of $5\times10^{-10}$ and -0.5 respectively.

\paragraph{Charge Exchange Reactions – Dust}
If dust grains are involved in a charge exchange reaction, we adjust the rate coefficient according to
\begin{equation}
    k_{\mathrm{j,ex}}(q)=A\frac{\sqrt{\frac{T}{300 K}}}{n_{d,\mathrm{ref}}}\left(\frac{a}{a_{\mathrm{ref}}}\right)^2
\end{equation}
where $A$ is the first factor in the Arrhenius equation and taken from the results from \cite{1984ApJS...56..231L}, who calculated these rates for reference dust particle density $n_{\rm d,ref}\!=\!2.64\times10^{-12}\;\mathrm{cm^{-3}}$ and reference grain radius $a_{\rm ref}\!=\!0.1\,\mu$m.
\begin{table}
    \caption{Proton affinities $P_A$ and reaction enthalpies with negatively charged silicate grains $\Delta H_r$ of selected molecules M$^{(1)}$.}
    \vspace*{-5mm}
    \label{tab:proton_aff}
    \begin{center}
    \begin{tabular}{|c|c|c|c|}
        \hline     
        $\ce{M}$ & $\ce{MH}^+$ & $P_A(\ce{M})$ [eV] & $\Delta H_r$ [eV] \\
        \hline 
        &&&\\[-2.2ex]
        $\ce{C_3H_2}$ & $\ce{C_3H_3}^+$ & -9.587 & 1.196\\
        $\ce{NH3}$ & $\ce{NH4+}$ & -8.904 & 1.879\\
        $\ce{SiO}$ & $\ce{SiOH+}$ & -8.06 & 0.352\\
        $\ce{C4}$ & $\ce{C4H+}$ & -8.032 & 0.324\\
        $\ce{CS}$ & $\ce{HCS+}$ & -8.00 & 0.2933\\
        %\hline\multicolumn{3}{|c|}{\small threshold for dissociative recombination with \ce{Z-}}\\
        $\ce{CH3OH}$ &    $\ce{ CH3OH2+}$ & -7.82 & 0.112\\
        \hdashline
         &&&\\[-2.2ex]        
        $\ce{H2CS}$ & $\ce{H3CS+}$ & -7.587 &-0.121 \\
        $\ce{H2CO }$ &   $\ce{ H3CO+ }$ & -7.47 &-0.243\\
        $\ce{HCN}$ & $\ce{HCNH+}$ & -7.426 & -0.282\\
        $\ce{H2S }$ &   $\ce{H3S+}$ & -7.39 &-0.314\\
        $\ce{H2O}$ & $\ce{H3O+}$ & -7.173 & -0.535\\
        $\ce{HS}$ &     $\ce{H2S+}$ & -6.98 &-0.728 \\
        $\ce{C2H2}$ &    $\ce{C2H3+}$ & -6.59 &-1.119\\
        $\ce{CO}$ & $\ce{HCO+}$ & -6.1 &-1.608\\
        $\ce{CH4}$ & $\ce{CH5+}$ & -5.765 & -1.943\\
        $\ce{N2}$ & $\ce{HN2+}$ & -5.12 &-2.588\\     
        $\ce{H2}$ & $\ce{H3+}$ & -4.36 &-3.345\\
        \hline 
    \end{tabular}\\[0mm]
    \end{center}
    $^{(1)}$: \small The proton affinities are calculated as
    $P_A(\ce{M})=H_f^0(\ce{MH+})-H_f^0(\ce{M})-H_f^0(\ce{H+})$, where $H_f^0(\ce{MH^+})$, $H_f^0(\ce{M})$ and $H_f^0(\ce{H}^+)$ are the enthalpies of formation of the protonated molecule $\ce{MH}^+$, the neutral molecule $\ce{M}$ and the proton, outside any electric fields, respectively, see App.\,\ref{AppC}. The last column is the reaction enthalpy for a dissociative reaction of a protonated molecule with a negative dust grain $\rm Z^-+MH^+ \rightarrow Z+M+H$, see App.\,\ref{AppC}. The dashed line marks the threshold for reactions being considered in the code. Heat of formation data $H_f^0$ are taken from  \cite{1997A&AS..121..139M} and \cite{nist}.

\end{table}
\subsection{Dust Charge Moments}\label{subsec:moments}
In principle, each charge state $q$ of the grains in size bin $j$ could be included in the chemical rate network. However, since micron-sized grains can already collect thousands of elementary charges \citep{Stark2015, Tazaki2020}, this method would mean to include a couple of 10000 species, which is computationally very challenging.  
In order to avoid this problem, the dust grains and all their charge states are represented by three charge moments, called $Z_{{\rm m},j}^-$, $Z_{{\rm m},j}$ and $Z_{{\rm m},j}^+$ in each size bin $j$.
\begin{align}
    [Z_{{\rm m},j}^+] =&~ n_{{\rm d},j} \sum_{q=1}^{q_{\rm max}}  (+q)\,f_j(q)\label{eqn:Zplus}\\
    [Z_{{\rm m},j}^-] =&~ n_{{\rm d},j} \!\!\!\sum_{q=-q_{\rm max}}^{-1}\!\!\! (-q)\,f_j(q)\label{eqn:Zminus}\\
    [Z_{{\rm m},j}] =&~ n_{{\rm d},j}\,q_{\rm max} \,-\, [Z_{{\rm m},j}^-] \,-\, [Z_{{\rm m},j}^+] 
    \label{eqn:Zneutral}\ ,
\end{align}
where $q_{\rm max}$ is the maximum number of elementary charges on the dust grains, both positive or negative, and $n_{{\rm d},j}$ the dust number density of bin j.  The moments $Z_{{\rm m},j}^+$ and $Z_{{\rm m},j}^-$ express the total number of positive and negative charges on the dust grains in size bin $j$ per volume, respectively.  The neutral moment $Z_{{\rm m},j}$ is defined in such a way that it becomes zero when all grains have maximum charge $q_{\rm max}$ or $-q_{\rm max}$, and stays positive for any other charge distribution function $f_j(q)$.  The charge distribution function is normalized to one, i.e.~$\sum f_j(q)=1$.
According to Eq.\,(\ref{eqn:Zneutral}), we can identify a constant quantity,
\begin{equation}
  \epsilon_{Z_j} = \frac{q_{\rm max}\,n_{{\rm d},j}}{\nH}
    = \frac{[Z_{\mathrm{m,j}}^+] + [Z_{\mathrm{m,j}}] + [Z_{\mathrm{m,j}}^-]}{\nH} \ ,
\end{equation}
which we use in {\sc ProDiMo} to define a quasi-element abundance for the dust grains in size bin $j$. Here $\nH$ refers to the total number density of hydrogen nuclei.
The mean charge of the grains in size bin $j$ is
\begin{equation}
  \langle q_j\rangle 
  ~=\sum_{-q_{\rm max}}^{-1}q\,f_j(q) 
    + \sum_{1}^{q_{\rm max}}q\,f_j(q)
  ~=~ \frac{[Z_{{\rm m},j}^+] - [Z_{{\rm m},j}^-]}{n_{{\rm d},j}}
\end{equation}

\begin{table*}[!ht]
    \centering
    \caption{Glossary of symbols with units used in this paper.}
    \label{tab:glossar}
    \resizebox{16cm}{!}{
    \begin{tabular}{|c|c|l|}
        \hline
        symbol & unit & explanation\\
        \hline
        &&\\[-2.2ex]
         $n_{\rm d}$ & $\rm cm^{-3}$
            & total dust number density \\
         $a$ & cm
            & radius of a dust grain\\
         $f(a)$ & $\rm cm^{-1}$
            & dust size distribution function, normalised by $\int f(a)\,da = 1$\\
         $n_{{\rm d},j}$ & $\rm cm^{-3}$
            & dust number density in size bin $j$\\
         $\bar{a}_j$ & cm
            & mean radius of dust grains in size bin $j$\\
        \hline    
        &&\\[-2.2ex]
         $M$ &
            & representation of a (charged) molecule\\
         $\ce{Z}$, $\ce{Z}^+$, $\ce{Z}^-$ & 
            & representation of positively charged, neutral, and negatively charged dust grains \\
         $[Z]$ & $\rm cm^{-3}$
            & volume concentration of a species, here of the neutral grains\\ 
        \hline
        &&\\[-2.2ex]
         $q$& $-$ 
            & charge of a grain in elementary charges\\
         $q_{\rm max}$& $-$
            & maximum amount of positive/negative charges on a grain\\
         $f_j(q)$ & $-$
           & discrete charge distribution function in size bin $j$, normalised by $\sum_q f_j(q) = 1$\\
         $Q$& $\rm cm^{-3}$ 
            & total amount of charges on all dust grains per volume, $\sum_j n_{{\rm d},j} \sum_q q\,f_j(q) $\\
         $\langle q_j\rangle$ & $-$
            & mean charge of dust grains in size bin $j$, $\sum_q q f_j(q)$\\
        \hline
        &&\\[-2.2ex]
         $k_{\rm ph}(q)$ & $\rm s^{-1}$
            & rate coefficient for photoionisation\\
         $k_{\rm e}(q)$ & $\rm s^{-1} cm^{-3}$ 
            & rate coefficient for electron attachment\\
         $k^M_{\rm ex}(q)$& $\rm s^{-1} cm^{-3}$ 
            & rate coefficient for charge exchange with molecule $M$\\
        \hline
        &&\\[-2.2ex]
         $Z_{{\rm m},j}^+$, $Z_{{\rm m},j}$, $Z_{{\rm m},j}^-$ & 
            & representation of the dust charge moments in size bin $j$\\
         $[Z_{{\rm m},j}]$ & $\rm cm^{-3}$
            & volume concentration of a dust charge moment in size bin $j$, here the neutral moment\\
        \hline
        &&\\[-2.2ex]
         $k_{{\rm m},j,{\rm ph}}$ & $\rm s^{-1}$
            & rate coefficient for photoionisation of neutral moment $Z_{{\rm m},j}$\\
         $k_{{\rm m},j,{\rm ph}}^-$ & $\rm s^{-1}$ 
            & rate coefficient for photoionisation of negative moment $Z^-_{{\rm m},j}$\\         
         $k_{{\rm m},j,{\rm e}}$ & $\rm s^{-1} cm^{-3}$
            & rate coefficient for electron attachment to neutral moment $Z_{{\rm m},j}$\\
         $k_{{\rm m},j,{\rm e}}^+$ & $\rm s^{-1} cm^{-3}$
            & rate coefficient for electron attachment to positive moment $Z^+_{{\rm m},j}$\\
         $k^M_{{\rm m},j,{\rm ex}}$ & $\rm s^{-1} cm^{-3}$
            & rate coefficient of charge exchange between neutral moment $Z_{{\rm m},j}$ and molecule $M$\\ 
         $k^{M,-}_{{\rm m},j,{\rm ex}}$& $\rm s^{-1} cm^{-3}$
            & rate coefficient of charge exchange between negative moment $Z^-_{{\rm m},j}$ and molecule $M$\\
         \hline
    \end{tabular}}\\
\end{table*}

\subsection{Solving For The Charge Distribution Function}
\label{sec:linearChain}

\begin{figure}
    \centering
    \includegraphics[height=2cm]{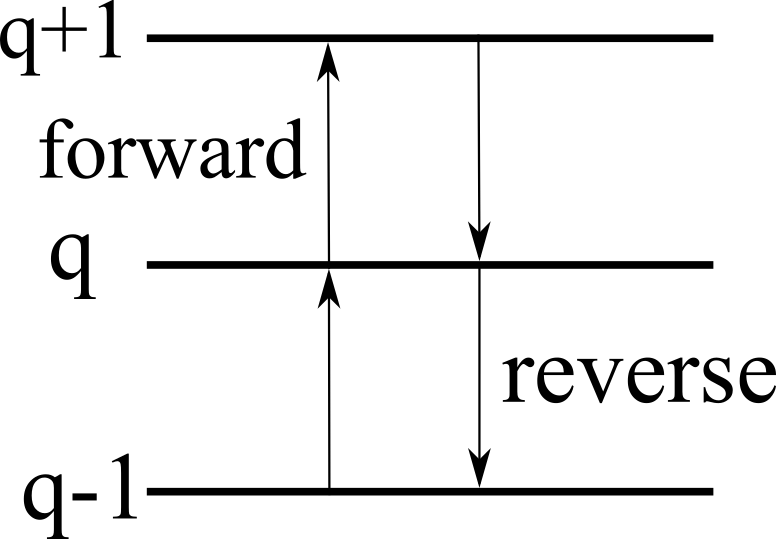}
    \caption{Sketch of the linear chain of reactions that populate and depopulate the charge states $q$. A forward reaction adds a positive charge, a reverse reaction adds a negative charge.}
    \label{fig:linearchain}
\end{figure}

To determine the discrete charge distribution function $f_j(q)$ for all charge states $q$ in every size bin $j$, we employ a method called linear chains of reactions. We note that we assume the charge distribution function to be in a steady state. The general idea is that in our model, the charge states $q$ of a grain can only change by one increment, either forward to $q+1$ or reverse to $q-1$. This is illustrated in Fig.~\ref{fig:linearchain}.

The processes that add a charge, $q\!\to\!q+1$, are photoionization (called photoejection for neutral grains and photodetachment for negative grains) and charge exchange between a dust grain and a molecular cation. The forward rates $[1/s]$ are calculated in the following way,
\begin{equation}
    \text{forward}_j(q) ~=~ k_{j,{\rm ph}}(q) ~+ \!\!\!\sum_{\text{cations}\,M} 
    \!\!\! n_M\,k^M_{j,{\rm ex}}(q) \ .
\end{equation}
Here, $n_M$ is the volume density of the molecule $M$ that interacts in a charge exchange reaction with the dust grain $Z$. 

Processes that decrease the charge, $q+1\!\to q$, include the recombination with electrons and charge exchange reactions with molecular anions. The reverse rates are calculated as
\begin{equation}
    \text{reverse}_j(q) ~=~ n_e\,k_{j,{\rm e}}(q) ~+ \!\!\!\sum_{\text{anions}\,M}
    \!\!\! n_{M}\,k^M_{j,{\rm ex}}(q) \ .   
\end{equation}
The charge distribution function $f_j(q)$ can now be determined, in each size bin separately, in a step-down approach. We put $f_j(q_{\mathrm{max}})=1$ and perform a sequence of steps, where we calculate $f_j(q)$ from $f_j(q\!+\!1)$, until $q\!=\!-q_{\rm max}$ is reached. During each step, we use
\begin{equation}
    f_j(q) ~=~ f_j(q\!+\!1) \, \frac{\text{reverse}_j(q\!+\!1)}{\text{forward}_j(q)} \ .
\end{equation}
Eventually, we calculate the normalization constant as
\begin{equation}
    f_{j,{\rm norm}} = \sum_{-q_{\mathrm{max}}}^{q_{\mathrm{max}}} f_j(q)
\end{equation}
and normalize as $f_j(q)\!\to\!f_j(q)/f_{j,{\rm norm}}$.

\begin{figure*}
        \includegraphics[width=.49\textwidth]{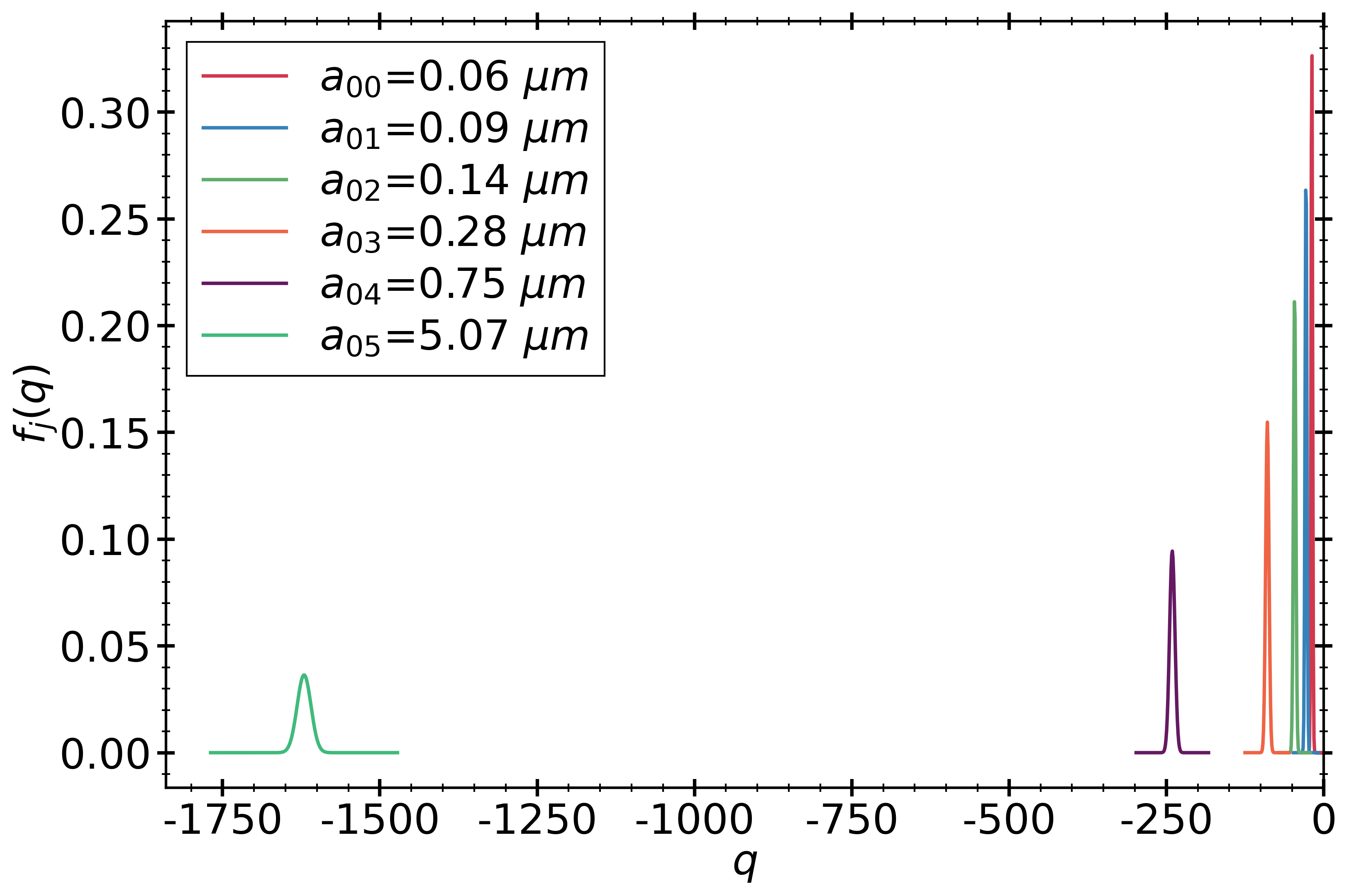}
        \includegraphics[width=.49\textwidth]{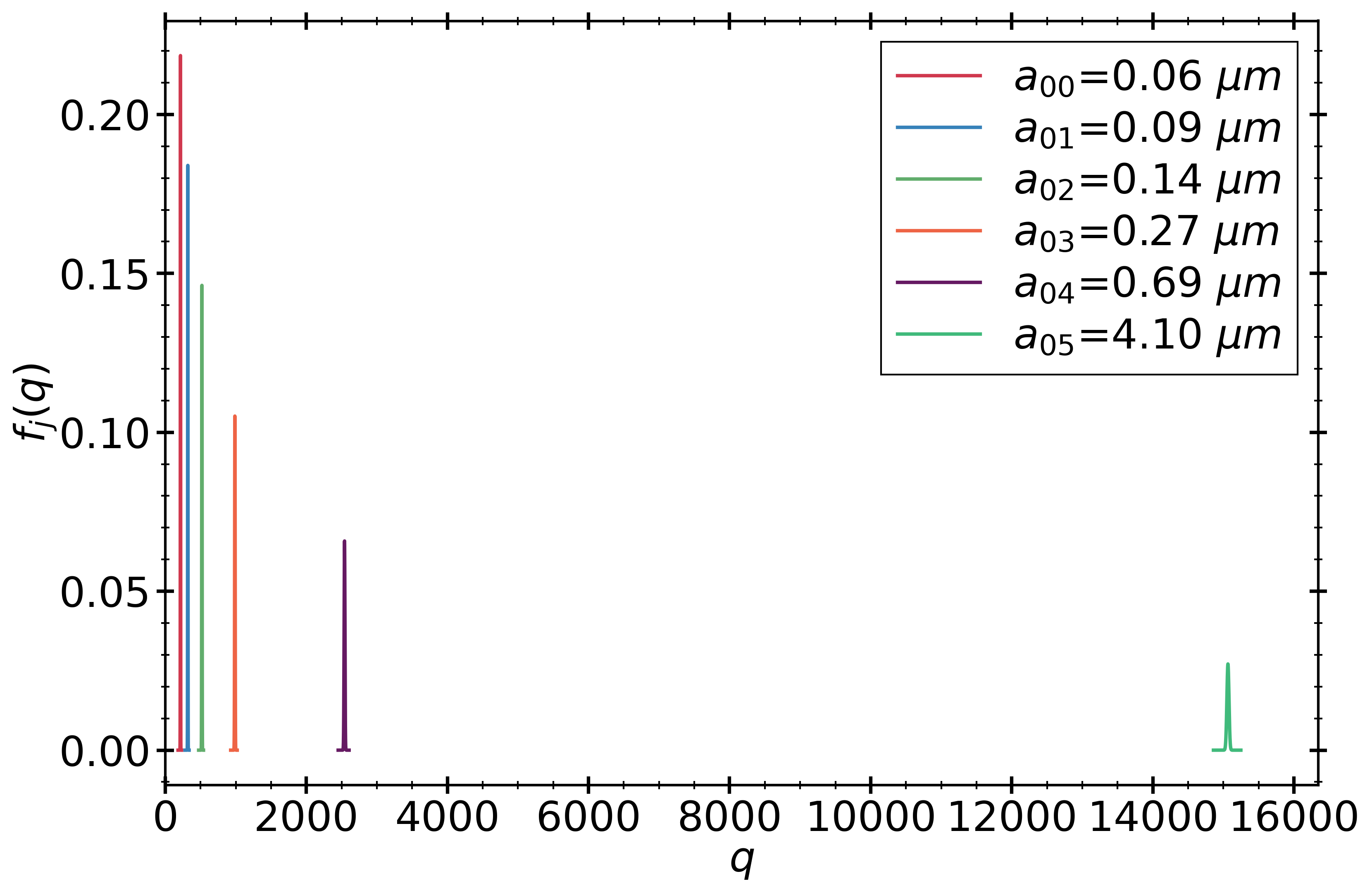}
        \includegraphics[width=.49\textwidth]{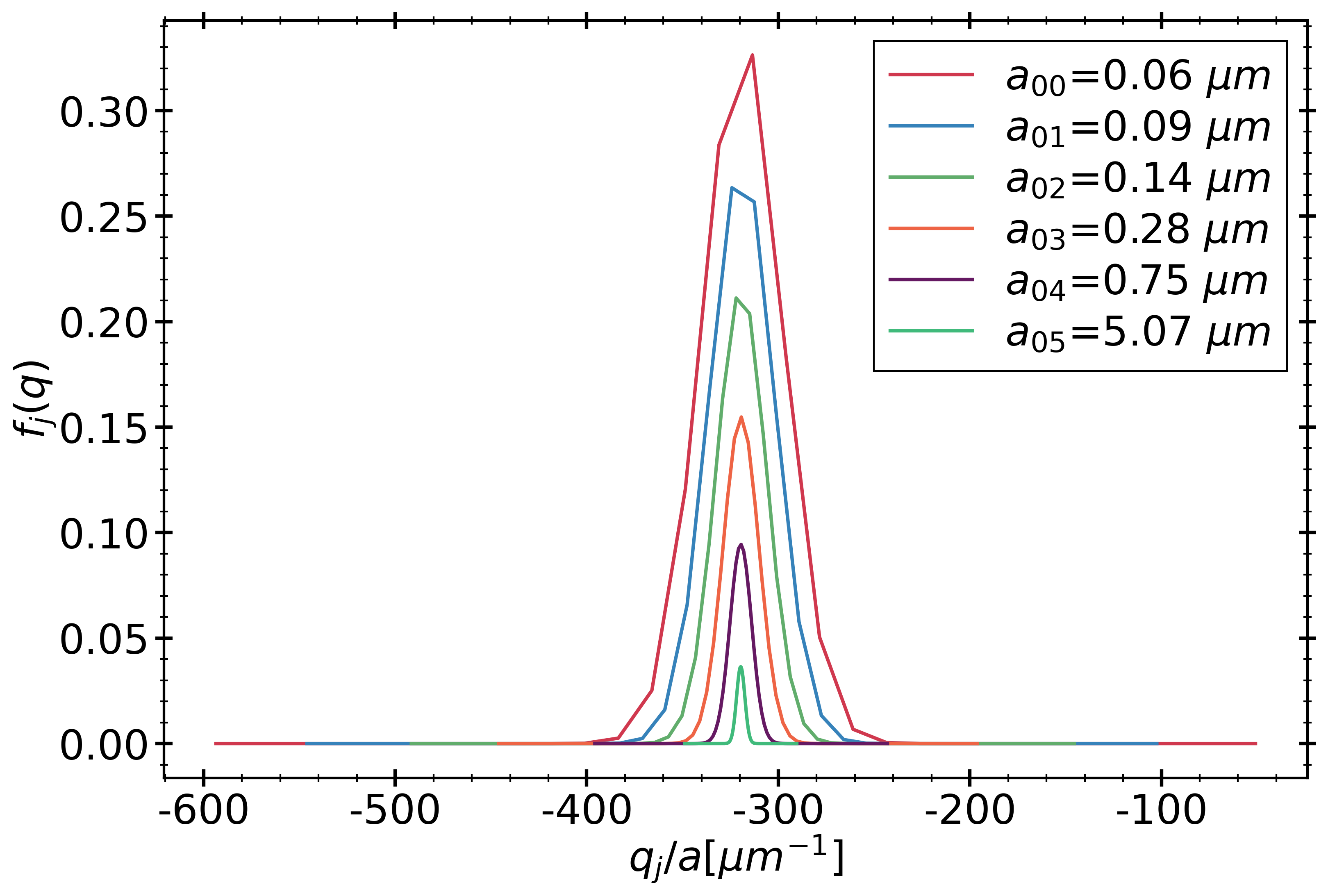}
        \includegraphics[width=.49\textwidth]{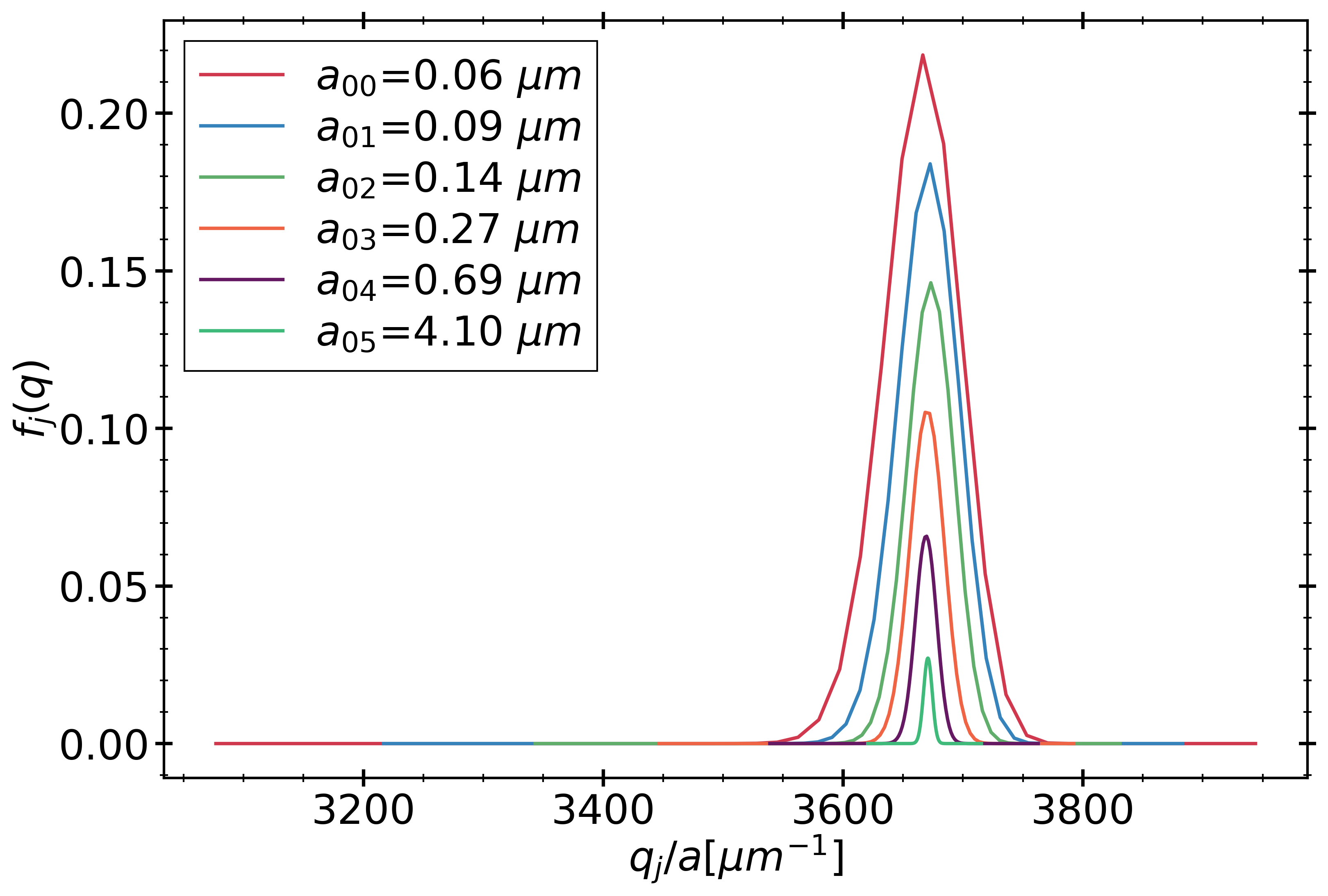}
        \caption{Overview of the charge distribution function $f_j(q)$. The two upper panels show the charge distribution function in relation to the charge held by each dust grain bin $q$. The lower panels also show the charge distribution function but in relation to total charge per dust grain bin size $Q/a$. The left two panels show the result at $r=0.1$ au and in the midplane at $z/r=0$ for each of the six different bins. The right two panels show the results also at $r=0.1$, but higher up in the disk at, $z/r=0.3$ Additionally we show the size of each bin represented by $a$.}
        \label{fig:f_q}
\end{figure*}

The results only depend on the UV to optical photon fluxes, the electron density $n_e$ and the molecular ion densities $M$.  In Fig.~\ref{fig:f_q} we show the resulting $f_j(q)$ for six dust size bins at two positions in the disc. The left side of the figure shows a point in the midplane at $r=0.1$\,au and $z=0$, and the right side a point in the upper areas of the disk with also $r=0.1$ au but $z=0.29$ au.

In the midplane, all grains charge up negatively, because there are no UV-photons here, and the rate coefficients for electron attachment are larger than the rate coefficients for charge exchange with molecular ions. In addition, grains charge up negatively, because free electrons have a higher mean thermal speed than molecular ions, i.e., $\sqrt{\frac{8 \rm k_b T}{\pi m_e}} \gg \sqrt{\frac{8 \rm k_b T}{\pi m_M}}.$ The lower part figure shows that the resulting equilibrium charge scales about linear with grain size, because it is the electric potential $q/a$ that enters into the rate coefficients.  For example, all grains will continue to charge up negatively, until the rate for electron attachment, with decreases exponentially with $q/a$, balances the rates for charge exchange with molecular ions, which also depend on $q/a$. As a result, we get similar $q/a$ values for all grain sizes.This behavior can also be seen and has been discussed in \citep{2009ApJ...698.1122O}. There they showed that the shape of the charge distribution function is approximately Gaussian, with its width scaling with the square root of the grain size (compare our Fig. \ref{fig:f_q}).

\begin{figure}[h!]
    \centering
    \includegraphics[width=.49\textwidth]{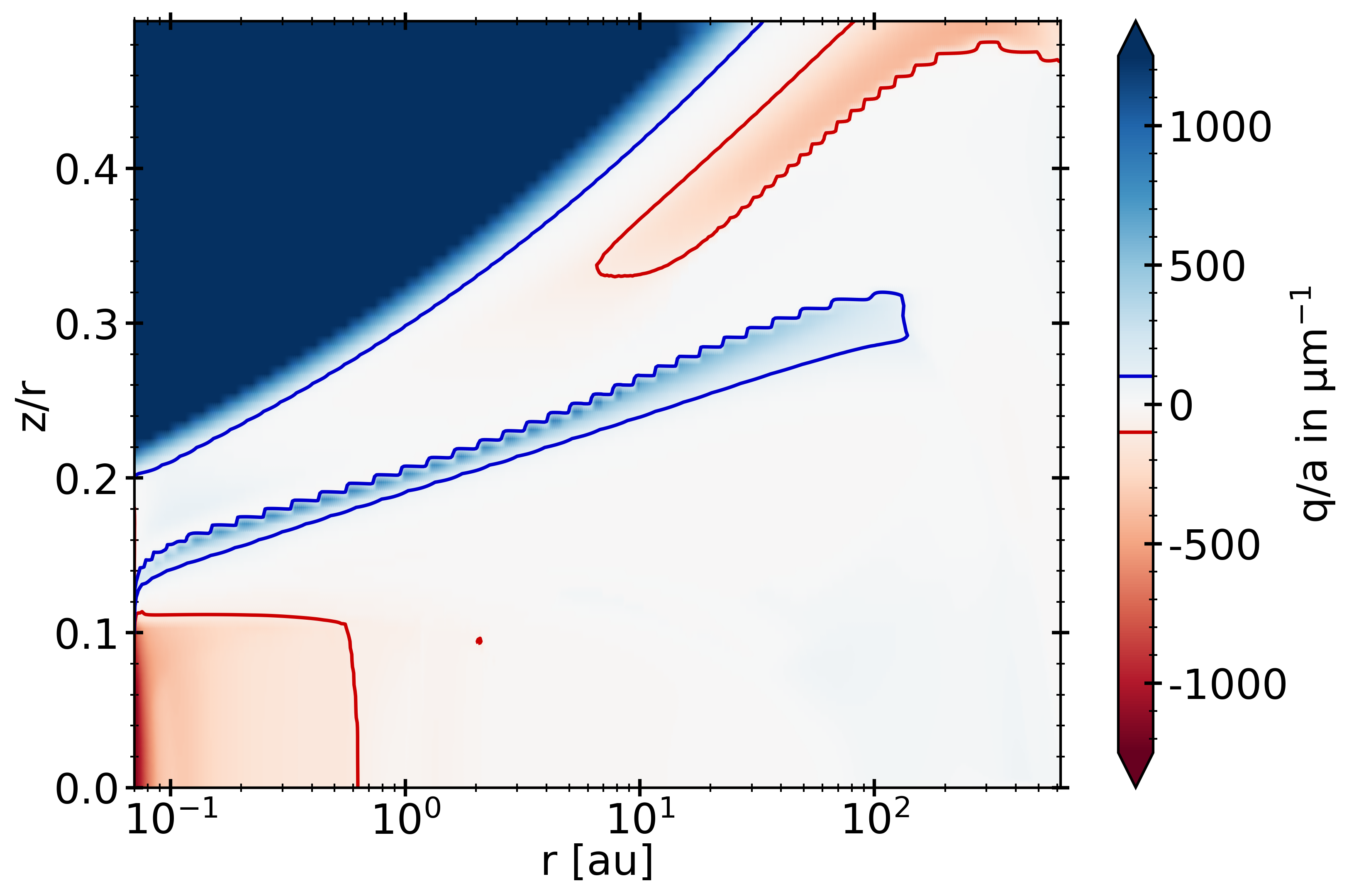}
    \caption{This plot illustrates how the charge per dust grain radius (q/a) [$\mu m ^{-1}$] changes within the whole disk. For large areas of the disk, the dust is mostly neutral. In the upper areas that are largely affected by photodissociation we find that dust charges very positively (> 100, blue contour lines). In the areas of the midplane closest to the star, we find that the dust charges very negatively (< -100, red contour lines).}
    \label{fig:Z_over_a_avg}
\end{figure}

An overview of the behavior of the charge distribution function for the whole simulation, instead of examples at single points in the simulation (as seen in Fig \ref{fig:f_q}), can be seen in Fig \ref{fig:Z_over_a_avg}.

\subsection{Effective Rates In The Chemical Rate Network}
\label{subsec:moment_rate_coeff}

In order to solve the chemical network with the dust charge moments as species, as introduced in Sect.~\ref{subsec:moments}, we need to derive the effective rate coefficients for the moments.  We will only demonstrate the derivations for the most relevant processes here.  For a full description of all reactions that concern dust grains and their moment representation, we refer to Appendix~C of \citet{Thi2019}. For photoejection and photodetachment

\begin{align}
    Z_{{\rm m},j} + h\nu &~\rightarrow~ Z_{{\rm m},j}^+ + e^- \\
    Z_{{\rm m},j}^- + h\nu &~\rightarrow~ Z_{{\rm m},j} + e^-
\end{align}
we have
\begin{align}
    \frac{d[Z_{{\rm m},j}^+]}{dt} &=~
    k_{{\rm m},j,{\rm ph}}\,[Z_{{\rm m},j}] ~=~
    n_{{\rm d},j} \sum_{0}^{q_{\rm max}-1} k_{j,{\rm ph}}(q)\,f_j(q) 
    \nonumber\\
    \frac{d[Z_{{\rm m},j}]}{dt} &=~
    k_{{\rm m},j,{\rm ph}}^-\,[Z_{{\rm m},j}^-] ~=~
    n_{{\rm d},j} \sum_{-q_{\rm max}}^{_1} k_{j,{\rm ph}}(q)\,f_j(q)
    \nonumber
\end{align}
We note that these effective rates represent all photo-reactions from charge state ($q$) to charge state ($q+1$), not just for the neutral and negatively charged grains. To gain the effective rate coefficient, we divide by $[Z_{{\rm m},j}]$ and $[Z_{{\rm m},j}^-]$, respectively.  
\begin{align}
    k_{{\rm m},j,{\rm ph}} &=~ 
    \frac{n_{{\rm d},j}}{[Z_{{\rm m},j}]} 
    \sum_0^{q_{\rm max}-1} k_{j,{\rm ph}}(q)\,f_j(q) 
    \label{rate:kph_neu}\\
    k_{{\rm m},j,{\rm ph}}^- &=~
    \frac{n_{{\rm d},j}}{[Z_{{\rm m},j}^-]} 
    \sum_{-q_{\rm max}}^{-1} k_{j,{\rm ph}}(q)\,f_j(q)   
    \label{rate:kph_min}
\end{align}
In the code, once $f_j(q)$ has been determined as explained in Sect.~\ref{sec:linearChain}, we compute $[Z^+_{{\rm m},j}]$, $[Z_{{\rm m},j}]$ and $[Z^-_{{\rm m},j}]$ according to Eqs.~(\ref{eqn:Zplus}) to (\ref{eqn:Zneutral}), and then determine the effective photorates from $f_j(q)$ by Eqs.~(\ref{rate:kph_neu}) and (\ref{rate:kph_min}). The rate coefficients are hence fully determined by the discrete charge distribution functions $f_j(q)$.

For the electron attachment, two new rate coefficients have to be derived. The derivation is analogous to the photoprocesses, but one has to consider the different charge states of the grains. 
\begin{align}
    Z_{{\rm m},j}   + e^- &~\rightarrow~ Z_{{\rm m},j}^- \\
    Z_{{\rm m},j}^+ + e^- &~\rightarrow~ Z_{{\rm m},j} 
\end{align}
As the positively charged grains enhance the electron attachment, we need to describe their behavior with the rate coefficient $k_{\mathrm{m,j,e}}^+$. Neutral grains do not show this kind of enhancement, therefore we define another rate coefficient $k_{\mathrm{m,j,e}}$. We will also include the contribution of negatively charged grains to the rate for the neutral grains, even though their contribution to the overall rate coefficient will be low. For positive charged grains and neutral grains, with the added contribution of negatively charged grains, we can define
\begin{align}
    \frac{d[Z_{{\rm m},j}^-]}{dt} &= 
    k_{{\rm m},j,{\rm e}} [Z_{{\rm m},j}] 
    = n_{{\rm d},j}
    \left(k_{j,{\rm e}}(0)f_j(0) + \sum_{-q_{\rm max}+1}^{-1}
    k_{j,{\rm e}}(q)f_j(q)]\right) 
    \nonumber\\
    \frac{d[Z_{{\rm m},j}]}{dt} &=
    k_{{\rm m},j,{\rm e}}^+[Z_{{\rm m},j}^+]
    = n_{{\rm d},j}
    \sum_1^{q_{\rm max}} 
    k_{j,{\rm e}}(q)f_j(q)
    \nonumber
\end{align}
here $k_{j,{\rm e}}(q)$ are the rate coefficients for the individual charge states defined by Eq.\,(\ref{eqn:rate_coeff_elec}). The effective rate coefficients for electron attachment are, hence,
\begin{align}
    k_{{\rm m},j,{\rm e}} &=
    \frac{n_{{\rm d},j}}{[Z_{{\rm m},j}]}
    \left(k_{j,{\rm e}}(0)f_j(0) 
    +\sum_{-q_{\rm max}+1}^{-1}
    k_{j,{\rm e}}(q)f_j(q)\right)\\
    k_{{\rm m},j,{\rm e}}^+&=
    \frac{n_{{\rm d},j}}{[Z_{{\rm m},j}^+]} 
    \sum_{1}^{q_{\rm max}} 
    k_{j,{\rm e}}(q)f_j(q)
\end{align}
The charge exchange reactions between dust grains and molecules also have to be adjusted. We consider the rates for neutral grains and negative grains separately.
\begin{align}
    &Z_{{\rm m},j}   + M^+ \rightarrow Z_{{\rm m},j}^+ + M\\
    &Z_{{\rm m},j}^- + M^+ \rightarrow Z_{{\rm m},j}   + M
\end{align}
This results in the following charge exchange rates
\begin{equation}
     \frac{d[Z_{{\rm m},j}^+]}{dt} =
    k_{{\rm m},j,{\rm ex}}^M [Z_{{\rm m},j}][M^+]
    =n_{{\rm d},j} [M^+] \sum_{0}^{q_{\rm max}-1}
    k_{j,{\rm ex}}^M(q) f_j(q)
\end{equation}
\begin{equation}
    \frac{d[Z_{{\rm m},j}]}{dt} =
    k_{{\rm m},j,{\rm ex}}^{M,-} [Z_{{\rm m},j}^-][M^+]
    =n_{{\rm d},j} [M^+] \sum_{-q_{\rm max}}^{-1}
    k_{j,{\rm ex}}^M(q) f_j(q)
\end{equation}
from which we derive the rate coefficients for charge exchange
with molecule $M^+$ 
\begin{align}
    k_{{\rm m},j,{\rm ex}}^M
    &=\frac{n_{{\rm d},j}}{[Z_{{\rm m},j}]}
    \sum_{0}^{q_{\rm max}-1}
    k_{j,{\rm ex}}^M(q) f_j(q)\\
    k_{{\rm m},j,{\rm ex}}^{M,-}
    &=\frac{n_{{\rm d},j}}{[Z_{{\rm m},j}^-]}
    \sum_{-q_{\rm max}}^{-1}
    k_{j,{\rm ex}}^M(q) f_j(q) \ .
\end{align}
With these effective rate coefficients for the dust charge moments, we can now either solve our rate network system for the time-independent solution, or can advance the ordinary system of first-order differential equations (ODE-system) in time from an initial vector of particle densities.
Appendix~\ref{sec:solve} explains how to deal with the problem that, because $f_j(q)$ depends on the electron and molecular ion densities, the rate coefficients depend on particle densities.

\subsection{Cosmic Ray Implementation}\label{subsec:CR}
As explained in Sect.~\ref{sec:results}, we expect lightning to occur, if at all, in the dense and shielded midplane regions of the disk close to the star, where cosmic rays are the most relevant source for ionization. It is therefore important to use a model for the cosmic ray penetration that is as realistic as possible.  

We use the description of cosmic ray attenuation and ionization following \citet{2009A&A...501..619P}, as implemented by \citet{2017A&A...603A..96R,2018A&A...609A..91R}. According to this model, the H$_2$ cosmic ray ionization rate $\zeta_{\mathrm{cr}}$ $s^{-1}$ is given by
\begin{equation}
    \zeta_{\mathrm{cr}} = \left\{\begin{array}{ll}
    \zeta_{\mathrm{low}} 
    & \mbox{, \small $\NH\!<\!10^{19}\,$cm$^{-2}$}\\
    \displaystyle
    \frac{\zeta_{\mathrm{low}}\,\zeta_{\mathrm{high}}} {\zeta_{\mathrm{high}}
    \left(\frac{\NH}{10^{20}{\rm cm^{-2}}}\right)^{\rm a} 
    +\zeta_{\mathrm{low}}
    \left(\exp\frac{\Sigma}{\Sigma_0}-1\right)}\hspace*{-3mm}
    & \mbox{, \small otherwise}
    \end{array}\right.\hspace*{-3mm}
    \label{eq:CR}
\end{equation}
where $\NH$ [$\rm cm^{-2}$] is the vertical hydrogen nuclei column density and $\Sigma\!\approx\!(1.4\,{\rm amu})\,\NH$ [$\rm g/cm^2$] the vertical mass column density.  $\zeta_{\mathrm{low}}$, $\zeta_{\mathrm{high}}$, $\Sigma_0$ and $a$ are fitting parameters. These fitting parameters originate from the fitting of  cosmic ray spectra  and account for different attenuation's at different column densities. $\zeta_{low}$ is used to fit the spectra for low column densities between smaller than $10^{19}\; \mathrm{cm}^{-2}$  where the attenuation can be described as a power law, and $\zeta_{\mathrm{high}}$ is used for column densities higher than $10^{19}\; \mathrm{cm}^{-2}$ where the attenuation can be described with an exponential term. Additionally Eq. \ref{eq:CR} is applied twice in our models, once to account for cosmic rays impacting the disk from above and once to account for cosmic rays impacting the disk from below, by adjusting $\NH=2\NH(r,0)-\NH (r,z)$ where $\NH(r)$ is the total vertical hydrogen nuclei column density at point r and $\NH(r,z)$ the local hydrogen nuclei column density at the point $(r,z)$ for which the cosmic ray ionization rate has to be determined. In our main simulation, we made use of the cosmic ray spectra named Solar Max in \cite{2013ApJ...772....5C}. We do not include stellar energetic particles (SEPs) in this work.

\section{Resulting dust charge and ionization properties}
\label{sec:results}

This section summarizes our results concerning the mean charge of the dust grains, how the grain charge distribution function depends on the position in the disk and on dust size, and how the dust charge is linked to the degree of ionization and molecular ions in the gas, with particular emphasis on the midplane regions. We first illustrate our standard case, where we only generate exchange reactions but no new protonation reactions and only consider exothermic reactions. A full discussion can be found in Appendix \ref{sec:charge_exchange} These results will later be used to discuss whether charge separation and lightning can occur in disks, see Sect.~\ref{sec:ChargeSep}.

\begin{table}[]
    \caption{Parameters of the {\sc ProDiMo} disk model used in this paper.} 
    \label{tab:parameters}
    \vspace*{-3mm}
    \resizebox{90mm}{!}{\begin{tabular}{|c|c|c|}
        \hline
        \multicolumn{3}{|c|}{}\\*[-2.1ex]
        \multicolumn{3}{|c|}{stellar and irradiation parameters}\\
        \hline 
        &&\\*[-2.1ex]
        stellar mass & $M_*$ & $0.7\,M_{\odot}$\\
        stellar luminosity & $L_*$ & $1\,L_{\odot}$ \\
        eff. temperature & $T_*$ & 4000\,K \\
        excess UV luminosity\,$/L_\star$ & $f_{\mathrm{UV}}$ & 0.01\\
        UV power law exponent & $p_{\mathrm{UV}}$ & 1.3 \\
        X ray luminosity & $L_{\mathrm{Xray}}$ & $1(+30)\,\rm erg\,s^{-1}$ \\
        X ray emission temperature & $T_{\mathrm{Xray}}$ & $2(+7)$\,K\\
        interstellar UV field strength & $\chi_{\mathrm{ISM}}$ & 1 \\
        fitting parameter for $\zeta_{\mathrm{cr}}$ & $\zeta_{\mathrm{low}}$ & $2(-19)\,\rm s^{-1}$\\
        fitting parameter for $\zeta_{\mathrm{cr}}$ & $\zeta_{\mathrm{high}}$ & $8(-19)\,\rm s^{-1}$\\
        fitting parameter for $\zeta_{\mathrm{cr}}$ & $\Sigma_0$ &  230 $\rm cm^{-2}$\\
        fitting parameter for $\zeta_{\mathrm{cr}}$ & a &  $-0.01$\\
        \hline
        \multicolumn{3}{|c|}{}\\*[-2.1ex]
        \multicolumn{3}{|c|}{disk mass and shape parameters}\\
        \hline
        &&\\[-2.1ex]
        disk mass & $M_{\mathrm{disk}}$ & 0.01 $M_{\odot}$ \\ 
        inner radius & $R_{\mathrm{in}}$ & 0.07 au \\ 
        %12 & Outer Radius & $R_{\mathrm{out}}$ & 600 au \\
        tapering-off radius\,$^{(1)}$ & $\Rtap$ & 100 au \\
        column density exponent & $\epsilon$ & 1.0 \\ 
        reference radius & $r_0$ & 100 au \\ 
        scale height at $r_0$ & $H_0$ & 10.0 au \\ 
        flaring power & $\beta$ & 1.15 \\ 
        \hline
        \multicolumn{3}{|c|}{}\\*[-2.1ex]
        \multicolumn{3}{|c|}{dust material and size parameters}\\
        \hline
        &&\\[-2.1ex]
        dust to gas ratio & dust to gas & 0.01 \\ 
        dust composition & $\mathrm{Mg}_{0.7}\mathrm{Fe}_{0.3}\mathrm{SiO}_3$ & 60\% \\
        & amorph. graphite & 15\% \\
        & porosity & 25\% \\
        dust mass density & $\rho_{\mathrm{gr}}$ & 2.094 $\mathrm{g}/\mathrm{cm}^3$ \\ 
        min.~dust radius & $a_{\mathrm{min}}$ & $0.05\rm\,\mu$m \\ 
        max.~dust radius & $a_{\mathrm{max}}$ & $3\rm\,$mm \\ 
        dust size dist.~exponent & $a_{\rm pow}$ & 3.5 \\ 
        no.~of dust size dist.~points & $\Nsize$ & 100 \\
        dust settling parameter & $\alpha_{\mathrm{settle}}$ & 0.01 \\ 
        \hline
        \multicolumn{3}{|c|}{}\\*[-2.1ex]
        \multicolumn{3}{|c|}{dust parameters for the chemistry}\\
        \hline
        no.~of dust size bins & $\Nbin$ & 6 \\
        dust bin fitting indices & $\kappa$, $\zeta$ & 1.0, 2.0 \\
        max.~dust charge per radius & $q_{\rm max}/a$ & 4000 $\mu{\rm m}^{-1}$ \\ 
        \hline
        \multicolumn{3}{|c|}{}\\*[-2.1ex]
        \multicolumn{3}{|c|}{simulation parameters}\\
        \hline
        &&\\[-2.1ex]
        Grid points in $r$-direction & NXX & 500 \\ 
        Grid points in $z$-direction & NZZ & 100 \\ 
        chemical heating efficiency & $\gamma^{\rm chem}$ & 0.2 \\
        \hline
    \end{tabular}}\\*[1mm]
    \footnotesize
    Notation $a(b)$ means $a\times 10^b$.
    
    $^{(1)}$: The outer radius $R_{\rm out}\!\approx\!620\,$au is adjusted automatically to obtain a vertical hydrogen nuclei column density of $N_{\langle\rm H\rangle}(R_{\rm out})=10^{\,20}\rm\,cm^{-2}$. 
\end{table}

\subsection{Electron concentration}

Figure \ref{fig:elec_dens} shows the resulting electron concentration $n_{\rm e}/\nH$ as function of position $(r,z)$ in our disk model. 
On the basis of these results, we introduce 
six different disk regions A-F, see Fig.~\ref{fig:elec_dens}, where the physical and chemical processes leading to ionization and dust charge are qualitatively different in each case.

\subsection{Regions F, E and D --- the plasma regions}\label{subsec:photons}

\begin{figure}
    \hspace*{-2mm}
    \includegraphics[width=.5\textwidth]{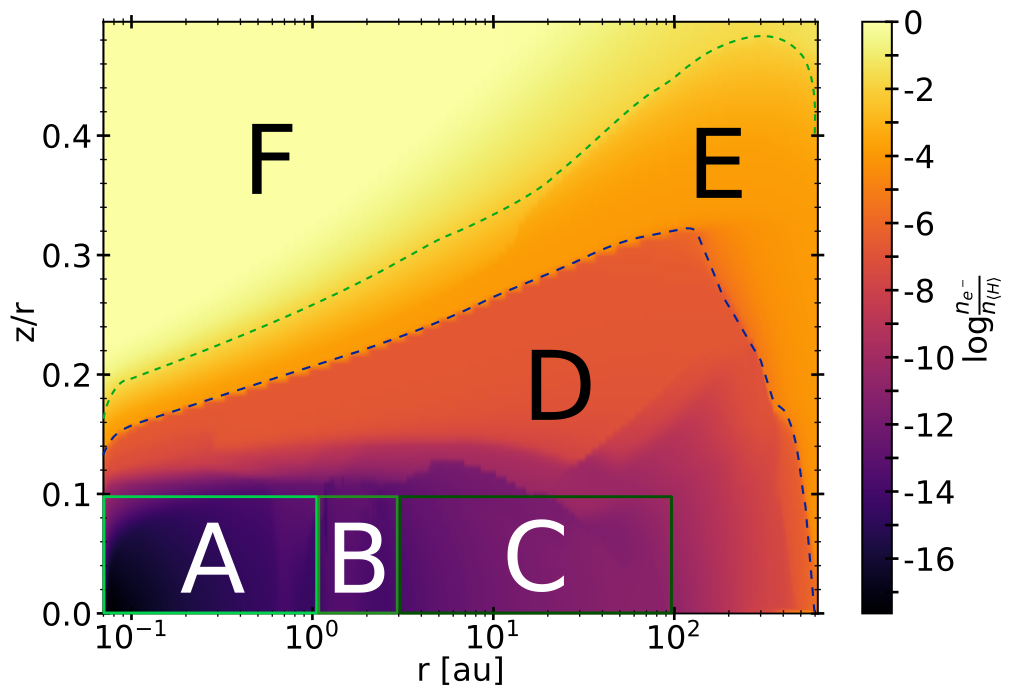}\\*[-5mm]
    \caption{Electron concentration $n_{\rm e}/\nH$ as function of position in the disk.  We highlight areas A-F, where the character of the chemical processes leading to grain charge and gas ionization are different, see text.}

    \label{fig:elec_dens}
\end{figure}

Regions D-F in Fig.~\ref{fig:elec_dens} are radiation dominated regions, where the UV and X-ray photons create a large degree of ionization along with positive dust charges.  We find that region~F is dominated by the ionization of H and \ce{H2}, region~E by the ionization of atomic carbon and region~D by the ionization of atomic sulfur. According to the assumed element abundances of H, C and S in our disk model, the degree of ionization in regions F, E and D is about 1, $10^{-4}$ and $10^{-7}$, respectively. 
The grains charge up positively via photo-effect, $\ce{Z} + h\nu \to \ce{Z+} + \ce{e-}$, and their charge is balanced by electron recombination $\ce{Z+} + \ce{e-} \to \ce{Z}$. The total number of charges on the grains, however, is insignificant in comparison to the number of free electrons and positive ions in the gas phase. Any dynamical displacement of the charged grains would easily be balanced out by slight motions of the free electrons in the gas, making charge separations very unlikely to occur in these plasma regions, see further discussion in Sect.~\ref{sec:ChargeSep}. 

\begin{figure}
    \hspace*{-2mm}
    \includegraphics[width=.5\textwidth]{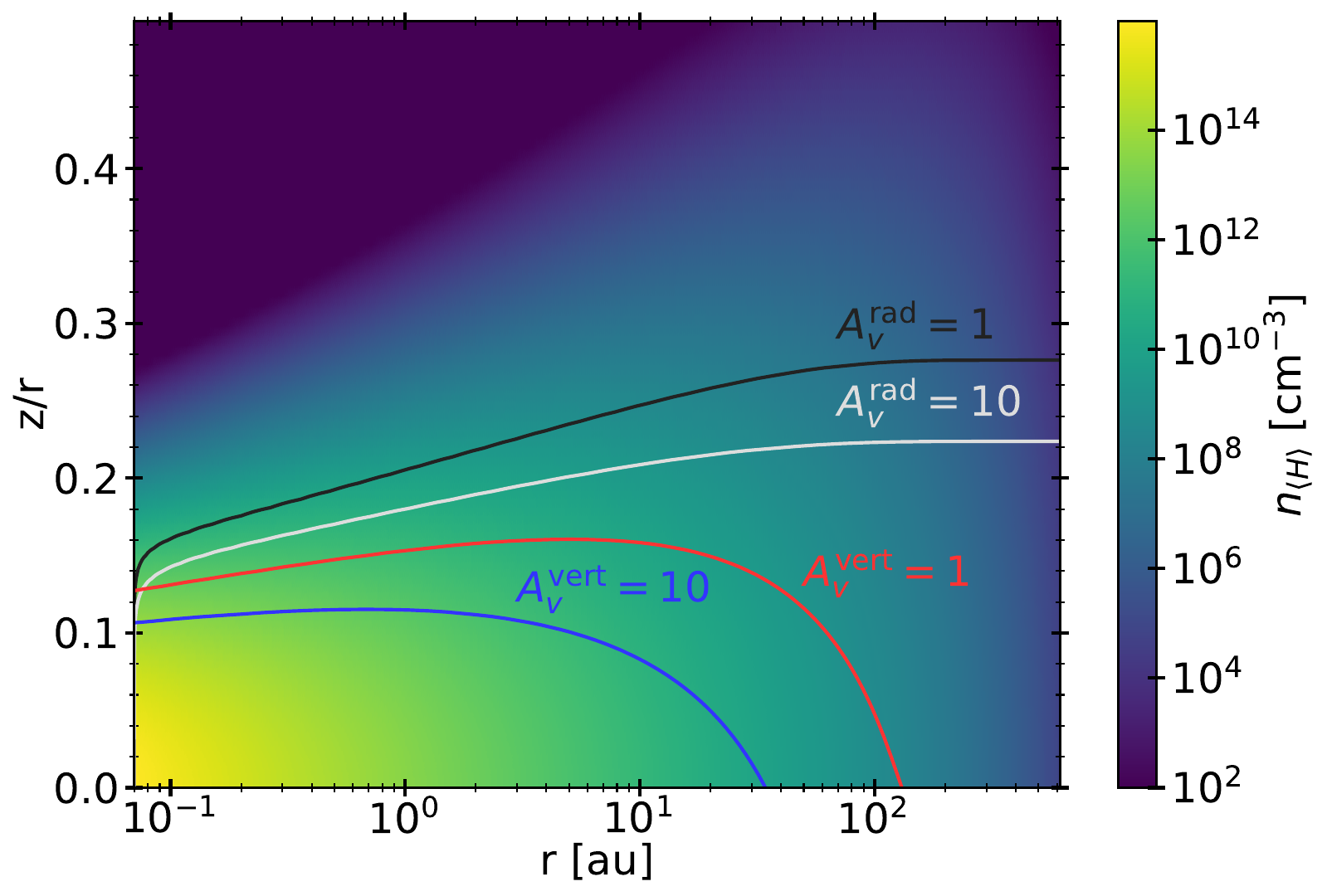}\\*[-5mm]
    \caption{Hydrogen nuclei density $\nH\,[\rm cm^{-3}]$ as function of radius $r$ and height over the midplane $z$ in our disk model. The different colored lines represent the visual extinctions in the radial direction (white and black) measured from the star outwards and vertical direction (red and blue) measured from the surface of the disk towards the midplane.}
    \label{fig:dens}
\end{figure}
\begin{figure}
    \hspace*{-2mm}
    \includegraphics[width=.5\textwidth]{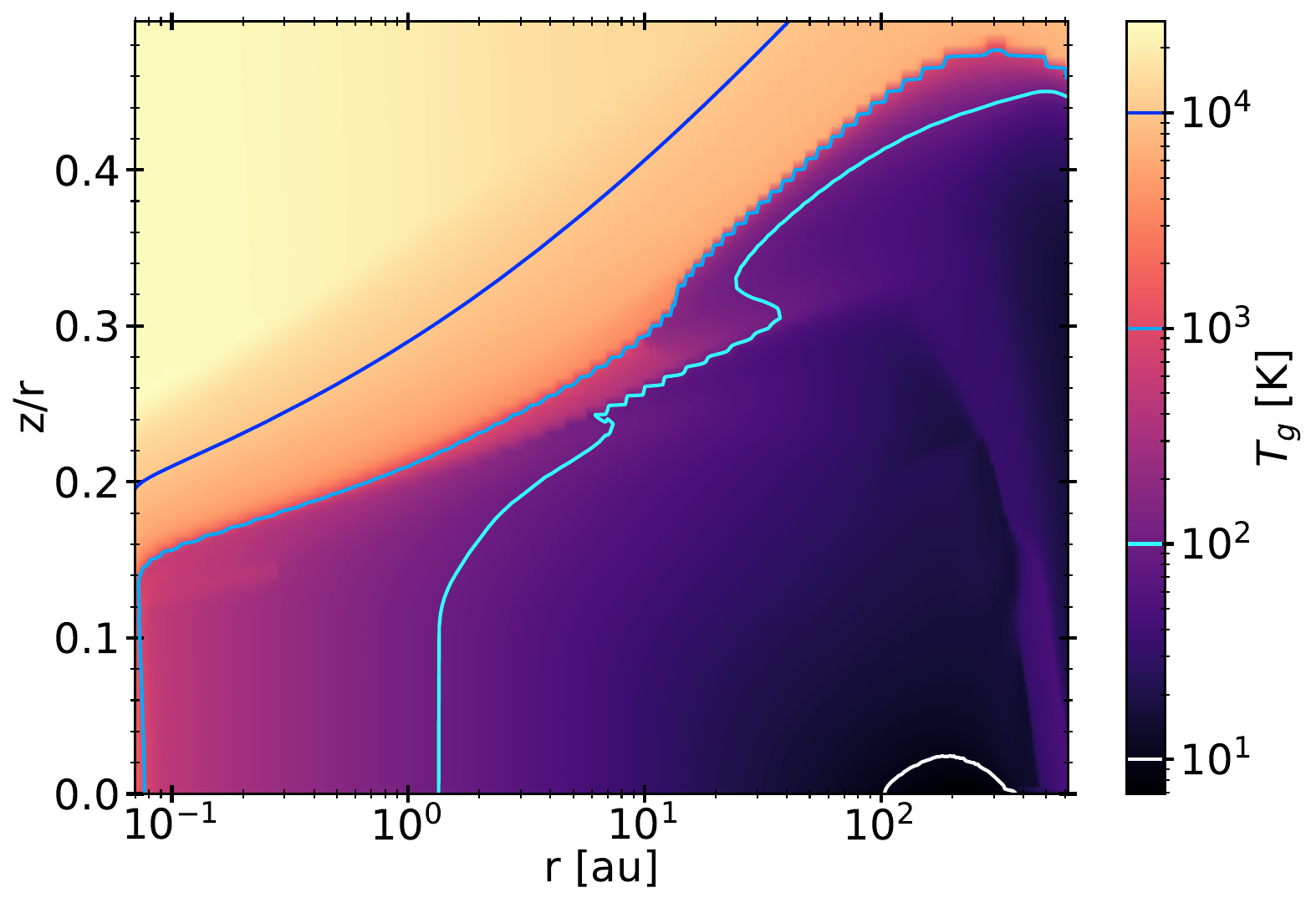}\\*[-5mm]
    \caption{Calculated gas temperature structure in the disk model $T_{\rm g}(r,z)$. Colored lines show different orders of magnitude in Kelvin.}
    \label{fig:temp}
\end{figure}

\subsection{Region A --- the dust dominated region}\label{subsec:dustregime}

We define region~A as the disk region where the number of negative charges on the dust grains is balanced by the abundance of molecular cations.  Free electrons are very rare in this region (degree of ionization $10^{-18}\,...\,10^{-14}$) and unimportant for the charge balance.  In our disk model, region~A is the area that stretches out from just behind the inner rim to about $r\!\approx\!1$\,au, and $z/r\la 0.1$. The outer boundary coincides with the location where water and ammonia ice emerge (snowline), and the upper boundary is roughly given by a vertical visual extinction of $A_V^{\rm ver}\!=\!10$, which makes sure that UV photons cannot penetrate into region~A.

\begin{figure}
    \hspace*{-2mm}
    \includegraphics[width=.5\textwidth]{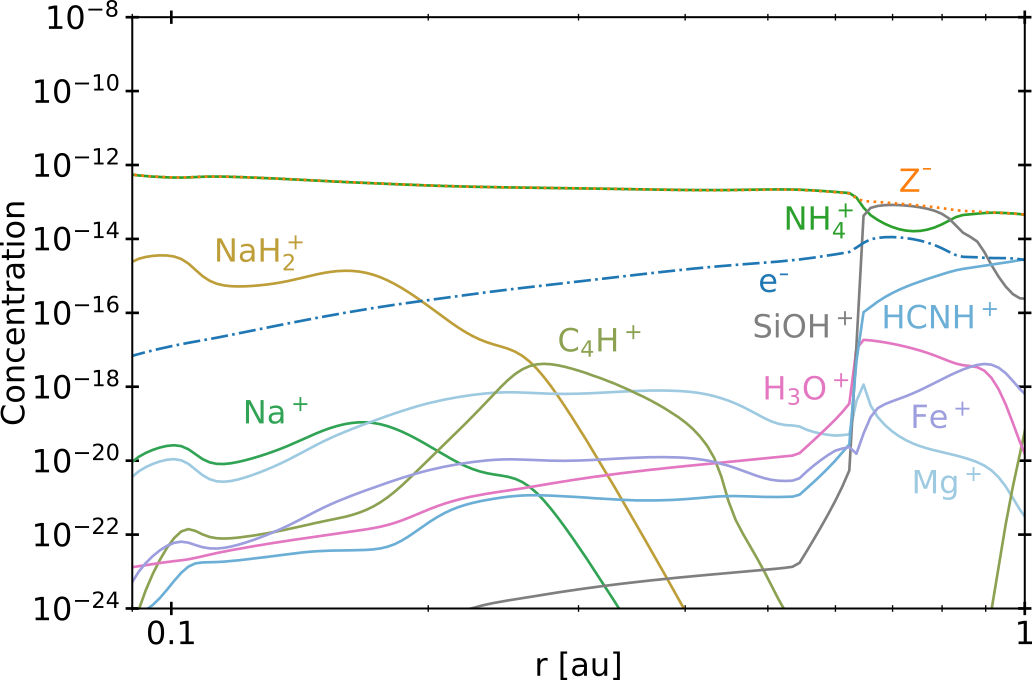}\\*[-5mm]
    \caption{Concentrations $n_i/\nH$ of selected chemical species important for the charge balance in the midplane in region~A. 
    The dotted blue line represents the concentration of negative charges on all dust grains $Z^-=\sum_j[Z^-_{{\rm m},j}]$. The negative charges have accumulated on grain surfaces, whereas free electrons e$^-$ are less important here. }
    \label{fig:species_low_r}

\end{figure}

\begin{figure}
    \centering
    \includegraphics[width=.5\textwidth]{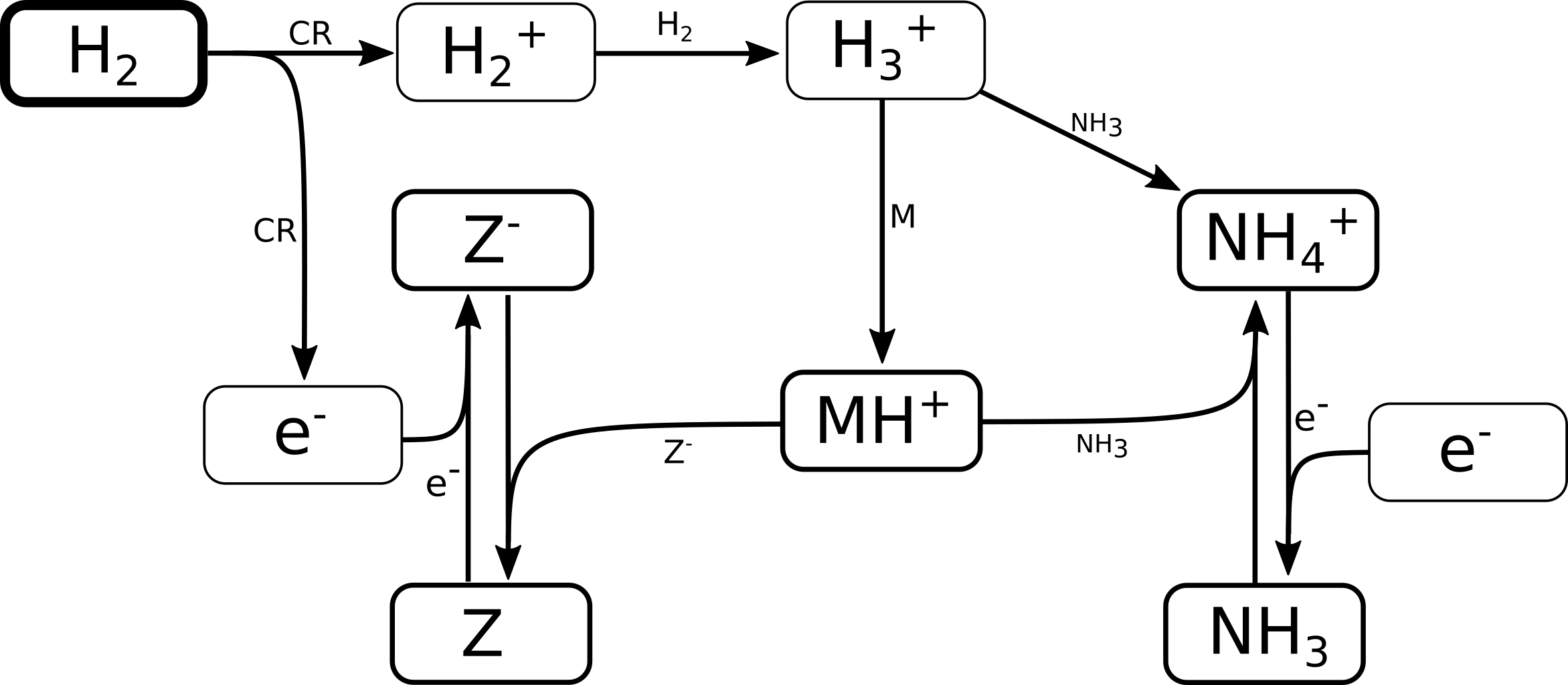}
    \caption{Reaction diagram showing how the gas is ionized, and the grains obtain negative charges in region~A.}
    %Overview of the reactions network of the midplane for $r$ at 1 au and less. The left hand column represents the chain of reactions that happen due to the ionization of molecular hydrogen via cosmic rays (CR) and refers to the reactions shown in equations \ref{eqn:I}, \ref{eqn:II} and \ref{eqn:III}. The last arrow illustrates the chain of reaction that leads to the observed abundance of $\mathrm{NH}_4^+$ referring to equation \ref{eqn:IV}. The top part represent the channel that the electrons take after the ionization via Cosmic rays. They react predominately with dust grains (Eqn. \ref{eqn:V}). The central rectangle represents the backwards reactions that the $\mathrm{NH}_4^+$ take with electrons and the recombination reactions of dust grains and molecules. The border thickness represents the relative abundances of the species compared to each other, with thicker borders representing more abundant species. \pw{$\rm Z^- + MH^+$ is incorrect.  No proton exchange proceses here. $\rm NH_4^+ + e^- \to M$ is a bit unclear.}
    
    \label{fig:reac_scheme}
\end{figure}

Figure~\ref{fig:species_low_r} shows that we find \ce{NH4+} to be by far the most important molecular cation in region~A. The chemical processes that lead to this kind of charge balance in region~A is multistaged and illustrated in Fig.~\ref{fig:reac_scheme}. 

Since region~A is entirely shielded from UV photons and X-rays, cosmic rays are found to be the only relevant ionisation source, in particular 
\begin{equation}
\rm H_2 ~+~ CR ~\to~ H_2^+ + e^- \ .
\end{equation}
While the free electrons are quickly picked up by the dust grains via ~$\rm Z + e^- \to Z^-$, the \ce{H2+} molecules, after fast reactions with \ce{H2}, form the molecular cation \ce{H3+}, which has a relatively low proton affinity, see Table~\ref{tab:proton_aff}.  Therefore, the surplus protons in \ce{H3+} are quickly passed on to other abundant molecules, creating more complex molecular cations, such as \ce{H3O+}, \ce{HCNH+} and \ce{NH4+}.  Further proton exchange reactions with abundant neutrals tend to increase the abundances of the protonated molecules which have the highest proton affinities.

The \ce{NH4+} molecule has an extremely high proton affinity of 8.9\,eV which means that it cannot recombine and dissociate on grain surfaces, i.e.,\ the reaction $\rm Z^- + NH_4^+ \to Z + NH_3 + H$ is energetically forbidden.  This creates a dead end, in which the concentration of \ce{NH4+} is enhanced until an equilibrium is established with the dissociative recombination reaction $\rm NH_4^+ + e^- \to NH_3 + H$ with the extremely rare free electrons.

However, the other protonated molecules below the threshold shown in Table~\ref{tab:proton_aff}, denoted by \ce{MH+} in Fig.~\ref{fig:reac_scheme}, can dissociatively recombine on the surface of the negatively charged dust grains, which creates a charging balance for the dust grains via the following two reactions
\begin{align}
    \ce{Z} ~+~ \ce{e-} &~\rightarrow~ \ce{Z}^- 
    \label{eq:Zattach}\ ,\\
    \ce{Z-} ~+~ \ce{MH+} &~\rightarrow~ 
    \ce{Z} ~+~ \ce{M} ~+~ \ce{H} \ ,\\
    \ce{Z-} ~+~ \ce{A+} &~\rightarrow~
    \ce{Z} ~+~ \ce{A} \ ,
%    \ce{Z-} ~+~ \ce{HCNH+} &~\rightarrow~ 
%    \ce{Z} ~+~ \ce{HCN} ~+~ \ce{H} \ \mbox{, and}\\
%    \ce{Z-} ~+~ \ce{H3O+} &~\rightarrow~ 
%    \ce{Z} ~+~ \ce{H2O} ~+~ \ce{H} \ .
\end{align}
where M are neutral abundant molecules and \ce{MH+} are their protonated counterparts, in particular \ce{H3O+} and \ce{HCNH+} in region~A.  There are also some simple charge exchange reactions with atomic ions \ce{A+}, such as \ce{Mg+} and \ce{Fe+}, whereas the electron affinity of \ce{Na+} of 5.14\,eV is too small to detach an electron from a silicate grain on impact (work function 8\,eV). We therefore have eliminated the reaction $\rm Z^- + Na^+ \to Z + Na$ from our databases.
The latter, with added reactions for, \ce{NaH2+} should be moved into the chemical rate descriptions.
None of the chemical reaction rates visualized in Fig.~\ref{fig:reac_scheme} involve activation barriers, which is important to remember when using this reaction scheme to discuss how the charge balance between \ce{Z-}, \ce{MH+}, \ce{NH4+} and \ce{e-} reacts to any changes in density, CRI-rate, or temperature.  However, reaction (\ref{eq:Zattach}) depends on the number of negative charges already collected by the grains, which is hence a result of the model.

Figure \ref{fig:species_low_r} shows that \ce{NH4+} remains the most abundant molecular cation up to a radial distance of about 0.7\,au, which coincides with the formation of ammonia ice in the midplane of our disk model.  At this point, \ce{NH3} is no longer an abundant molecule, and protonated silicon monoxide \ce{SiOH+} takes over the role of proton keeper from \ce{NH4+}. \ce{SiOH+} has a proton affinity of $8.1\,$eV, which means that it also cannot recombine dissociatively on negatively charged dust grain surfaces. Eventually, SiO freezes out as well, at 0.9\,au, and we transit into region~B.

\subsection{Region B --- the intermediate region}

\begin{figure}
    \hspace*{-2mm}
    \includegraphics[width=.5\textwidth]{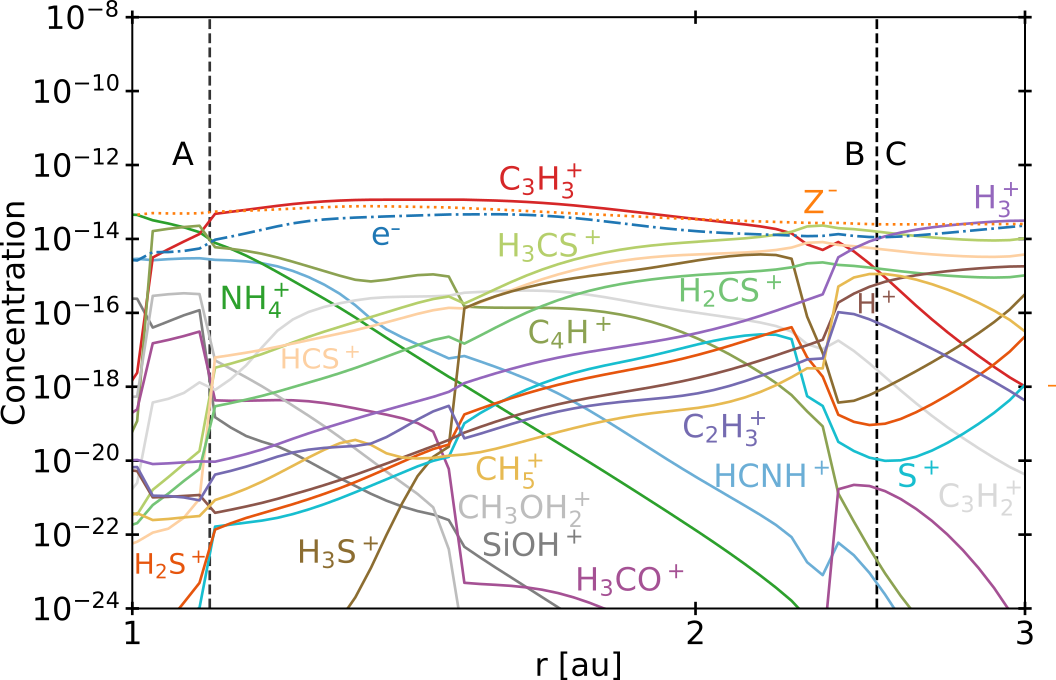}\\*[-5mm]
    \caption{Same plot as Fig. \ref{fig:species_low_r} but for region B. The dashed black lines illustrates the transition between the regions A and B and B and C respectively.}
    \label{fig:species_med_r}
    %The solid green line represents the $\mathrm{NH}_4^+$ molecules. The dashed red line represents the electron abundance. The solid dark yellow line represents the $\mathrm{SiOH}^+$ molecules.
\end{figure}

We identify region B as the disk region where the concentrations of (1) negative charges on dust grains, (2) free electrons, and (3) molecular cations are all of the same order $\approx\!10^{-13}$.  This happens in a transition region between about 1\,au and 3\,au in our disk model, see Fig.~\ref{fig:species_med_r}. Again, we require $A_V^{\rm ver}\!\ga\!10$ ruling out photoionization, which corresponds to $z/r\la 0.1$ in our disk model.

\begin{figure}
    \centering
    \includegraphics[width=.49\textwidth]{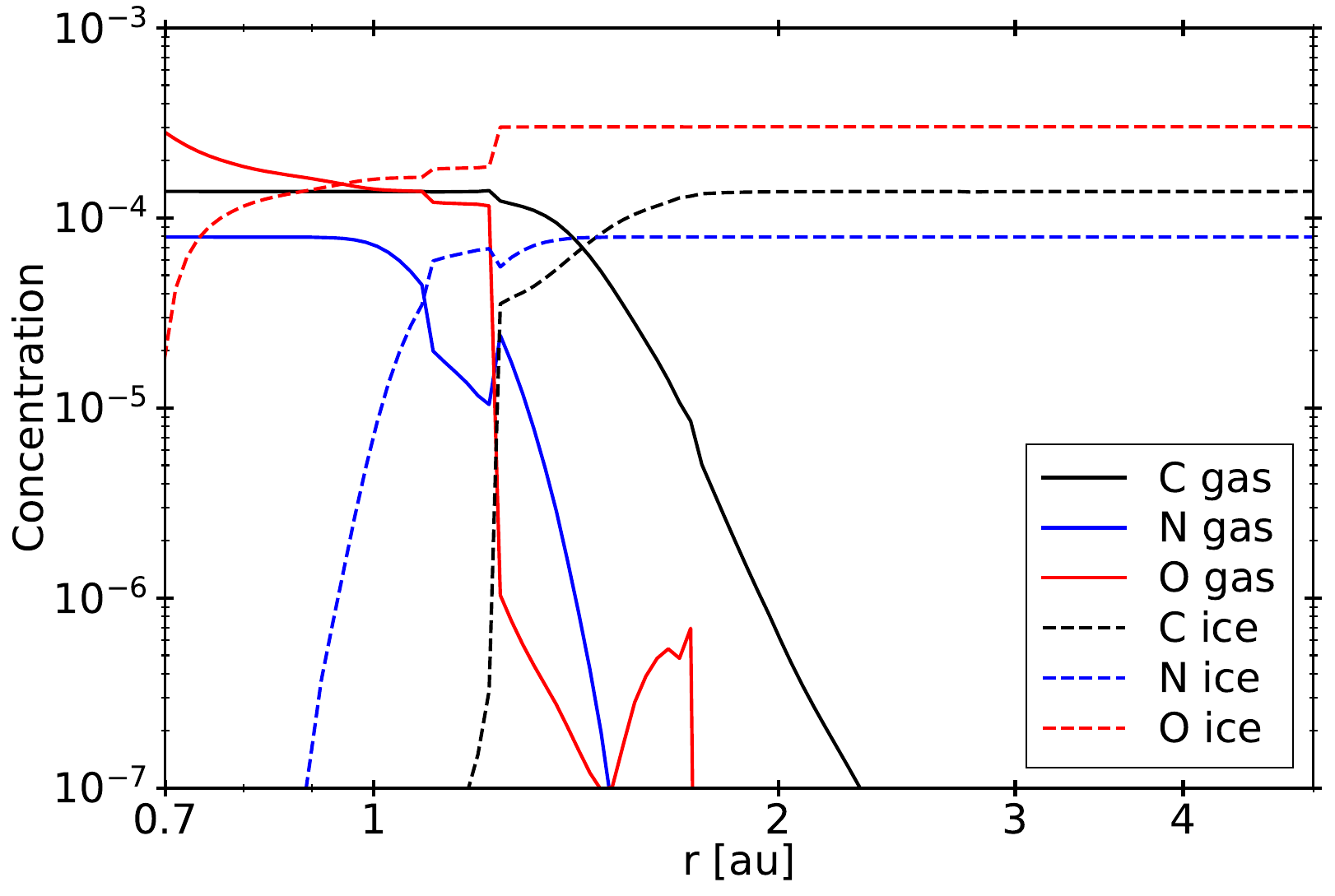}
    \caption{Total abundances of oxygen, carbon, and nitrogen in gas molecules and ice phases.}
    %Comparison of the phases, either gas or ice, in which Carbon, Nitrogen and Oxygen can be found in dependence of the radius $r$ [au]. The plots show the sum of all Carbon, Nitrogen and Oxygen bearing gas phase species in solid lines and the sum of all Carbon, Nitrogen and Oxygen bearing ice phase species in dashed lines.}
    \label{fig:ices}
\end{figure}

Figure~\ref{fig:ices} shows the stepwise freeze-out of oxygen, nitrogen, and carbon into ices with outwards falling temperature, which happens mostly in region~B, starting with water ice, then ammonia ice around 1\,au, and eventually several hydro-carbon ice phases such as \ce{C3H2\#}, \ce{C2H5\#} and \ce{C2H4\#} further out.  At the end of this ice formation zone, at a distance of about 3\,au, the midplane is virtually devoid of any molecules other than \ce{H2}, He and noble gases, and some sulfur molecules such as \ce{H2S}, \ce{CS} and \ce{H2CS} .   

Therefore, the transition region~B is carbon-rich, with oxygen and nitrogen already being strongly depleted by ice formation. Among the remaining hydro-carbon molecules in region~B, we find Cyclopropenylidene \ce{C3H2} to have the highest proton affinity, see Table~\ref{tab:proton_aff}.  Consequently, the reaction diagram in Fig.~\ref{fig:reac_scheme} just need to be slightly modified for region~B by (1) replacing \ce{NH4+} by \ce{C3H3+}, which takes over its role as proton keeper, and (2) identifying M with \ce{H2S} are \ce{CS} and \ce{H2CS}.

\subsection{Region C --- the metal-poor region}

In region C, there are almost no molecules other than \ce{H2} left in the gas phase, because almost all oxygen, carbon, nitrogen, and sulfur is in the ice. Since the various electron recombination rates all scale with $n^2$, but the cosmic ray ionization rate scales with $n$, the electron density is steadily increasing towards larger radii, whereas the concentration of negative charges on dust grains \ce{Z^-} stays about constant. In fact, we use $n_{\rm e}=\rm [Z^-]$ to set the boundary between regions B and C, which happens at about 3\,au in our model.

Consequently, we see a charge equilibrium between free electrons and \ce{H3+} in region~C, with some traces of \ce{H+}, whereas the negatively charged dust grains become increasingly unimportant for the charge equilibrium. Towards the outer edge of region~C, the disk becomes vertically transparent again, some interstellar UV and scattered UV starlight reaches the midplane, and we find increasing amounts of ionized sulfur in the gas phase.  At a radial distance of about 100\,au in our model, \ce{H+} and \ce{S+} become more abundant than \ce{H3+} and we enter region~D.

\begin{figure}
    \hspace*{-2mm}
    \includegraphics[width=.5\textwidth]{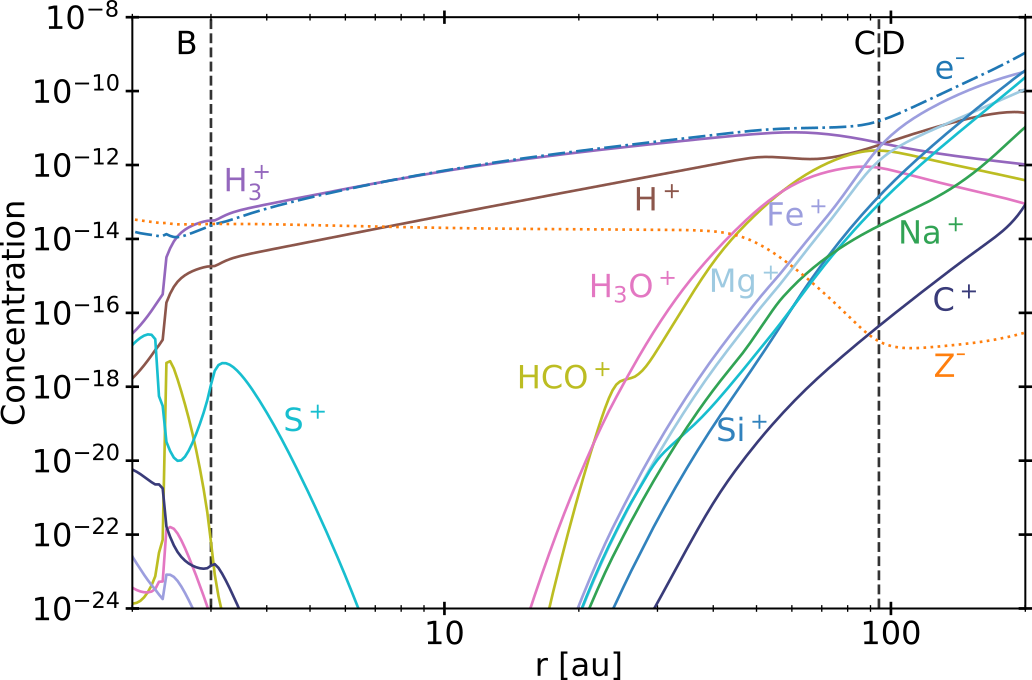}
    \caption{Same as Fig.~\ref{fig:species_low_r} but with the species that dominate the charge balance in regions~C and D. The vertical dashed lines indicate the transitions between regions B and C, and between C and D.}
    \label{fig:species_high_r}
\end{figure}

\subsection{Dependence of grain charge on physical parameters}
\label{subsec:disc_charging_grains}

In order to investigate the charging behavior of dust grains in more detail, we want to investigate the trends, seen in Fig. \ref{fig:f_q}, further and see how they react to the changing of parameters that influence dust, gas and electron abundance. In order to do this, we change five specific parameters and see how the affected species react to these changes. We will change the cosmic ray intensity $\zeta$, the minimum dust grain size $a_{\min}$, the total gas density $\rho$, the dust to gas ratio and the dust temperature $T_{\rm d}$. We do this by changing the parameters compared to our standard case (Tab. \ref{tab:parameters}) and solve the chemistry only at the point in our simulation with $r$=0.1 au and $z/r$=0 in the midplane. An overview of the parameter space for each different changed parameter can be found in Tab. \ref{tab:timedepend}. We only ever change one parameter, every other unchanged parameter is equivalent to our standard case.

\begin{figure*}
    \centering
    \includegraphics[width=.49\textwidth]{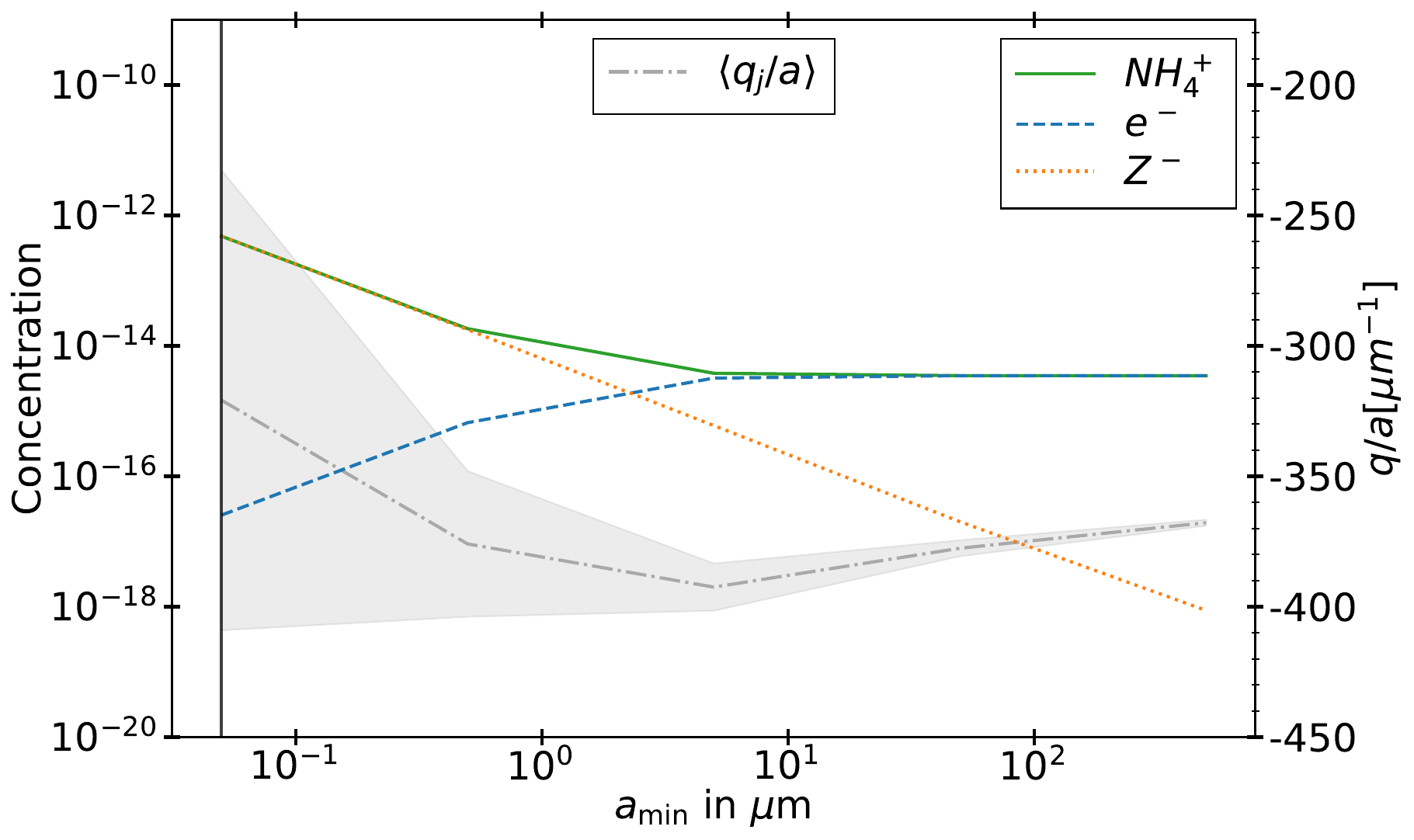}
    \includegraphics[width=.49\textwidth]{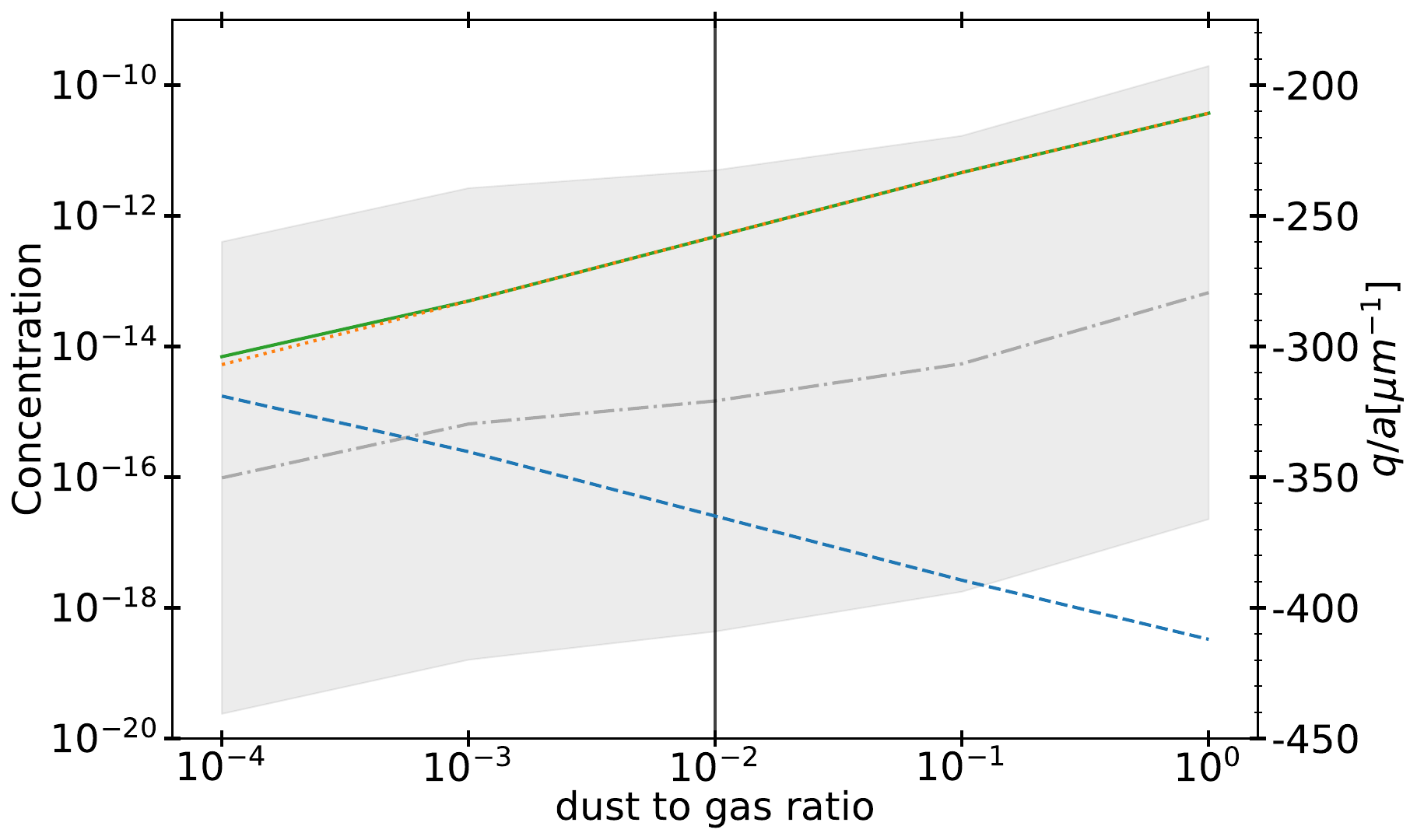}
    \includegraphics[width=.49\textwidth]{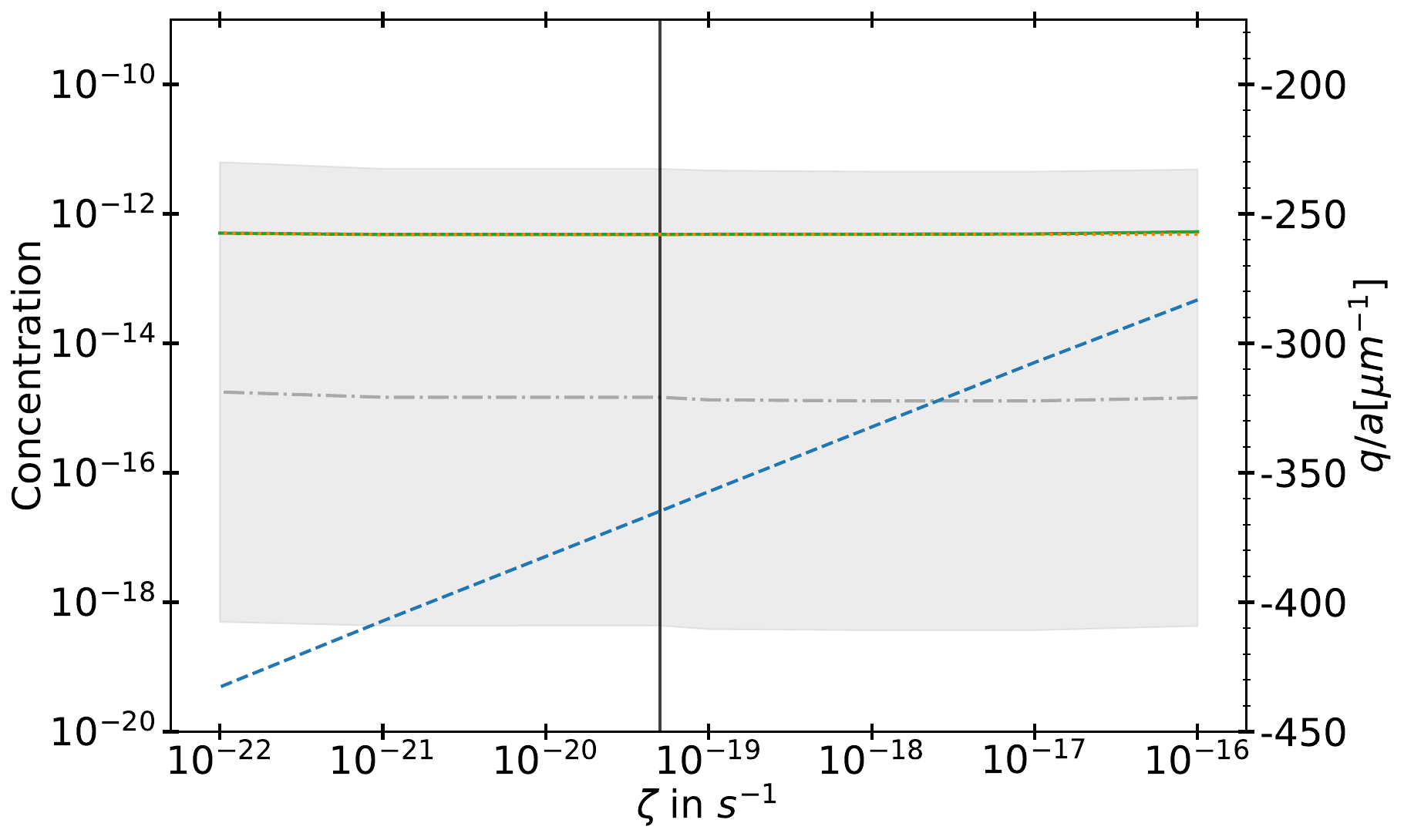} 
    \includegraphics[width=.49\textwidth]{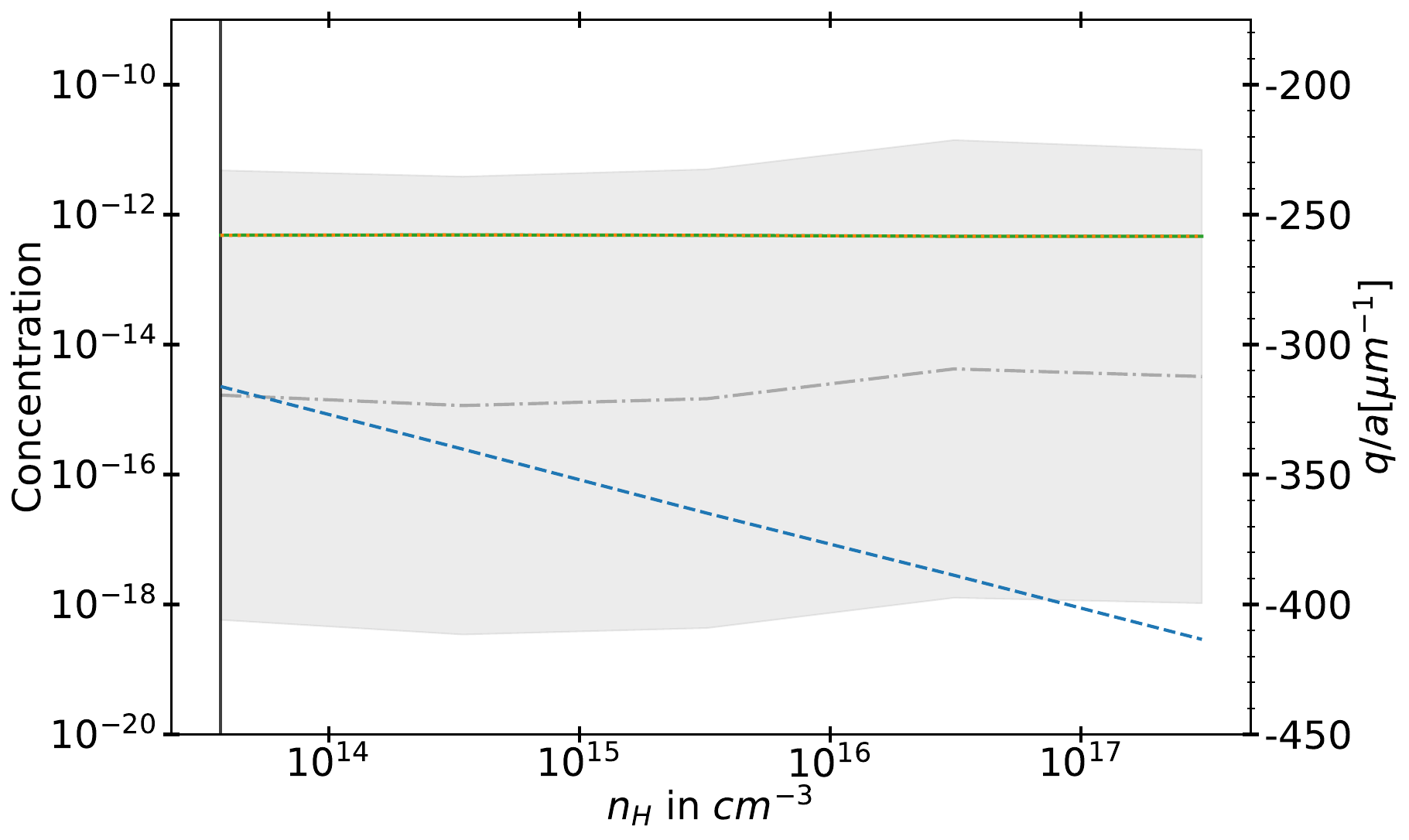}  
    \includegraphics[width=.49\textwidth]{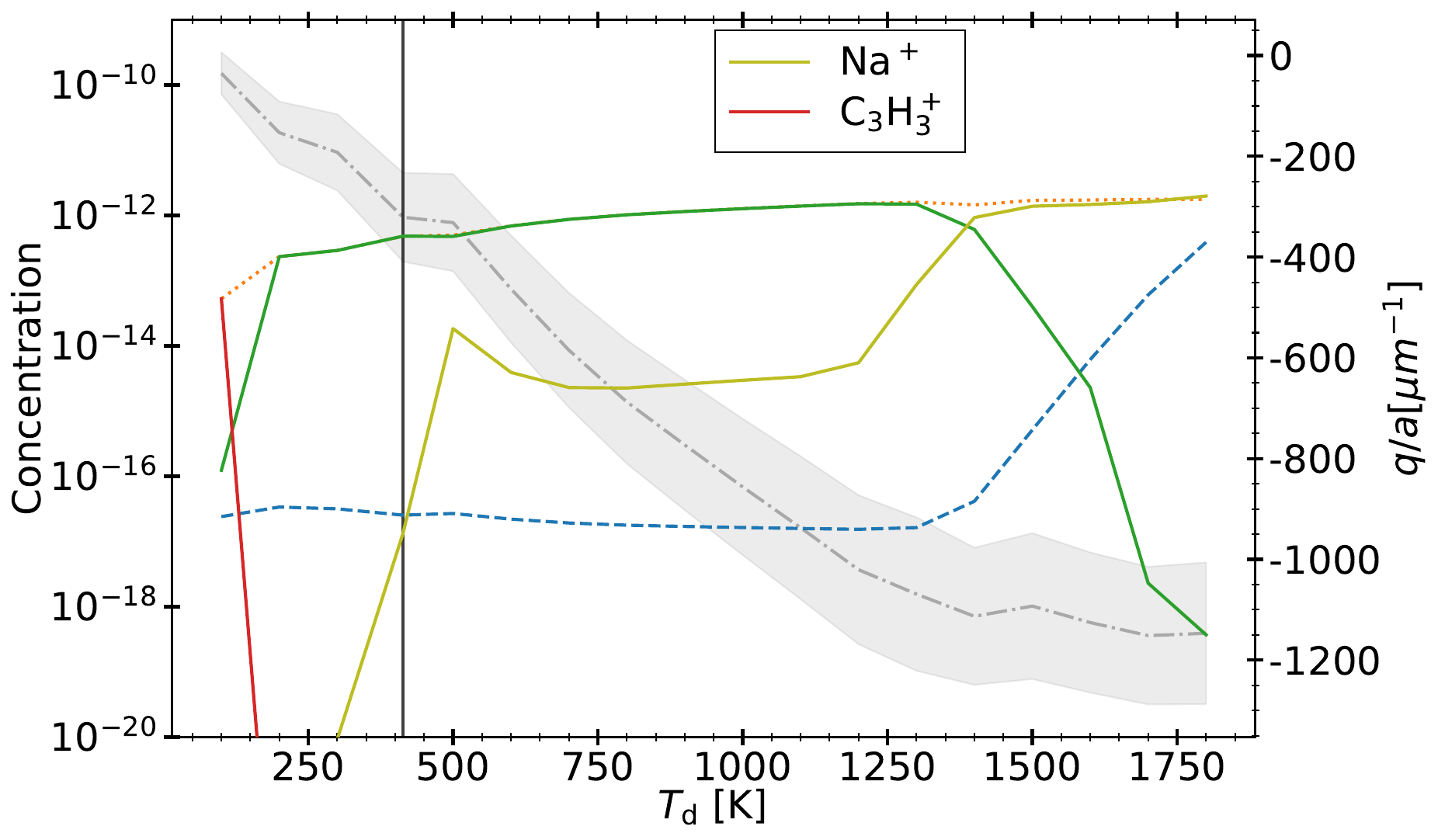} 
    \caption{Results for the parameter analysis. For all panels the solid green line represents the abundance of $\mathrm{NH}_4^+$, the dotted orange line represents negatively charged dust grains abundance and the dashed blue line represents the abundance of the electrons. In addition the last plot also shows the abundance of $\rm C_3H_3^+$ in red and $\rm Na^+$ in yellow green. All abundances are represented in units of hydrogen atom abundance, seen on the left Y-axis. The grey dash-dotted line represents the mean of the  amount of charge a dust grain carries relative to its size in $\mu m$, $q_j/a$. The shaded area around this line represents the average, the of standard deviations of $\sigma_j$. In all plots, we also plotted a vertical black line, which represents the conditions of the large simulation from which this set of simulations originates from. \\ \textbf{Upper Left Panel}: These are the results for the runs where we modify $a_{\min}$.\\ \textbf{Upper Right Panel}: These are the results where we modify the dust to gas ratio.\\ \textbf{Middle Left Panel}: These are the results where we modify the simulations with a constant cosmic ray ionization rate. \\\textbf{Middle Right Panel}: These are the results where we modify the disk mass in order to increase the total gas density. \\\textbf{Lower Panel}: These are the result where we modify the dust temperature.}
    \label{fig:timedepend}
\end{figure*}

The main results of this section can be found in Fig. \ref{fig:timedepend}. In these plots we investigate the abundances of the three species which are most influential to the charge balance at the investigated point and how the charge distribution function, as seen in the lower parts of Fig. \ref{fig:f_q} compared to $q_j/a$, changes when varying the different parameters. In order to see changes in the charge distribution functions compared to, $q_j/a$ we create these plots again for every dustbin of the different simulations. In order to make a comparison feasible, we take an average of the peaks of these charge distribution functions, since Fig. \ref{fig:f_q} already revealed a very uniform behavior between the different dust bins. In addition we calculate the standard deviations of the different $q_j/a$ values $\sigma_j$. These standard deviations have been averaged as well, and this average can be seen in Fig. \ref{fig:timedepend} as the gray shaded areas.
We now describe the different subplots of Fig. \ref{fig:timedepend} for each parameter in more detail.
\begin{table}[]
    \centering
    \caption{Parameters that we varied to test the charging behavior of grains $^{(1)}$.}
    \begin{tabular}{|c|c|c|c|c|}
    \hline
         $a_{min}$ [$\mu m$] & dust to gas & $\zeta$ [$s^{-1}$] & $M_{\mathrm{disk}}$ [$M_{\odot}$] & $T_{\rm d}$ [K] \\
         \hline
         \textbf{0.05} & 0.0001 & $10^{-16}$& 0.0001 & 100\\
         0.5 & 0.001 & $10^{-17}$& 0.001 & 200\\
         5 & \textbf{0.01} & $10^{-18}$& \textbf{0.01} & 300\\
         50 & 0.1 & $10^{-19}$& 0.1 & \textbf{413}\\
         500 & 1 & $\mathbf{5\times10^{-20}}$& 1 & 500\\ 
          &  & $10^{-21}$& & \vdots\\ 
          &  & $10^{-22}$& & 1800\\          
         \hline
    \end{tabular}
    \footnotesize
    $^{(1)}$:The bold values are the default values of our standard simulation.  
    \label{tab:timedepend}
\end{table}
\paragraph{Minimal dust size}
For testing the influence of the minimum dust size on the dust charging behavior we tested 5 different minimum dust sizes starting with $a_{\min}=0.05\;\mu m$, which is the default value of our standard simulation, and increase it by factors of 10 each step until we reach $a_{\min}=500\;\mu m$. The first big trend one sees is the decrease of the abundance of the negatively charged dust grains. This is the case due to the smaller surface area a few large dust grains have, compared to many smaller ones, since one of the main contributor to the abundance of negatively charged dust grains is the total surface area. This result in lower rate coefficients and therefore less negatively charged dust grains, as they react less with electrons. This also explains the two other trends we see, an increase in electron abundance and a decrease in $\mathrm{NH}_4^+$ abundance until a saturation is reached at $a_{\min}=50\;\mu m$. As explained before, this is due to the lower amount of dust grains. Dust grains are the main reaction partner with electrons, resulting in the normally high amount of negatively charged dust grains. If less dust grains react with electrons, one gets more free electrons.

These free electrons are also part of the main destruction path for the $\mathrm{NH}_4^+$, which explains the decrease of the $\mathrm{NH}_4^+$ molecules. Both, the electrons and the $\mathrm{NH}_4^+$ abundances converge towards each other because the dust grain abundance becomes fully negligible with a sufficiently high minimum dust grains size.
\paragraph{Dust to gas ratio}
We test the influence of different dust to gas ratios by simulating 5 different dust to gas ratios, starting with $10^{-4}$, again going in steps of a factor of ten up to a dust to gas ratio of 1. The clear trend that can be seen is the decrease of the electron abundance with increasing dust to gas ratios. To understand these results, it is central to remember the role dust grains play in the midplane chemistry. Dust grains are the main reaction partner with free electrons. Therefore, increasing the amount of dust grains by increasing the dust to gas ratio, decreased the amount of electrons. Secondly, as mentioned in the last paragraph, the main destruction pathway of $\mathrm{NH}_4^+$ molecules is via the reaction with free electrons. Hence, a reduction of electrons, via having more dust grains, results in an increase of the $\mathrm{NH}_4^+$ molecules as well.
\paragraph{Cosmic Rays}
In order to investigate the impact differing cosmic ray ionization rates have on our results, we choose to simulate six different cosmic ray ionization rate from $\zeta=10^{-16}$ to $\zeta=10^{-22}$. Note that we assume a constant cosmic ray ionization rate instead of the method we choose for our large simulations. This has been done to make it easier to compare results. At the point we are investigating, the cosmic ray ionization rate for our standard case is $5\times10^{-20}$.

The strongest trend one can see is the strong relation between the abundance of the electrons and the cosmic ray ionization rate. The trend is that the lower the cosmic ray ionization rate, the lower the amount of electrons. This is caused by the fact that the main source for electrons in these highly shielded regions is the ionization of molecular hydrogen via cosmic rays. We also see no change in the abundance of $\mathrm{NH}_4^+$ and negative charged dust grains. This is the case because the cosmic ray ionization rate does also influence how many $\mathrm{H}_3^+$ molecules are available. A decrease in cosmic ray ionization rate also results in less molecular hydrogen ionization. This results in a constant abundance of both negative dust grains and  $\mathrm{NH}_4^+$ since we see a decrease of the species which are influential for both the creation and destruction of negatively charged dust grains and $\mathrm{NH}_4^+$. For negatively charged dust grains, the important creation and destruction species are electrons for the creation and protonated molecules for the destruction. For $\mathrm{NH}_4^+$ it is vice versa. 
\paragraph{Gas density}
In order to see if changing the gas density changes our result, we simulated different models with a differing disk mass. As the amount of gas in our model cannot be directly increased or decreased, we had to change the disk mass instead, which indirectly increases or decreased the gas density. We simulated 5 different disk masses, varying from $10^{-4} M_{\odot}$ up to $1 M_{\odot}$. Note however that we plot against gas density and not disk mass. The result are very similar to the ones we get from the runs where we vary the cosmic ray ionization rate and this is the case as increasing the gas density also increases the shielding and hence decreases the amount of cosmic ray ionization and therefore free electrons. Therefore, the effects mentioned in the previous section mostly hold true for this part as well.
\paragraph{Dust temperature}
We further want to investigate how changing the dust temperature could influence the charging behavior of the grains. We did this by changing the dust temperature from 100 K to 1800 K in steps of 100 K. Note that we also changed the gas temperature at the investigated spot to be equal to the dust temperature at the beginning. The findings can be summarized into four different areas.

At the very first point at 100 K we can see one clear deviation from the standard picture in the concentration of the $\mathrm{NH}_4^+$ molecules. They are less abundant by 2 orders of magnitude compared to the normal case. This is due to the fact that at these lower temperatures we see a similar case to Region B in our larger simulations, where nitrogen bearing species are frozen out, therefore reducing nitrogen abundances in the gas. Similarly to Region B also is that $\ce{C_3H_3}^+$ is now the most abundant positive species as carbon is not frozen out yet and therefore abundant in the gas phase.

For temperatures between 200 K to 1300 K the concentrations of the different species resembles the standard case with two exceptions. Firstly, we find that the ability of the dust grains to gather electrons scales linearly with their temperature. Secondly, due to the dust grains being able to carry more negative charges, we find a slight increase in the concentrations of the negatively charged dust grains and a decrease in the electron concentration. This also results in an increase in the concentration of the $\mathrm{NH}_4^+$ molecules.

From 1400 K to 1800 K we can see $\rm Na^{+}$ becoming a dominant species. This is due to the creation reactions of $\rm Na^{+}$ becoming more efficient. The main creation reaction that becomes more efficient is the collisional ionization of $\rm Na$ via $\rm H_2$
\begin{equation}
    \rm Na + \rm H_2 \rightarrow Na^+ + H_2 + e^-.
    \label{eqn:react_Na}
\end{equation}
This reaction is endothermic in nature, but the barrier of $dH_f/k_b \approx 60000$ is still low enough to be overcome more and more often at Temperatures > 1400 K. This results in $\rm Na^{+}$ becoming the most abundant positive species after 1400 K and after 1600 K having a similar abundance as the dust grains. Additionally we observe an increase in the electron concentration, as one of the products of reaction in Eq. \ref{eqn:react_Na} are electrons. This increase of electrons results in an overall reduction of the $\mathrm{NH}_4^+$ molecules, as the main destruction reaction for $\mathrm{NH}_4^+$ is the recombination with electrons.

\section{A simple model for turbulent charge separation}
\label{sec:ChargeSep}

\subsection{Turbulence Induced Electric Fields}
\label{subsec:turbulence}
\begin{figure*}[!t]
\centering
\includegraphics[width=.6\textwidth]{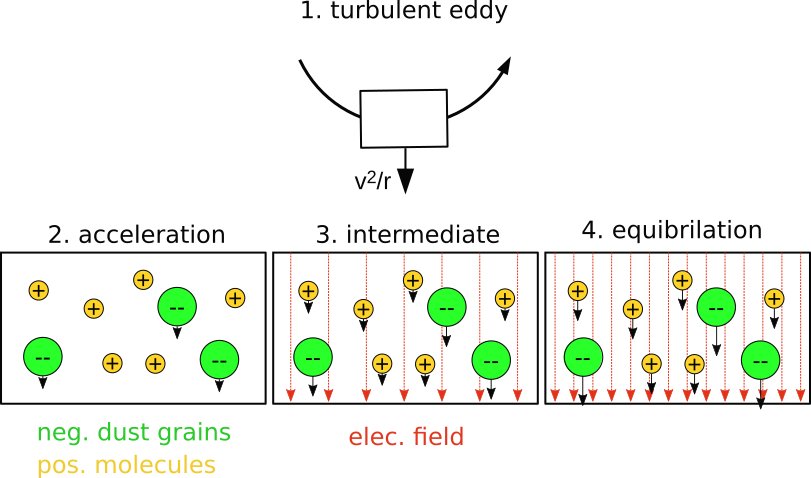}
\includegraphics[width=.3\textwidth]{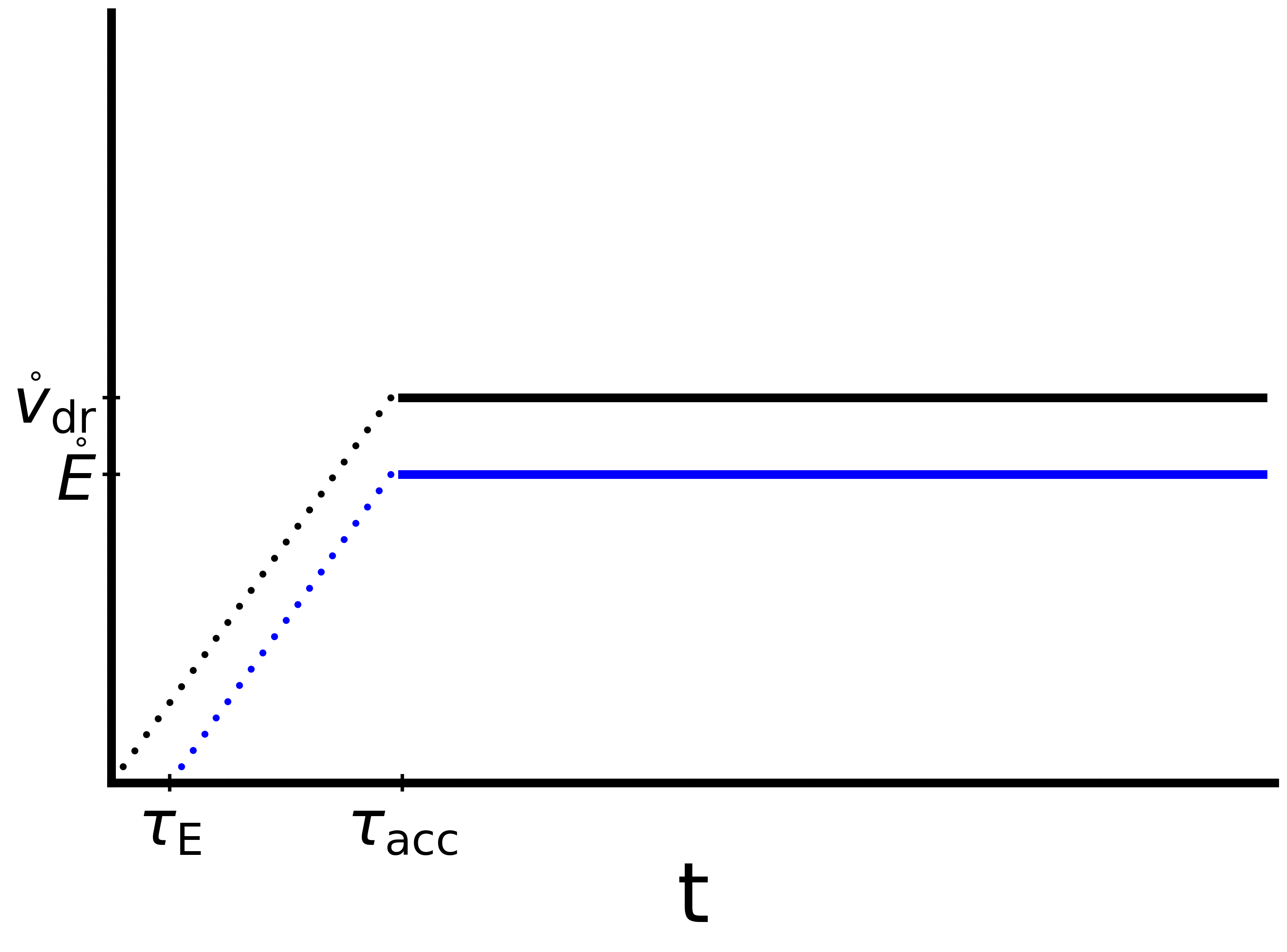}

\caption{Sketch of our electrification model. The left-hand pictures show how the process evolves overtime for the related particles, and the right-hand side shows the time evolution of the drift velocity, of the related particles, and the electric fields.\\
\textbf{Left:} Illustration of the different phases of our electrification model. Firstly, the centrifugal force in a turbulent eddy separates the charges, until an electric field builds up, which causes the molecular cations to follow the negatively charged grains.\\
\textbf{Right:} Example plot of the length of timescales for acceleration and equibrilation processes showed in the left illustration. The dotted line illustrates that either the dust grains have not reached the drift velocity (Black) or that the electric field is not build up yet (Blue). The units are arbitrary, as this plot is just made to make the model more understandable.}
\label{fig:elec_model}
\end{figure*}

In this section, we develop a simple model to estimate the maximum electric field $E$ that can arise when a mixture of negatively charged grains and molecular cations is shaken by turbulence.  Figure~\ref{fig:elec_model} shows a physical sketch of the situation. We use the Richardson-Kolmogorov approach for turbulence \citep{Richardson1926} \citep{1941DoSSR..30..301K} as superposition of turbulent eddies $k$ with various spatial scales $\ell_k$, timescales $\tau_k$, and characteristic velocities $v_k\!=\!\ell_k/\tau_k$. In the inertial range
\begin{equation}
  v_k \propto \left(\dot\epsilon\;\ell_k\right)^{1/3}  
  \quad\quad\mbox{$(\ell_\eta < \ell_k < \ell_L)$}\ ,
  \label{eqn:energy_dissipation_rate_I}
\end{equation}
is valid, where $\dot\epsilon$ is the energy dissipation rate towards smaller and smaller turbulent scales, $\ell_\eta$ and $\ell_L$ are the smallest (thermalization) and largest (driving) turbulent length scales, respectively.
In the co-moving frame of a curved gas flow related to the eddy $k$, the dust grains are accelerated by the centrifugal force $m\,v_k^2/\ell_k$, until an equilibrium with the frictional force $F_{\rm fric}$ in the gas is established.  
This problem is well-known for constant gravity $g$, see e.g.,\ \citet{2003A&A...399..297W}.  
After an initial acceleration timescale $\tau_{\rm acc}$, the grains of mass $m$ and size $a$ will reach a constant drift velocity $\mathring{v}_{\rm dr}$, also known as the final fall speed. 
We take over the results from Woitke \& Helling for the case of a subsonic flow and large Knudsen numbers, also known as the Epstein regime.  
After replacing $g$ by $v_k^2/\ell_k$, the results read
\begin{equation}
    \tau_{\rm acc} = \frac{a\,\rho_{\rm m}}{\rho\,v_{\rm th}} 
    \quad,\quad
    F_{\rm fric} = \frac{m\,v_{\rm dr}}{\tau_{\rm acc}} 
    \quad,\quad
    \mathring{v}_{\rm dr} = \frac{v_k^2}{\ell_k}\,\tau_{\rm acc}
    \label{eq:graindrift}
\end{equation}
where $\tau_{\rm acc}$ is the acceleration (or stopping) time, $\rho_{\rm m}$ is the dust material density $\approx\rm\!2\,g/cm^3$, and $v_{\rm th}\!=\!\sqrt{8kT/(\pi\bar{m})}$ is the thermal velocity with $\bar{m}$ being the mean molecular weight of the gas particles.

The drift of the charged grains leads to a charge separation, which causes an electric field. How exactly this field will look like is complicated and will depend on the geometry and the superposition of the electric currents caused by the drifting grains in the various turbulent eddies. However, we can estimate the maximum electric field that can be generated by a single long-lived eddy by considering the case ($\tau_E,\tau_{\rm acc}$) < t < $t_k$, where $\tau_E$ is the electric field build up timescale and $t_k$ the eddy turnover time.  We assume the electric field builds up in a coherent manner, from larger to smaller eddies, but, as shown later by Eq. \ref{eqn:efield_elec} and Eq. \ref{eqn:cent_a}, the main contribution will be coming from smaller eddies.

In this case, after the dust grains got accelerated by the eddies (Fig. \ref{fig:elec_model} Panel 2. acceleration), an electric field will continue to build up (Fig. \ref{fig:elec_model} Panel 3. intermediate) until the molecular cations in the gas follow the drifting grains because of their mobility in the $E$-field created by the grains. We call this equilibration (Fig.~\ref{fig:elec_model} Panel 4. equilibration), which is characterized by a vanishing electric current
\begin{equation}
  j_{\rm el} ~=~ \sum_j\Big([Z_{{\rm m},j}^+] - [Z_{{\rm m},j}^-]\Big)
              \,\mathring{v}_{{\rm dr},j}
  - n_{\rm e} v_{\rm e} + n_{\rm I} v_{\rm I} ~=~0 \ .
  \label{eqn:current}
\end{equation}
As before, $j$ is an index for the dust size bins. $[Z_{{\rm m},j}^+]$, $[Z_{{\rm m},j}^-]$ are the concentrations of the dust charge moments [charges cm$^{-3}$], see Eqs.\,(\ref{eqn:Zplus}) and (\ref{eqn:Zminus}), and $n_{\rm e}$ and $n_{\rm I}$ are the electron and molecular cation densities [charges cm$^{-3}$].

The drift velocities of the molecular cations and electrons are given by
\begin{align}
    v_{\rm I} &= -\mu_{\rm I}\,E 
    \label{mobility_I} \\
    v_{\rm e} &= \mu_{\rm e}\,E
    \label{mobility_e}
\end{align}
where $\mu_{\rm I}$ and $\mu_{\rm e}$ are the mobilities of the molecular cations and free electrons, respectively, in units 
$\rm cm^{2}\,s^{-1}\,V^{-1}$. The $E$-field is measured in $\rm V/cm$.
The different signs in Eqs.\,(\ref{mobility_I}) and (\ref{mobility_e}) are because of the opposite directions of the drift velocities of molecular cations and electrons in a given electric field.

If the turbulent eddy under consideration lives long enough for the drifting grains to build up the maximum, “equilibrated” electric field, it follows from Eq.\,(\ref{eqn:current}) that
\begin{equation}
    E = \frac{\sum_j\Big([Z_{{\rm m},j}^+] - [Z_{{\rm m},j}^-]\Big)
              \,\mathring{v}_{{\rm dr},j}}
             {n_{\rm I}\,\mu_{\rm I} + n_{\rm e}\,\mu_{\rm e}}.
    \label{eqn:efield_elec}
\end{equation}
From Eq.\,(\ref{eqn:efield_elec}) we can immediately see which turbulent flows can potentially produce large $E$-fields.
\begin{itemize}
\item Normal plasmas have $n_{\rm e}\!\approx\!n_{\rm I}\!\gg\![Z^+],[Z^-]$, in which case we don't expect large turbulence-induced electric fields.
\item Dusty plasmas, like the one we expect in region~A in the disk, have $[Z^-]\!\approx\!n_{\rm I}\!\gg\!n_{\rm e}\!\gg\![Z^+]$.  The induced electric fields in this case may be large when $\mathring{v}_{\rm dr}$ is large, $\mu_{\rm I}$ is small and $n_{\rm e}$ is sufficiently small to play no role, in which case $n_{\rm I}$ and $[Z^-]\!=\!\sum_j [Z_{{\rm m},j}^-]$ cancel, and we find
\begin{equation}
   E = \frac{\langle \mathring{v}_{\rm dr}\rangle}{\mu_{\rm I}} \ ,
\end{equation}
where $\langle \mathring{v}_{\rm dr}\rangle = 
\sum_j [Z_{{\rm m},j}^-]\,v_{{\rm dr},j}\;\big/\,\sum_j [Z_{{\rm m},j}^-]$ is the charge-mean grain drift velocity.
\item There is another case when $[Z^-]\!\approx\![Z^+]\!\gg\!n_{\rm I},n_{\rm e}$.  In that case, which we call 'region~X' (not found in our present disk models), large $E$-fields might be produced when many small grains are strongly charged negatively and many large grains are strongly charged positively, or vice versa, such that the nominator in Eq.\,(\ref{eqn:efield_elec}) remains large despite charge neutrality, in an isolating gas with few charged gas particles. We note that tribolelectric charging, which is not included in our current disk models, might be able to turn parts of region~A into a region~X. \\
We would like to mention at this point that there are alternative approaches to how we considered ion and electron mobilities. In, \citep{2015ApJ...800...47O} they present analytic expressions for ion and electron drift velocities in a \ce{H2} gas.
\end{itemize}
\noindent In order to estimate the duration of the equilibration phase in Fig.\,\ref{fig:elec_model}, henceforth called the $E$-field build-up timescale $\tau_{\rm E}$, we consider the travel time of the grains
\begin{equation}
  \tau_{\rm E} = \frac{d}{\mathring{v}_{\rm dr}}
  \label{eq:taubuild}
\end{equation}
where $d$ is the charge separation distance, i.e.,\ the distance between the moving grains and the moving molecular ions causing the electric field $E$. To estimate that distance, we consider the most simply geometry of a parallel plate capacitor (with equations in SI units):
\begin{align}
  E_{\rm SI} &= \frac{Q_{\rm SI}}{\epsilon_0\,A} \\
  Q_{\rm SI} &= e\;[Z^-]_{\rm SI}\,A\,d_{\rm SI} \\
  \Rightarrow\quad
  d_{\rm SI} &= \frac{\epsilon_0\,E_{\rm SI}}{e\;[Z^-]_{\rm SI}} \ ,
  \label{eq:dist}
\end{align}
\noindent where $\epsilon_0$ is the electric vacuum permittivity and $e$ the electron charge in SI units, $E_{\rm SI}\!=\!10^2\,E$ the maximum electric field in [V/m], $A$ a horizontal area $\rm[m^2]$ that cancels, $[Z^-]_{\rm SI}\!=\!10^6\,[Z^-]$ the concentration of negative charges on dust grains in $\rm[m^{-3}]$, $d_{\rm SI}\!=\!10^{-2}\,d$ the charge separation distance in [m], and $Q_{\rm SI}$ the total negative charge on the grains in volume $d_{\rm SI}\times A$ in [C].  Equations~(\ref{eq:graindrift}), (\ref{eq:taubuild}) and (\ref{eq:dist}) provide an estimate for the time required to clear a volume $d\times A$ from dust grains of size $a$ by centrifugal forces in a turbulent eddy $k$, which results in an overpopulation of negative charges at the lower boundary, and missing negative charges at the upper boundary, which together build up the maximum electric field.

Thus, we have three different timescales to consider, which may depend on the considered particle size $a$ and the selected turbulent eddy $k$.  In most cases we find
\begin{equation}
  \tau_k \gg \tau_{\rm acc}(a) \gg  \tau_{\rm E}(a,k) 
  \label{eqn:timescales}
\end{equation}
to be valid, i.e.,\ the turbulent eddy lives long enough to allow the grains to accelerate quickly and then drift for long enough distances to cause the maximum electric field.  If the relation (\ref{eqn:timescales}) does not hold, Eq.\,(\ref{eqn:efield_elec}) is not valid.

For the ion mobility $\mu_{\rm I}$, we consider the \ce{NH4+} cation as we find this molecule to be the most abundant gas charge carrier in region~A.

\citet{2014IJMSp.370..101A} presented measurements for the mobility of \ce{NH4+} in a \ce{N2} gas at standard pressure and temperature, $P_0\!=\!760\,$Torr and $T_0\!=\!273\,$K, which we denote by  $\mu^0_{\rm NH_4^+,N_2}\!=\!2.22\rm\,cm^2 s^{-1} V^{-1}$. As the most abundant species in disks is, \ce{H2} we need to adjust this value.  This is done by multiplying with the ratio of the reduced masses of \ce{N2} and \ce{H2} with \ce{NH4+}, denoted with $\bar{m}_{\rm N_2}$ and $\bar{m}_{\rm H_2}$,  respectively, and scaling with the gas particle density \citep[see Eq.\,4 in][]{2014IJMSp.370..101A}
\begin{equation}
    \mu_{\rm I} = \mu^0_{\rm NH_4^+,N_2}
                \left(\frac{\bar{m}_{\rm N_2}}{\bar{m}_{\rm H_2}}\right)^{1/2}
                \frac{P_0}{P} \frac{T}{T_0} \ .
    \label{eqn:mob_nh4}
\end{equation}
The electron mobility $\mu_{\rm e}$ is calculated from the drift velocity measurements for electrons of \cite{1965PPS....85.1283R} and calculated as
\begin{equation}
    \mu_{e,0}=\frac{v_{\rm dr,e}}{E}
\end{equation}
with $v_{\rm dr,e}$ measured in dry air to be
\begin{equation}
    v_{\rm dr,e}=256\times 10^3 \frac{E}{E_0}\frac{P_0}{P}\frac{T}{T_0} [cm/s],
\end{equation}
where $P_0$ and $T_0$ are the same as for the ion mobility and $E_0\!=\rm\!75\,V\,cm^{-1}$. {This formula results in an electron mobility that is about a factor of 200 larger than the mobility of \ce{NH4+}.} As the original measurements are done in dry air we would have to adjust our mobility calculations in the same way we did for ion mobility, but due to only minor differences in reduced masses when considering the electron mass, we omit a similar correction. 

According to our charge separation model, the electric fields that can potentially be caused by turbulence are proportional to the centrifugal accelerations $v_k^2/\ell_k$ that the turbulent eddies can provide. Following Eq.\,(\ref{eqn:energy_dissipation_rate_I}) we find,
 
\begin{equation}
    \frac{v_k^2}{l_k} \propto l_k^{\,-1/3}
    \label{eqn:cent_a} \ ,
\end{equation}
i.e.,\ the smallest eddies cause the largest centrifugal forces.
%where $v_k$ is the velocity of the eddy, $l_k$ the size of the eddy, and $\dot\epsilon$ the energy dissipation rate per unit mass, which is constant between each modes. 
We therefore consider the Kolmogorov timescale $t_{\eta}$, i.e.\ the turn-over timescale of the smallest eddy in the inertial subrange. We determine $\tau_{\eta}$, following \cite{2007A&A...466..413O}, by 
\begin{equation}
    \tau_{\eta}=\frac{\tau_L}{\sqrt{\mathcal{R}e}}
\end{equation}
where $\tau_L$ is the turn-over timescale of the largest eddy in our system and $\mathcal{R}e$ is the Reynolds number, defined by
\begin{equation}
    \mathcal{R}e=\frac{v_L \ell_L}{\nu_{\rm kin}}
\end{equation}
where $v_L$ is the velocity of the largest eddy, $\ell_L$ its size and $\nu_{\rm kin}$ the kinematic viscosity. The kinematic viscosity is taken from \citet{2003A&A...399..297W}, using their formula for the dynamic viscosity, we get $\nu_{\rm kin} = \nicefrac{1}{3}\,v_{\rm th} \bar{\ell}$.
For the size and velocity of the largest eddy we assume  $\ell_L=\sqrt{\alpha} H_p$ and 
$v_L=\sqrt{\alpha} c_s$, see \citet{2007A&A...461..215O}, where $H_p$ is the pressure scale height, which is set in our simulations as 
\begin{equation}
    H_p(r)=10\,{\rm au} \left(\frac{r}{100\,{\rm au}}\right)^{1.15}.
\end{equation}
The resulting size of the smallest turbulent eddy $\ell_\eta$ is considerably larger than the gas mean free path $\bar{\ell}$, by about 4 orders of magnitude, but also much smaller than scale height, by about 8 orders of magnitude.

\begin{figure}
    \hspace*{-2mm}
    \includegraphics[width=92mm]{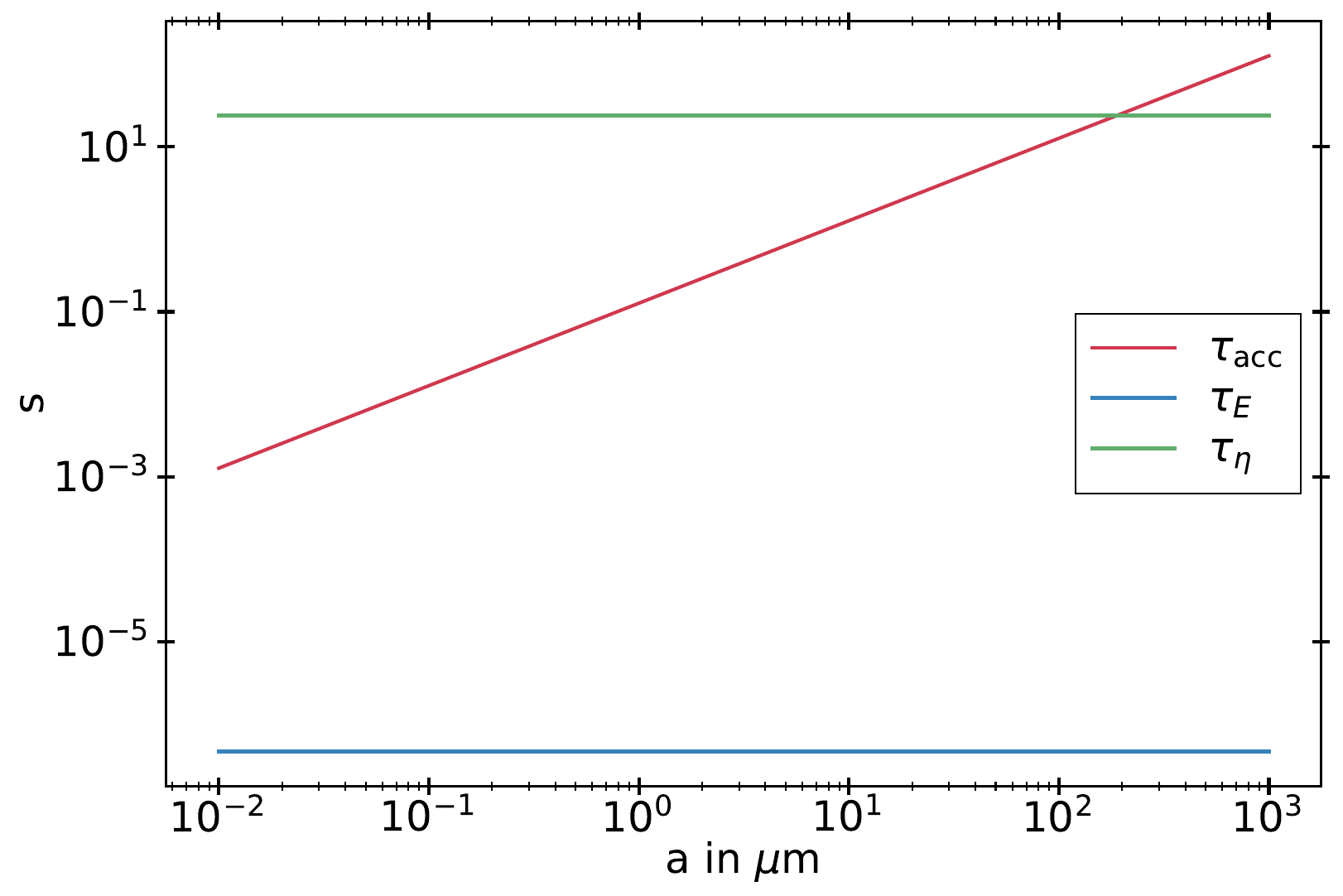}\\*[-6mm]
    \caption{Comparison of the different timescales considered in our turbulence implementation.}
    \label{fig:turbulent_times}
\end{figure}

Since we want to maximize the effect of the turbulence, we have to choose the $\tau_k$ that maximizes the centrifugal acceleration. Additionally we have to choose a time that allows the electrical fields to develop. Therefore, we follow the principle that we choose the shortest time that is still long enough for the fields to develop. This is illustrated by Fig. \ref{fig:turbulent_times}. Following this, we can see that in most cases we can choose $\tau_k=\tau_{\eta}$. In cases with larger grains, $a > 10^{-1} \rm cm$, we would choose $\tau_k=\tau_{\rm acc}$, to ensure the grains have enough time to reach their drift velocity $\mathring{v}_{\rm dr}$.

With these results, we can now calculate what electric fields would result from the interaction of our dust and gas mixture in a turbulent eddy and answer the question if lightning could emerge with this mixture already. Yet we still have to consider how large these electric fields can be.
For lightning to emerge in any medium, the electric fields generated have to be large enough, for an electron cascade to occur. We follow the description and Eq. 56 from \citet{2009AIPC.1158..145M}. With this we can define a critical electric field $E_{\mathrm{crit}}$, which is the electric field at which an electron cascade can occur, because the electrons are accelerated enough to ionize the neutral components of the gas.
\begin{equation}
    E_{\mathrm{crit}}=\frac{\Delta W}{e \lambda}
    \label{eqn:crit_field}
\end{equation}
Here $\Delta W$ is the ionization energy, required to ionize the neutral components, $e$ the coulomb constant and $\lambda$ the mean free path. For $\Delta W$ we take the ionization energy of $\ce{H_2}$ with 15.4 eV. To determine, $\lambda$ we follow Eq. 10 of \citet{2003A&A...399..297W}.

\subsection{Electric Field Magnitudes}

\begin{figure*}
    \centering
    \includegraphics[width=.49\textwidth]{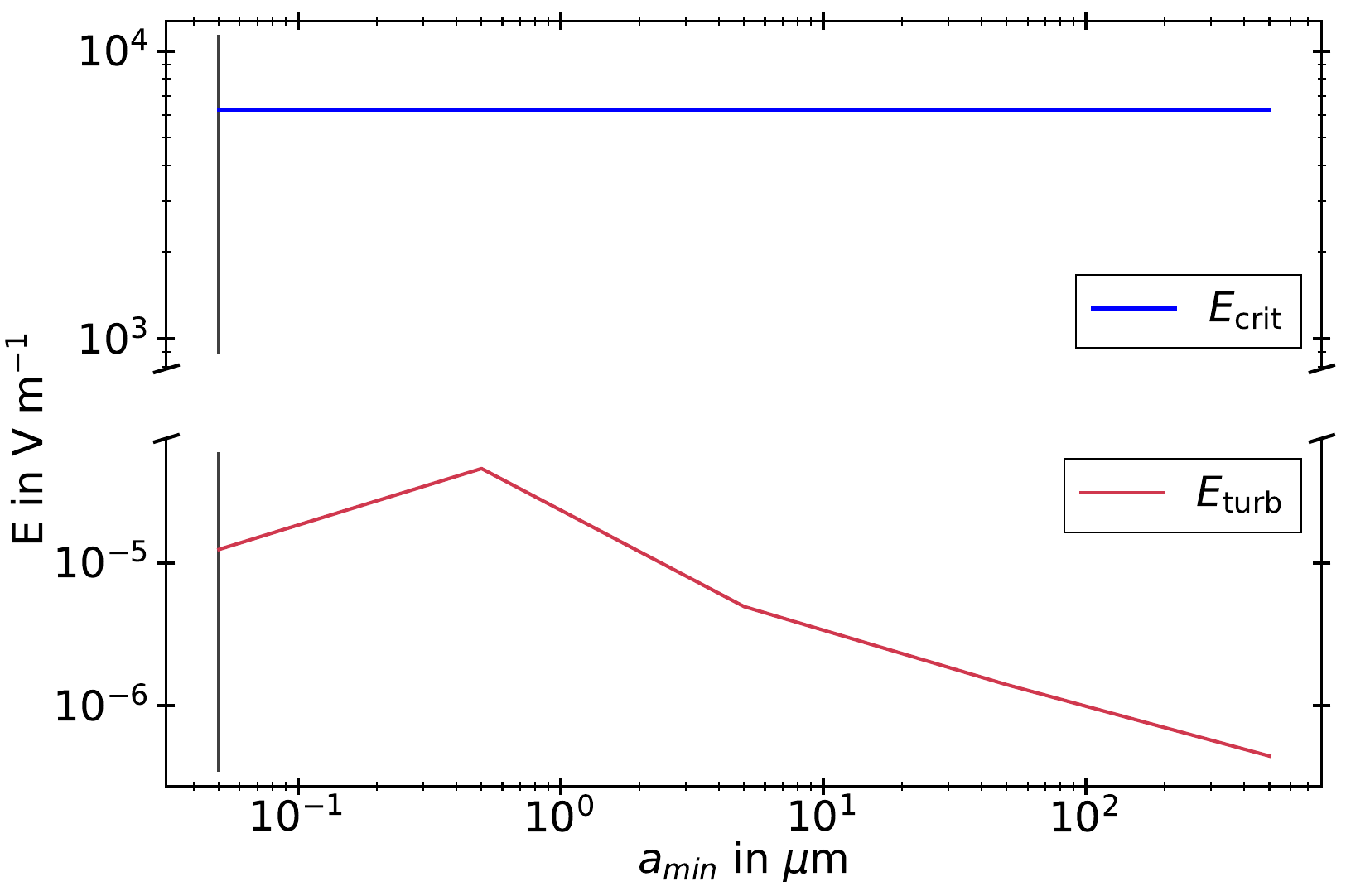}
    \includegraphics[width=.49\textwidth]{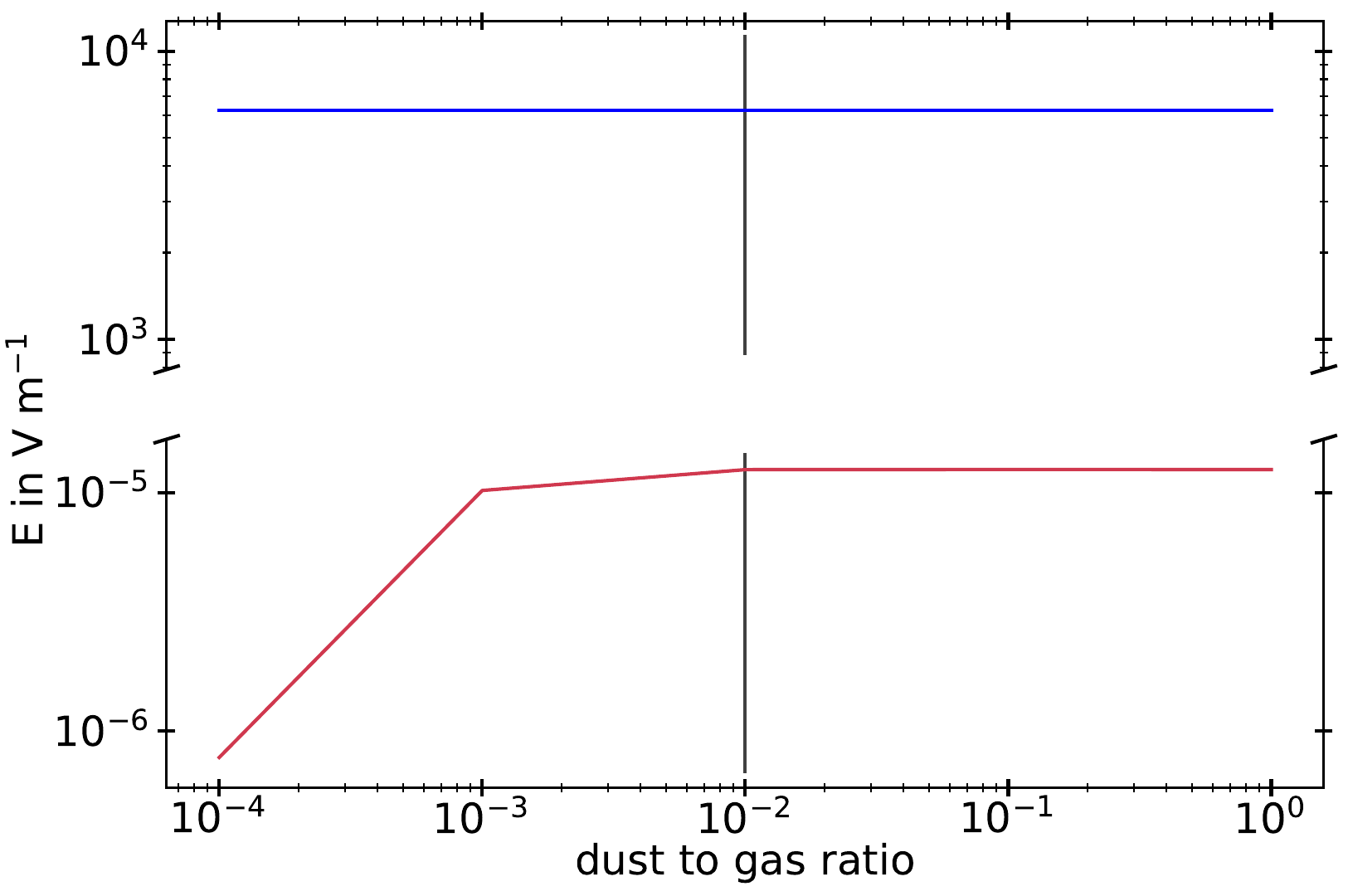}
    \includegraphics[width=.49\textwidth]{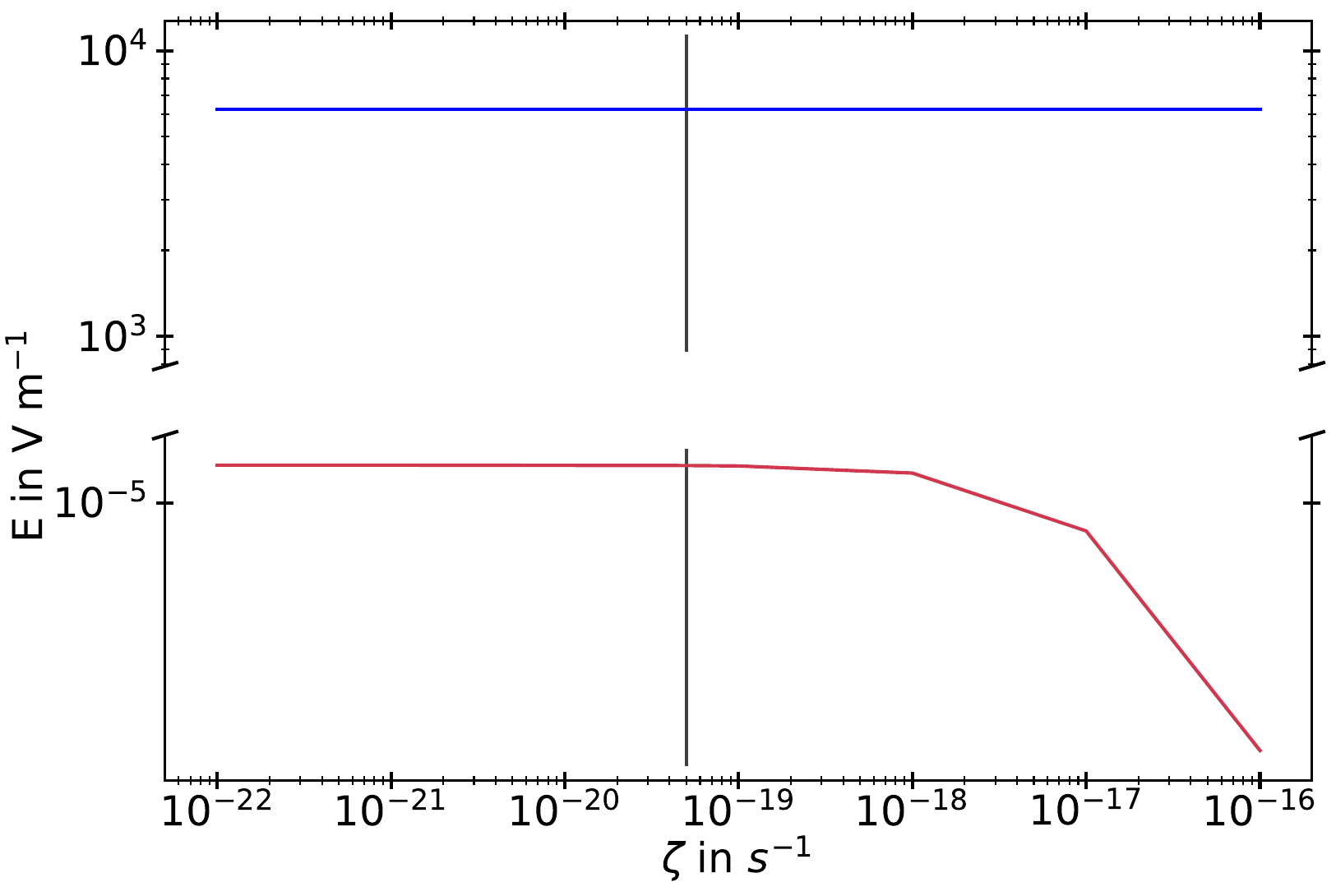} 
    \includegraphics[width=.49\textwidth]{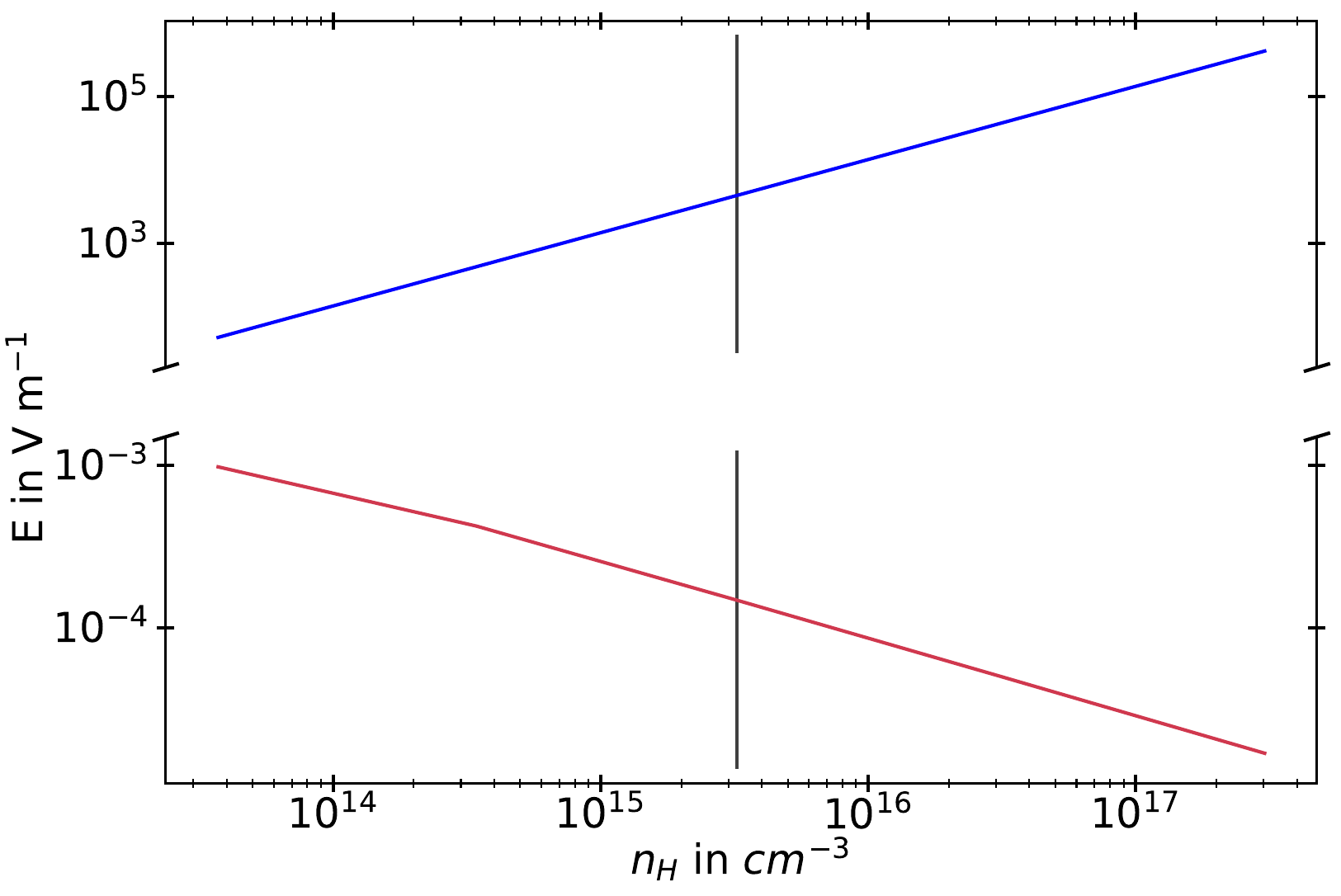}
    \includegraphics[width=.49\textwidth]{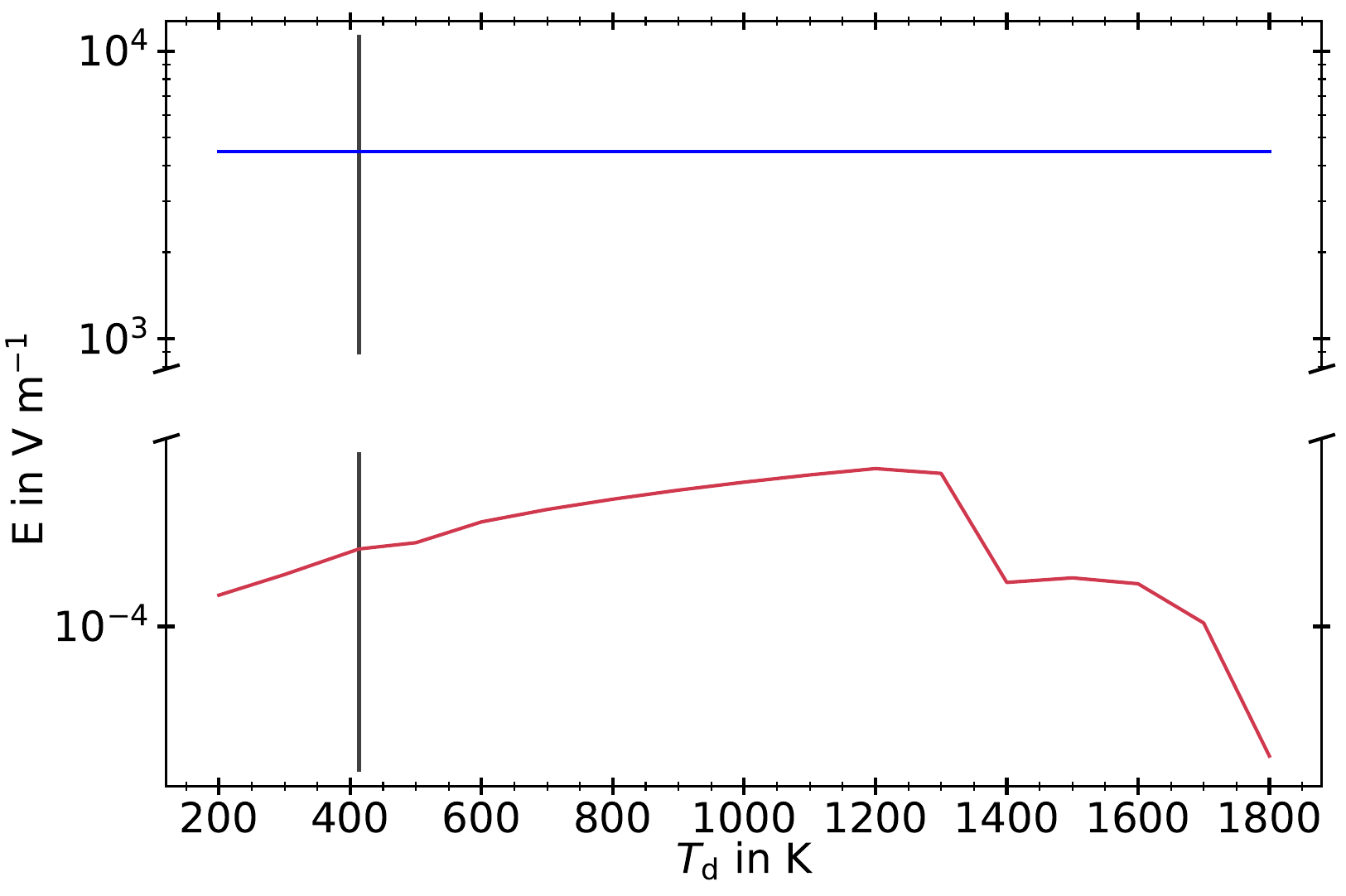} 
    \caption{The electric fields resulting from turbulence induced eddies for the different parameter variations shown in Sec \ref{subsec:disc_charging_grains}. The red line represents said electric fields and the blue line the critical electric field as from Eq. \ref{eqn:crit_field}. The solid black lines again represent the conditions from our standard simulation. Note that, for the sake of visibility, the Y-axis is interrupted as the differences between the critical fields and the turbulence induced fields are quite large.}
    \label{fig:electric_fields}
\end{figure*}

In this section, we investigate the magnitude of the previously discussed turbulence induced electric fields. We do this by taken the data from our simulations that we made for investigating the dust charge behavior from Sec. \ref{subsec:disc_charging_grains}. The resulting electric fields are shown in Fig.~\ref{fig:electric_fields}.

\paragraph{Minimal dust size}
We find that if one changes the minimal dust size, that one first sees an increase in electric field strength, due to the increase in dust size and its increase in drift velocity and therefore electric field strength. However, with a minimum dust grain size of larger than $0.5 \mu m$ a decrease in electric field strength occurs. This is due to the fact that larger dust grains are less abundant in our model. The way we set up the simulations, only very large dust grains are considered and therefore the dust abundance drops. According to Eq. \ref{eqn:efield_elec} this has of course negative impact on total electric field strength. 

\paragraph{Dust to gas ratio}
For increasing dust to gas ratios, we observe a steady increase in electric field strength until a saturation takes place at our standard value of $10^{-2}$. The reason for the low electric field values at lower dust to gas ratios is again the fact that less dust grains can contribute positively towards the creation of electric fields.

\paragraph{Cosmic rays}
The influence of cosmic rays on the electric fields is not very surprising due to their strong link to electron abundance. High cosmic ray ionization rates result in high electron abundances and therefore low electric fields due to the attenuating effect electrons have on the electric fields build up. The reverse is true for low cosmic ray ionization rates, however, this effect stagnates at around $10^{-19} [\mathrm{s}^{-1}]$.

\paragraph{Gas density}
One could expect that the gas density has a similar effect on the electric fields as the cosmic ray ionization rate, but this does not fully hold. While the general trend is the same, where high gas densities reduce the electron abundance, due to the larger shielding, there are more effects that one has to mention. Compare to the other tests, where the critical field kept constant, we find that the critical field rises with increasing gas densities. This is the case as with higher gas density, the mean free path increases, which increases the required energy for an electron avalanche to occur. We also find that for gas densities lower than our standard value, one can observe an enhancing factor for the electric fields. This is the case due to the lessened drag the dust grains feel in lower gas densities. This increases the drift velocities of the dust grains and can therefore enhance the charge separation and the electric fields.

However, one general trend hold true for all of the studied combination of parameters discussed above, which is that the resulting electric fields are several orders of magnitude smaller than the required critical field. Yet, we can still use this study to state which conditions are best for lightning to emerge. In conclusion of this section we therefore state, that for lightning to emerge in a protoplanetary disk, large and abundant dust grains in an area with low cosmic ray ionization rate and low gas density are the most likely parameter conditions.

\paragraph{Dust Temperature}
In order to calculate the electric field strength for the simulations where we varied the dust temperature we have to adjust our calculations of the electric field, as the most abundant positive ion for certain temperatures is $\rm Na^+$ and not $\rm NH_4^+$. We adjust Eq. (\ref{eqn:efield_elec}) to account for the concentration of $\rm Na^+$, $n_{\rm Na^+}$, and their mobility $\mu_{\rm Na^+}$
\begin{equation}
    E = \frac{\sum_j\Big([Z_{{\rm m},j}^+] - [Z_{{\rm m},j}^-]\Big)
              \,v_{{\rm dr},j}}
             {(n_{\rm I}\,\mu_{\rm I} +n_{\rm Na^+}\mu_{\rm Na^+}) + n_{\rm e}\,\mu_{\rm e}}.
    \label{eqn:efield_elec_2}
\end{equation}
For the mobility of the $\rm Na^+$ we calculate it similar to the $\rm NH_4^+$ 
\begin{equation}
    \mu_{\rm Na^+}=K_{0,\rm Na^+} \frac{P_0}{P} \frac{T}{T_0}
\end{equation}
where we take 17.5 $[cm^2\;V^{-1}\;s^{-1}]$ for $K_{0,\rm Na^+}$ according to \cite{1931PhRv...38..549L}, who measured the mobility of $\rm Na^+$ in $\rm H_2$. The two normalization factors $P_0$ and $T_0$ are chosen to be the same as for our $\rm NH_4^+$ calculations.

With these additional calculations in place, we can analyze the behavior of the electrical field at different temperatures. Note that in our analysis that we omit the first point at 100 K, as this would require taking $\rm C_3H_3^+$ into account with its mobility. We identify three different trends in the electric field curve. First we identify a steady increase in electric field strength from 200 K up to 1300 K. This can be explained by the steady increase we see in the concentration of negatively charged dust grains and the $\rm NH_4^+$, relative to the electron concentration. 

From 1400 K to 1600 K we see at first a drop by nearly a factor of two and then a plateau. This can be explained by the $\rm Na^+$ ions becoming the most relevant positive species. As their mobility is larger by nearly a factor of 10, it is harder for the dust grains to get high amounts of separation compared to the case with $\rm NH_4^+$.

Lastly, we can see a further and further drop in electric field strength until 1800 K. This can be explained with the electron concentration becoming relevant again. As shown in Section \ref{subsec:disc_charging_grains}, the dominant reaction creating $\rm Na^+$ also creates free electrons. Due to the very high mobility of the free electrons, they can strongly inhibit the electrical field build up. Their mobility is about 2 orders of magnitude higher than the ones we find for our positive molecules. Hence, their effect can be already seen when their concentration is still at only about 1 \% of the negative dust grains and positive molecules.

In conclusion, the picture that presents itself when we modify the dust temperature is more complicated than for our other examples. In isolation, where we would just observe a system of dust grains and one other positive species, one could conclude that higher temperatures show a positive impact on electrical field strength. But as shown in our example, this is an oversimplification and one should always consider what other effects and increase in temperature can have. 

\section{Discussion of the charge balance in Region A under consideration of ion attachment}
\label{sec:discussion}
At this point, we would like to discuss the charge balance in a case where a physical attachment between \ce{NH4+} and negative dust grains would be possible. %\pw{no, we are not discussing any chemical reaction here, just a physical attachment}.
Although the dissociative recombination reaction  $\ce{Z-} + \ce{NH4+} \rightarrow \ce{Z} + \ce{NH3} + \ce{H}$ is energetically
forbidden, the \ce{NH4+} ion could be trapped in the electrostatic potential of a strongly negatively charged grain, either after a collision with its surface, or after an inelastic collision with a gas particle close to the surface. This could occur when the electrostatic potential $e^2\frac{q}{a}$ is much larger than the molecule's thermal energy $kT$. At a temperature of 100\,K, this corresponds to a negative dust charge of $|q|/a \gg 6\rm\,\mu m^{-1}$. %\pw{Why did you change this paragraph, write $|q|/a \gg 6\rm\,\mu m^{-1}$, important is the $\gg$.}
In such cases, the \ce{NH4+} molecule would be expected to stick to the surface or to stay in close vicinity of the surface.

Such an attachment reaction could be formally written as
\begin{equation}
    \rm Z^- + NH_4^+ \leftrightarrow \mbox{(Z$^-$--\,NH$_4^+$)} \ .
\end{equation}
The compound Z$^-$--\,NH$_4^+$ would hold both the negative charges of the accumulated electrons and the positive charges of the accumulated molecular ions, and would therefore be considerably less negative in the far field, due to shielding by the \ce{NH4+}.

We assume that the overall effects of such a mechanism would not change the charge balance we find in Region A significantly, but could result in an increase of the non-shielded negative dust grain charge $q/a$. We would introduce a new destruction path for \ce{NH4+} with such an reaction, therefore decreasing the amount of \ce{NH4+} in the gas. This could increase the electron concentration in the charge balance, but the Z$^-$--NH$_4^+$ compound is less negatively charged than a dust grain \ce{Z-}. Which would result in more free electrons attaching to the compound compared to a dust grain. This process would increase the non-shielded grain charge $q/a$, until the shielded negative grain charge becomes similar to the $q/a$ calculated in this paper.
In conclusion, such a process would act as a sink of both NH$_4^+$ and free electrons, while increasing the non-shielded negative $q/a$.

However, in order to implement these effects into this work, one would need more data about the microphysical details of these processes, as such a reaction would not just need to be considered for \ce{NH4+}, but also every other positive molecule with a proton affinity higher than about 8 eV (Table \ref{tab:proton_aff}) and one would need a kinetic formulation of the back reactions. Therefore, an inclusion of this mechanism goes beyond the scope of this paper, but could be subject in a follow-up study.

\section{Summary and Conclusion}
We identified 6 different regions in our disk simulations with significantly different charging behavior of dust and gas. The three upper regions (D, E and F) are all photon-dominated, where the dust grains are positively charged, and the charge balance of the grains is characterized by an equilibrium between photoionization and electron attachment. The free electrons are generated by photoionization, leading to $\rm{H}^+$ in region~F, $\rm{C}^+$ in region~E and $\rm{S}^+$ in region D. 

In lower disk regions (A, B and C) we find a more complex picture with negatively charged grains. In these regions, the electron concentration is significantly lower due to the almost complete shielding from UV and X-rays.  This leaves cosmic ray ionization of molecular hydrogen as the most significant mechanism for electron production. The grain charge balance is characterized by an equilibrium between electron attachment and charge exchange reactions with molecular cations.

Region~C is metal poor due to ice-formation, and \ce{H3+} is the most abundant molecular cation. Region~B has a very complex hydrocarbon chemistry, as nitrogen and oxygen are already frozen out, but carbon is not.  The most abundant molecular cation is found to be \ce{C3H3+} due to its large proton affinity.  We find that negatively charged dust grains $Z^-$ become more abundant than free electron inward of about $r\!=\!3$\,au in our standard T\,Tauri disk model.

In this region~A, we find very low concentrations of free electrons between about $10^{-13}$ and $10^{-18}$. The low electron concentration is because of the free electrons attaching themselves very quickly to dust grains at the high gas densities in region~A. A charge balance is established between negatively charged dust grains $Z^-$ and positively charged molecular ions, whereas the free electrons do not play an important role for the charge balance in region~A.  \ce{NH4+} is found to be the by far most abundant molecular cation due to its very high proton affinity, and since region~A is too warm for \ce{NH3}-ice to form, \ce{NH4+} is available.

Additionally we find that the charge of a dust grain $q$ depends linear on its size $a$ in almost all cases and in all disk regions, i.e.\ $q/a\!=\!\rm const$. In disc region~A, we find $q/a\!\approx\!-350$ charges per micrometer to be a typical value.  This linear dependence is because of the electric grain potential $\propto\!q/a$, which enters the various photo-, electron attachment, and charge exchange rates.

We have studied how the quantities characterizing the charging behavior of gas and dust depend on certain gas, dust and cosmic ray properties in region~A. These properties are is the electron density $n_{\rm e}$, the molecular cation density $n_{\rm ion^+}$, the density of negative charges on grains $Z^-$, and $q/a$. In particular, we find that the electron concentration depends linear on the cosmic ray ionization rate in region~A, while this rate has no effect on $Z^-$ and $n_{\rm ion^+}$.  We find that at a cosmic ray ionization rate as low as $10^{-22}\rm\,s^{-1}$, the electron concentrations can drop to $<10^{-19}$. The gas density has the opposite effect, i.e.\ lower gas densities cause higher electron concentrations.

The minimum dust size and the dust to gas ratio
also have pronounced effects on $n_{\rm e}$, $n_{\rm ion^+}$ and $Z^-$, in particular higher dust/gas ratios cause lower electron concentrations, an effect that seems essential to understand ring formation in early disks via
disk instabilities (see \cite{2023arXiv230203430R}).

Higher gas temperatures favor more strongly negatively charged grains, as the higher thermal velocities of the free electrons allow them to still collide with the grains, even if they are already strongly charged negatively. However, at temperatures over about 1300\,K, thermal ionization of \ce{Na} leads to a sudden increase of the electron concentration.

Region~A is found to be the most likely site for lightning to occur in disks, because here the electron concentration is so low that the gas effectively becomes an insulator, and the dynamics of the charged grains can cause an electrification.

We developed a simple model for the electrification due to turbulence, in which a charge separation between the positive \ce{NH4+} molecules and the negative dust grains occurs, due to the centrifugal force in turbulent eddies. While this mechanism does not generate electric fields that are large enough to overcome the critical electric field for electron cascade in our standard T\,Tauri disk model, we still used it to explore where in the disk the best conditions for lightning might emerge. According to this electrification model, we favor low cosmic ray ionization rates, high dust/gas ratios, large dust grains and small gas densities.  The model also favors warm gas temperatures, but no higher than about 1300\,K where the thermal ionization of Na inhibits the build-up of electric fields via creating too many charged gas particles which enhance the gas conductivity.
%Additionally we find that dust grains overall, independent of the region, charge dependent of their size relatively more or less. But they still charge in a uniform way if one normalizes the total charge to their size $q/a$.

So far, we have not yet included triboelectric charging rates in our disk models. This effect, which generates negatively charged small grains and positively charged large grains by dust-dust collisions, could lead to a situation where both $Z^+$ and $Z^-$ are larger by orders of magnitude compared to our current results, whereas the concentrations of free electrons and molecular cations can remain very low. In that case (called \textit{Region X}), our turbulent electrification model (Eq.\,\ref{eqn:efield_elec}) would predict much larger electric fields which could potentially reach the
critical E-field for electron cascade, i.e.\ lightning.

In preliminary studies, we  noticed that the grain charge chemistry is very sensitive to the heat of formation of charged dust grains $H_f^0(Z^q)$, connected to the work function and the band gap.  As alluded to at the end of Sec.~\ref{subsec:dustchargereac}, the underlying physics here is complicated and the value rather uncertain, depending on material mixture, surface properties and ice-coating. We would like to investigate this question further in an upcoming study, where we would test different chemical potentials and show what effects this value might have on the midplane chemistry and potential emergence of lightning.
%a region where either the charge balance is totally dominated by dust grains, meaning the most abundant positive and negative species are dust grains, or just a region where we see a different charging behavior of the dust grains of different sizes.

\begin{acknowledgements}
  T.\,B., P.\,W., and U.\,G.\,J.\ acknowledge funding from the European Union H2020-MSCA-ITN-2019 under Grant Agreement no.\,860470 (CHAMELEON).
  U.G.J.\ further acknowledges funding from the Novo Nordisk Foundation Interdisciplinary Synergy Programme grant no. NNF19OC0057374.
\end{acknowledgements}

\bibliographystyle{aa}
\bibliography{library.bib}

\appendix
\section{On the impact of automatically generated charge exchange and protonation reactions}
\label{sec:charge_exchange}
In this section we want to investigate the effect that automatically generated charge exchange, protonation and endothermic reactions can have on our model. For this, we incorporate three different models (Tab. \ref{tab:models}), one where we allow no exchange, protonation or endothermic reactions to be generated and only use the network as provided, from now on called \textit{noEX}, one where we allow charge exchange reactions to be generated which is the case for our standard case discussed before and one where we allow charge exchange, protonation and endothermic reactions to be generated, from now on called \textit{EX}. Not that for endothermic reactions we set the barrier of the activation energy for the reactions to be considered to $5000\;K k_b$ which equals to 0.431 eV.

In order to illustrate the differences between the different models, we wanted to show a list of the different reactions considered for the different setups. However, as such a list would consider thousands of reactions, since our \textit{noEX} setup consists of 5530 reactions, the standard setup of 7009 and the \textit{EX} of 8672, we have to show the differences with example reactions, meaning reactions of a specific type not considered by a certain setup. We choose to show a curated list of reactions instead, which show what kind of reactions are considered additionally for the different setups (Tab. \ref{tab:models_react}).

\begin{table}[]
    \centering
    \caption{The three different models we considered when testing the influence of automatically generated reactions.}
    \begin{tabular}{|c|c|c|c|}
    \hline
        Reac. Type &  noEX & standard& EX \\
    \hline
        Charge Exchange & $\times$ & \checkmark & \checkmark\\
        Proton Exchange & $\times$ & $\times$ & \checkmark\\
        Endotherm & $\times$ & $\times$ & \checkmark\\
    \hline
    \end{tabular}

    \label{tab:models}
\end{table}
\begin{table*}[]
    \centering
    \caption{Example reactions present in the different models$^{(1)}$.}
    \begin{tabular}{|c|c|c|c|c|c|c|c|c|}
        \hline
        Nr. & Educts & Prod. & noEX & Standard & EX  & $\alpha$ & $\beta$ & $\gamma$ \\
        \hline
         1 & $\rm N^+$ + H& N + $\rm H^+$ & $\checkmark$ & $\checkmark$ & $\checkmark$ & 1.00E-12 & 0.00 & 0.0000 \\
         2 & Ar + $\rm H_3^+$ & $\rm Ar^+$ + H + $\rm H_2$ &    $\checkmark$ & $\checkmark$ & $\checkmark$ &      3.65E-10 & 0.00   &    0.0000\\
         3& H    +  $\rm CO_2^+$    &        $\rm H^+$  +    $\rm CO_2$ &   $\times$ & $\checkmark$ & $\checkmark$ &      5.00E-10 & 0.50  &     0.0000\\ 
         4&  $\rm H^-$   +   $\rm CO_2^+$   &         H    +   CO   +   O   &    $\times$ & $\checkmark$ & $\checkmark$ &        5.00E-10 & 0.00    &   0.0000             \\ 
 5& C   +    $\rm CH_4^+$      &      $\rm CH^+$   +  $\rm CH_3$          &   $\times$ & $\times$ & $\checkmark$ &        1.32E-09 & 0.00  &     0.0000 \\
 6& C    +   $\rm C_2H^+$    &      $\rm CH^+$   +  C2            &    $\times$ & $\times$ & $\checkmark$ &      1.32E-09 & 0.00   & 4317.7482 \\
 \hline
\end{tabular}\\
    \footnotesize
     $^{(1)}$:Reactions that are just in the noEX model (1 and 2) are taking from exterior references.\\
     Exchange reactions added in our standard model (3 and 4) are added as described in Sec. \ref{subsec:charge_exchange}.\\
     Additional reactions present in the EX model (5 and 6) are added again as explained in Sec. \ref{subsec:charge_exchange}.

    %\footnotetext[1],\footnotetext[2]}
    \label{tab:models_react}
\end{table*}
\subsection{noEx model}
For the noEx model in region A we find that firstly the whole region A starts a bit further in, due to $C_3H_4^+$ dipping at a later point. Secondly, we find that $C_4H^+$ is more abundant much further in.

For region B we can again find the trend that the region emerges more inward than in our standard case. Overall, we find that the chemistry very comparable. We however find that  species H2S+ and H2CS+ are more abundant and overcome our set threshold, whilst SiOH+ and H3CO+ are not overcoming this threshold and are therefore not abundant anymore.

For region C and outwards, we find no notable differences.
\begin{figure}
    \centering
   \includegraphics[width=.5\textwidth]{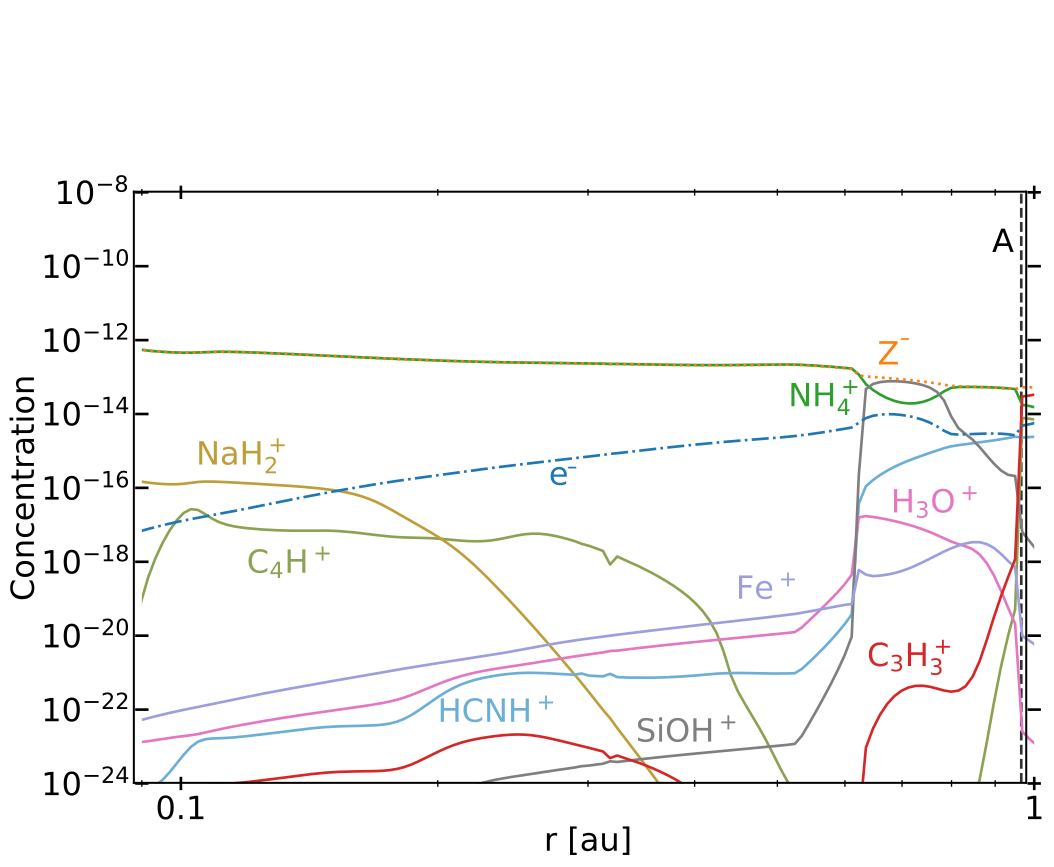}
    \caption{Results for noEX model in Region A. Compare Fig. \ref{fig:species_low_r} for the standard case.}
    \label{fig:species_low_r_noEX}
\end{figure}
\begin{figure}
    \centering
   \includegraphics[width=.5\textwidth]{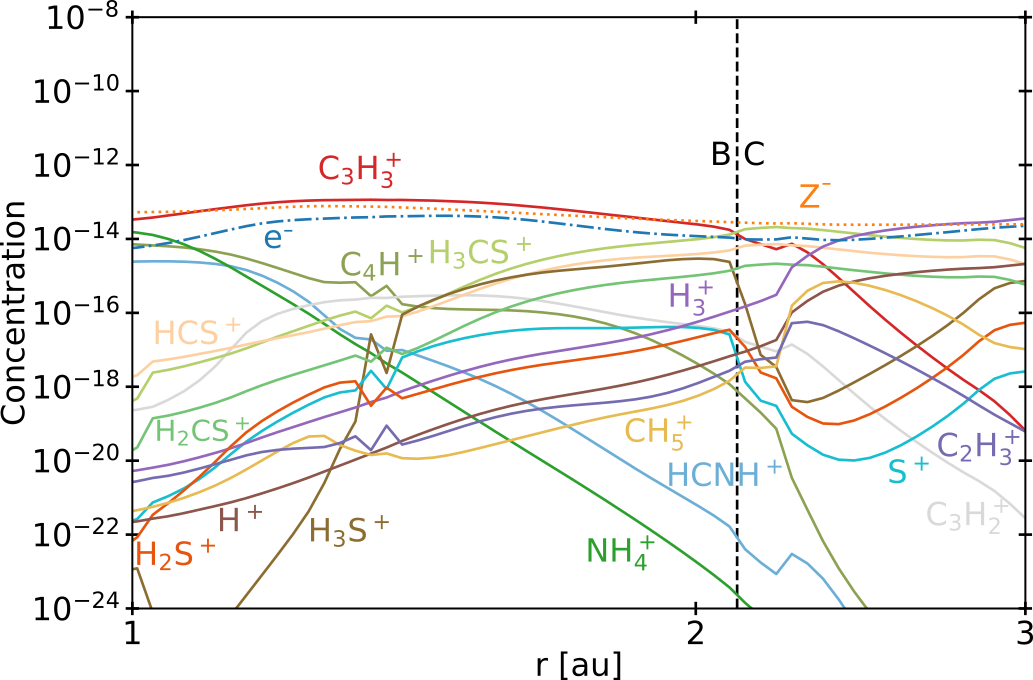}
    \caption{Results for noEX model in Region B. Compare Fig. \ref{fig:species_med_r} for the standard case.}
    \label{fig:species_med_r_noEX}
\end{figure}
\begin{figure}
    \centering
    \includegraphics[width=.5\textwidth]{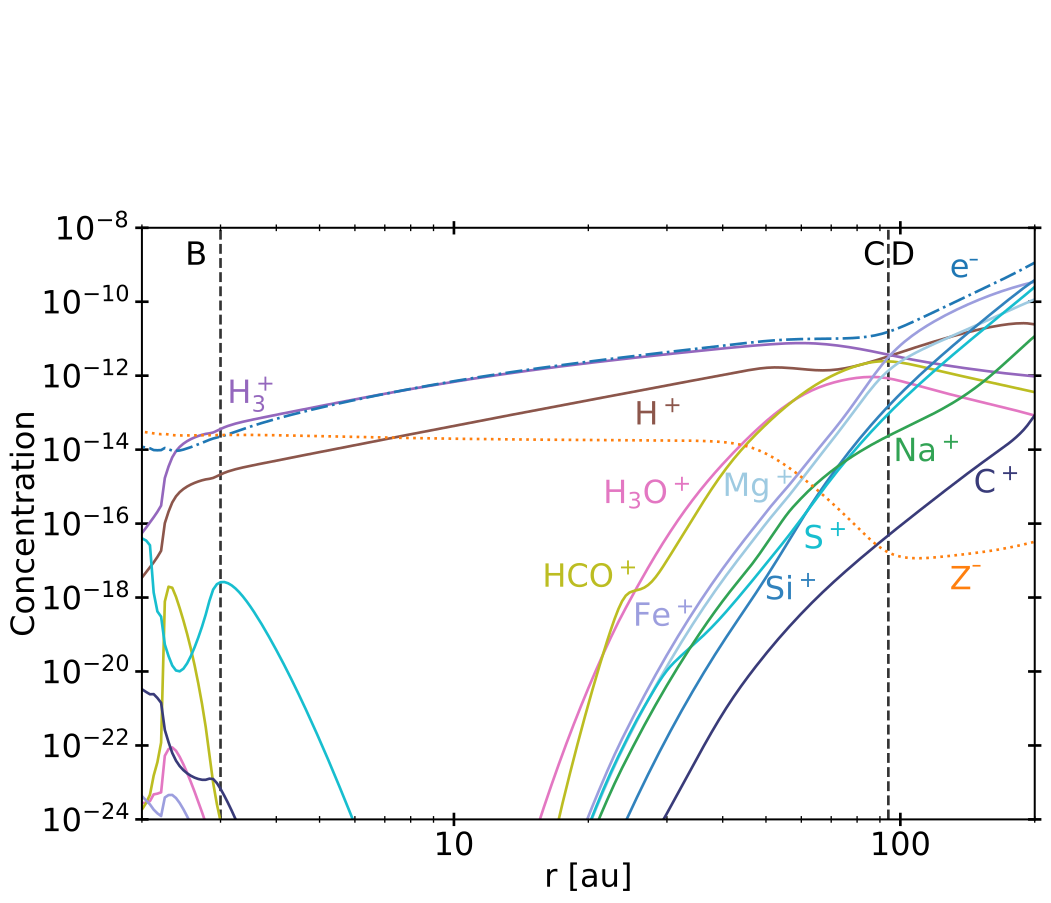}
    \caption{Results for noEX model in Region C. Compare Fig. \ref{fig:species_med_r} for the standard case.}
    \label{fig:species_high_r_noEX}
\end{figure}
\subsection{EX model}
For region A we find mainly that $Na^+$ becomes the second most abundant positive charge carrier and replaces $NaH2^+$ compared to our standard model. In addition we find that $Fe^+$ is no longer abundant enough to meet our set threshold. However, we find, protonated methanol $CH_3OH_2^+$ being abundant in the EX model.

For region B we find a much less complex chemistry. In particular $C_4H^+$ and $HCNH^+$, which are abundant in our standard model, do not show up anymore. We only find the sulfur bearing species, that are also abundant in the standard model, dominate as second and third most abundant cations after $\rm C_3H_3^+$.

For region C, we again do not find significant differences.
\begin{figure}
    \centering
    \includegraphics[width=.5\textwidth]{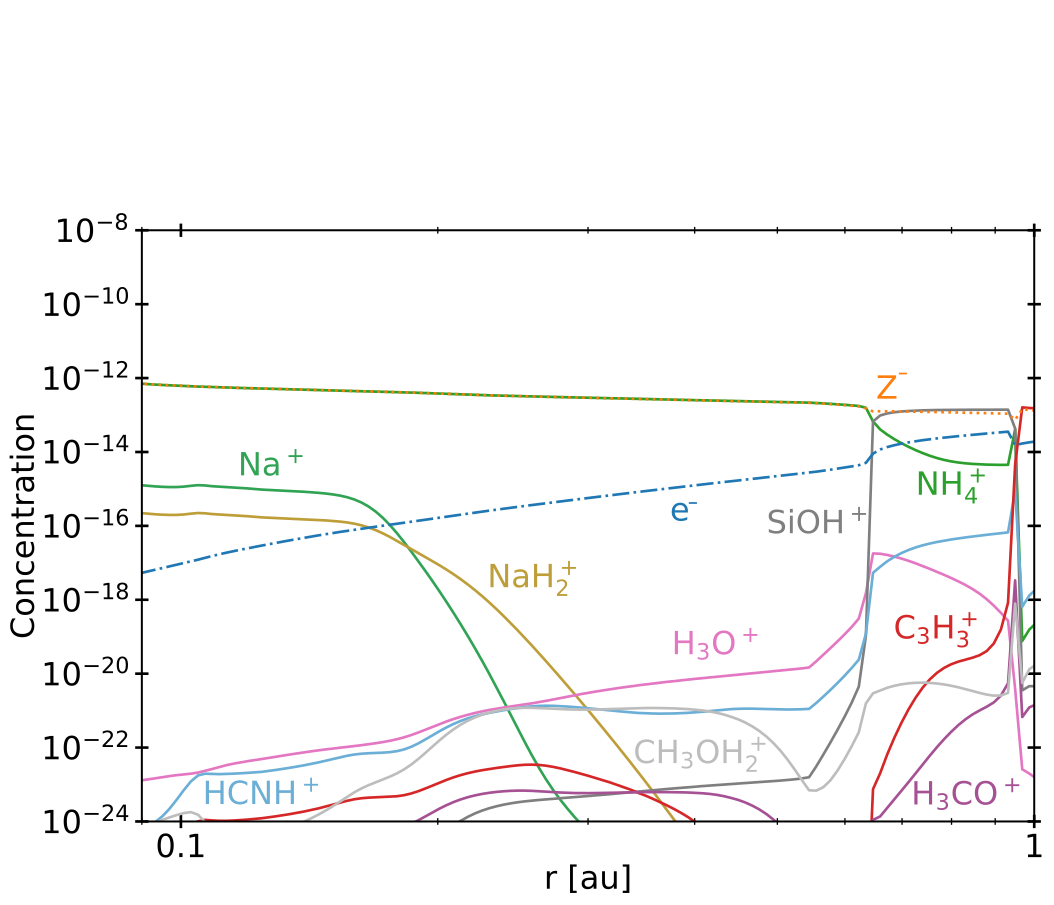}
    \caption{Results for EX model in Region A. Compare Fig. \ref{fig:species_low_r} for the standard case.}
    \label{fig:species_low_r_EX}
\end{figure}
\begin{figure}
    \centering
    \includegraphics[width=.5\textwidth]{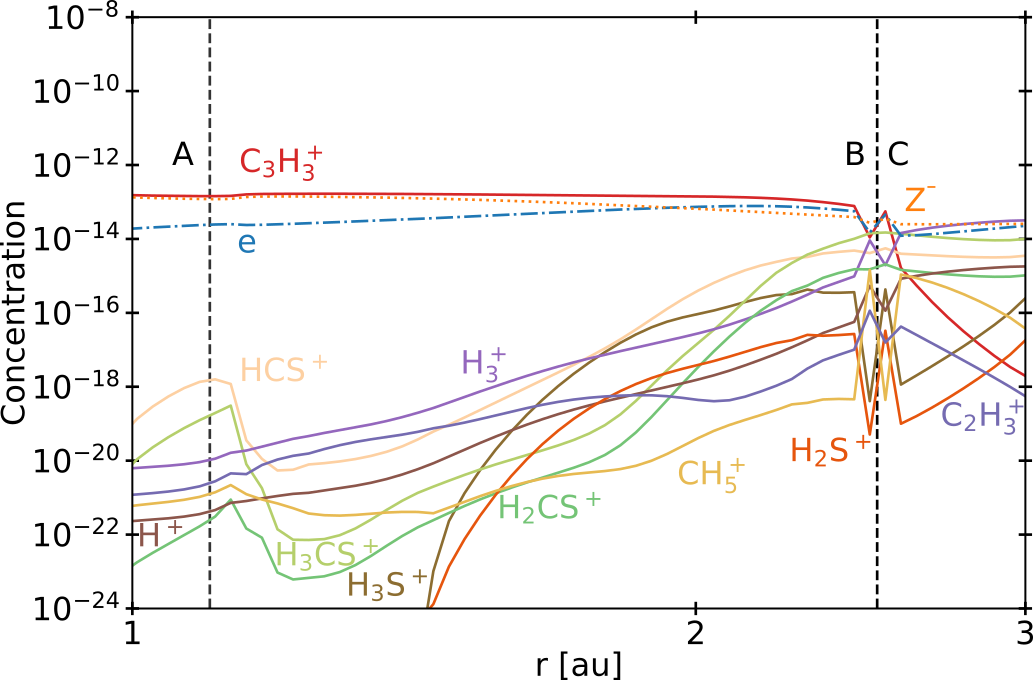}
    \caption{Results for EX model in Region B. Compare Fig. \ref{fig:species_med_r} for the standard case.}
    \label{fig:species_med_r_EX}
\end{figure}
\begin{figure}
    \centering
    \includegraphics[width=.5\textwidth]{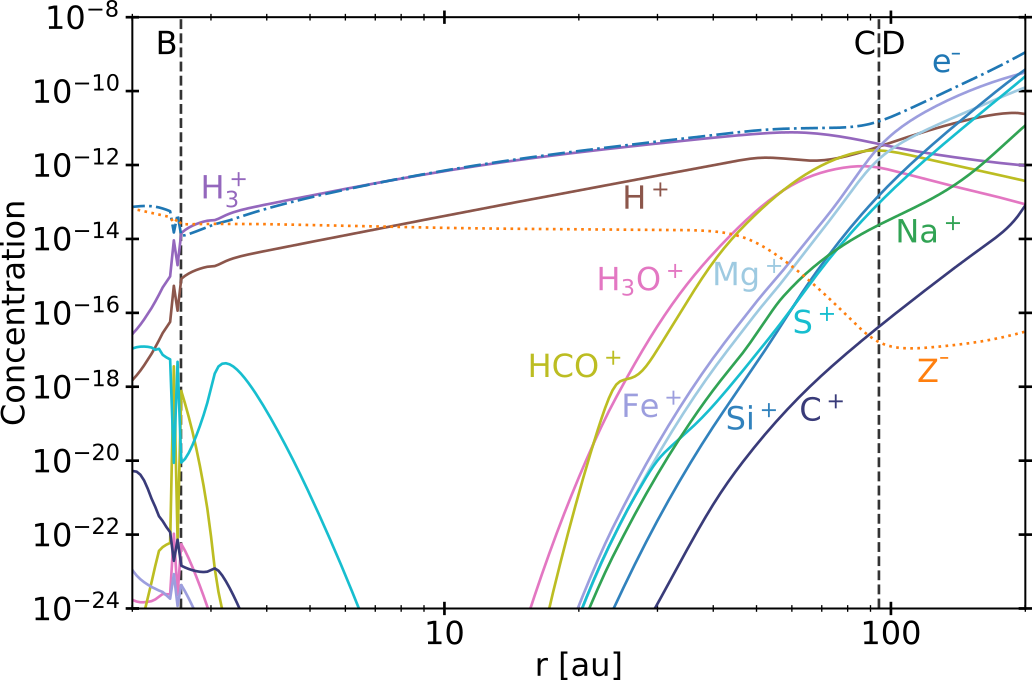}
    \caption{Results for EX model in Region C. Compare Fig. \ref{fig:species_high_r} for the standard case.}
    \label{fig:species_high_r_EX}
\end{figure}

\section{Solving the chemical rate network with dust charge moments}
\label{sec:solve}

The charge and size distribution of the dust grains are represented in our chemical models by three charge moments $Z_{{\rm m},j}^+$, $Z_{{\rm m},j}$, $Z_{{\rm m},j}^-$ in each dust size bin, using a small number of dust size bins $j\in\{1,\,...\,,J\}$.  In Sect.\,\ref{subsec:moment_rate_coeff} we have introduced the effective chemical rates for these moments for photoionisation, electron attachment and charge exchange reactions, which properly describe the overall effect of all grains on the gas and ice chemistry.  On the one hand side, the charge distribution function $f_j(q)$ must be known to calculate the effective rate coefficients. However, on the other hand side, as explained in Sect.~\ref{sec:linearChain}, the determination of $f_j(q)$ requires to know the photon flux and the concentrations of all chemical species including electrons and molecular anions and cations.  

Our formulation with dust charge moments and effective rates is still conformed with the basic shape of a coupled system of first-order ordinary differential equations (ODE system)  
\begin{equation}
  \frac{dn_i}{dt} = P_i - L_i = F_i(\vec{n}) \ ,
  \label{eq:ODE}
\end{equation}
which enables us to use numerical ODE solvers to advance the chemistry. $n_i$ is the particle density of chemical species $i$, $\vec{n}\!=\!\{n_1,\,...\,,n_I\}$ the solution vector, $P_i$ and $L_i$ are the total production and destruction rates, $F_i$ are the components of the right-hand side (RHS) vector, and $I$ is the number of chemical species which includes $3\times J$ dust moments.  From $\vec{n}$, we could in principle update $f_j(q)$ in each call of the RHS vector, then update the effective rate coefficients, and calculate all $F_i$.  

However, this would result in an extremely slow computational method, because the calculation of $f_j(q)$ requires the computation of tens of thousands of rates for all individual charging states.  In addition, we need implicit solvers which require the Jacobian derivative $\partial F_i/\partial n_j$, i.e.\ for each integration step one would need to carry out the determination of $f_j(q)$ a couple of 100 times. Normally, in astrochemistry, one can calculate the rate coefficients only once, prior to calling the ODE-solver, because the rate coefficients do not depend on the particle densities, and then compute $F_i$ and $\partial F_i/\partial n_j$ from these rate coefficients in very efficient, quick ways. In contrast, according to our formulation of the problem, the rate coefficients here do depend on the particle densities in complicated ways. 

We figured that there is a much more efficient way to solve our chemical equations, by means of an iterative approach.  To obtain a time-independent solution of our chemical equations, as used in this paper, we follow the following algorithm. 
\begin{itemize}
  \item[(1)] Provide an initial guess of the particle densities $\vec{n}$ based on the results obtained for the previously solved grid point,
  \item[(2)] compute $f_j(q)$ from $\vec{n}$,
  \item[(3)] compute all rate coefficients, including the effective rate coefficients from $f_j(q)$,
  \item[(4)] solve $\vec{F}(\vec{n}^{\rm new})=0$ for given rate coefficients to obtain the new solution vector $\vec{n}^{new}$,
  \item[(5)] compute the relative changes of electron density $n_{\rm e}$ and mean grain charges $\langle q_j\rangle$ in all size bins $j$ between iterations,
  \item[(6)] if these relative changes are $<\!10^{-5}$ and $<\!10^{-2}$, respectively, then take $\vec{n}^{new}$ as solution and stop the iteration,
  \item[(7)] provide a better guess for $\vec{n}$ and jump back to (2).
\end{itemize}
For step (7), we apply the following strategy. Each iteration is assigned a quality defined as $Q^{\rm new}=n_{\rm e}^{\rm new}-n_{\rm e}$. At first instance, we take $\vec{n}\leftarrow\vec{n}^{\rm new}$ as improved guess in step (7), i.e.\ we perform a $\Lambda$-iteration. In all subsequent calls, we use the previous and current electron densities and $Q$-values to provide a better guess of the solution vector 
\begin{eqnarray}
  0 &=& Q^{\rm new} + X\,(Q-Q^{\rm new}) \\
  \vec{n} &\leftarrow& \vec{n}^{\rm new} + X\,(\vec{n}-\vec{n}^{\rm new})
  \label{eq:nnew}
\end{eqnarray}
by determining the value for $X$ that nullifies $Q$ in case $Q$ is a linear function of $n_{\rm e}$, i.e., we perform a Newton step. Precisely speaking,
we put $0\!=\!Q^{\rm new} + \frac{\partial Q}{\partial n_{\rm e}}(n_{\rm e}^0-n_{\rm e}^{\rm new})$, where we find $\frac{\partial Q}{\partial n_{\rm e}}\!=\!(Q-Q^{\rm new})/(n_{\rm e}-n_{\rm e}^{\rm new})$ from the two previous iterations, to determine the next improved guess of the electron density $n_{\rm e}^0$, related to $X=(n_{\rm e}^0-n_{\rm e}^{\rm new})/(n_{\rm e}-n_{\rm e}^{\rm new})$. If the magnitude of the new quality $Q^{\rm new}$ is much smaller than the magnitude of the old quality $Q$, $X$ is close to zero, and we essentially perform another $\Lambda$-iteration.  The reader may verify that the resulting electron density according Eq.\,(\ref{eq:nnew}) is indeed $n_{\rm e}^0$. This numerical method typically converges within a couple of iterations. In a few cases, however, the method starts to oscillate. We therefore monitor the $Q$-values and save the $\vec{n}$-vectors that are encountered during the iteration.  Once we find negative and positive entries for $Q$, we have bracketed the solution, and can use the two iterations that produced the smallest positive and largest negative $Q$-values to provide a better next guess for $\vec{n}$ in step (7).

In case we require a time-dependent solution of Eq.\,(\ref{eq:ODE}), not performed in this paper, we use a different strategy.  The problem in this case is that, with constant rate coefficients, the solution may start to alternate between two charging states with every time step taken. For example, highly negatively charged grains do not take any further electrons, i.e.\ the $k_{{\rm m},j,{\rm e}}$ become small, which means that during the next time step, the grains charge up less negatively, and so on. Our solution here is to monitor the dependencies of the effective rate coefficients $k_r$ on electron density $n_{\rm e}$, and use the following correction power law 
\begin{equation}
  k_r = k_r^{\rm ref} \left(\frac{n_{\rm e}}{n_{\rm e}^{\rm ref}}\right)^{p_r}
  \label{eq:kr_correct}
\end{equation}
whenever the ODE-solver requires the RHS or the Jacobian.  $n_{\rm e}^{\rm ref}$ is a reference electron density from which $k_r^{\rm ref}$ was calculated, and $n_{\rm e}$ is the actual electron density occurring during the integration. $p_r$ is a rate coefficient correction power law index. The initial conditions are $\vec{n}=\vec{n}(t\!=\!0)$, $n_{\rm e}^{\rm ref}=n_{\rm e}(t\!=\!0)$ and $p_r=0$.
\begin{itemize}
  \item[(1)] Compute $f_j(q)$ from $\vec{n}$, and rate coefficients $k_r^{\rm ref}$ from $f_j(q)$,
  \item[(2)] advance the chemistry for time step $\Delta t$ from $\vec{n}$ with rate coefficients according to Eq.\,(\ref{eq:kr_correct}), resulting in $\vec{n}^{\rm new}(t+\Delta t)$,
  \item[(3)] compute new $f_j(q)$ and new rate coefficients $k_r^{\rm new}$ from $\vec{n}^{\rm new}$, 
  \item[(4)] update the rate coefficient correction indices as
  $p_r=(\log k_r^{\rm new}-\log k_r)/(\log n_{\rm e}^{\rm new}-\log n_{\rm e})$,
  \item[(5)] assess the uncertainty $\Delta \vec{n}$ based on the difference between $\vec{k}$ and $\vec{k}^{\rm new}$ and $\Delta t$,
  \item[(6)] limit the increase of the next time step $\Delta t$ when
  large relative uncertainties $\Delta n_i/n_i$ are found,
  \item[(7)] if $t+\Delta t$ reaches the requested total integration time, 
  take $\vec{n}^{\rm new}$ as solution and stop the iteration,
  \item[(8)] update $t\!=\!t+\Delta t$, $\vec{n}=\vec{n}^{\rm new}$, $n_{\rm e}^{\rm ref}\!=\!n_{\rm e}^{\rm new}$ and $k_r^{\rm ref}=k_r^{\rm new}$, and jump back to (2).
\end{itemize} 
This algorithm can slow down the computation of a $t$-dependent solution when strong oscillations would occur, resulting in large $\Delta n_i$, where smaller time steps are indeed required.  However, the algorithm manages to sufficiently damp these oscillations and will quickly re-increase the time step once the conditions have changed and the oscillations disappear.  
revers
\section{Dissociative molecular ion attachment reactions}
\label{AppC}
ike \ce{NH4+} in which the molecule dissociates
\begin{equation}
    Z^{-q}+\ce{NH4+} \rightarrow Z^{-(q-1)} + \ce{NH3} + \ce{H}
\end{equation}
where $Z^{-(q-1)}$ is a dust grain that is one time less negatively charged than $Z^{-q}$. The reaction enthalpy, which tells us whether a reaction is exothermic or endotherm, is given by
\begin{equation} 
    \Delta H_r=
    H_f(Z^{-(q-1)})
   +H_f(\ce{NH3}) 
   +H_f(\ce{H}) 
   -H_f(Z^{-q})
   -H_f(\ce{NH4+}) \ ,
    \label{eqn:delta_h_r}
\end{equation}
where $H_f$ is the heat of formation of a substance from a chemical standard state, which we set here as purely atomic in the absence of electric fields.  For neutral molecules, we use the atomization energies $H_f^0$ derived from the standard enthalpies at 0K from UMIST \citep{UMIST1}, but for $H_f(\ce{NH4+})$ we need to take into account the electric potential in the presence of the electric field of the dust grain at its surface
\begin{eqnarray}
   H_f(\ce{NH3})  &=& H_f^0(\ce{NH3})\\
   H_f(\ce{H})    &=& H_f^0(\ce{H})\\[-2mm]
   H_f(\ce{NH4+}) &=& H_f^0(\ce{NH4+})-\frac{qe^2}{a} \ .
\end{eqnarray}
The heat of formation of a negatively charged dust grains is
\begin{equation}
    H_f(Z^{-q}) = H_f({\rm atoms}\rightarrow Z^0) 
    + \sum_{i=1}^q H_f(Z^{-(i-1)} \rightarrow Z^{-i}) \ ,
\end{equation}
where $H_f({\rm atoms}\rightarrow Z^0)$ is the heat of formation of a dust grain from the atoms it is composed of.  The term $H_f(Z^{-(i-1)} \rightarrow  Z^{-i})$ describes the energy liberated when moving a single electron from infinity to the grain's surface (takes energy) and then attaching it to the surface (liberates energy $E_0$)
\begin{equation}
  H_f(Z^{-(i-1)} \rightarrow Z^{-i}) 
  = -E_0 + \frac{(i-1)e^2}{a}
\end{equation}
Using the Gaussian sum formula, the total formation enthalpy of a negatively charged grain is hence
\begin{equation}
  H_f(Z^{-q}) = H_f({\rm atoms}\rightarrow Z^0) 
  -q\,E_0 +\frac{e^2}{a}\frac{q}{2}(q-1) \ .
\end{equation}
From Eq.\,(\ref{eq:EA}) we can identify the constant $E_0=W_0-E_{\rm bg}$ where $W_0$ is the work function and $E_{\rm bg}$ the band gap, which we both assume to be independent of $q$.

The reaction enthalpy according to Eq.\,(\ref{eqn:delta_h_r}) is then given by
\begin{align}
  \Delta H_r 
    &= E_0 +\frac{e^2}{a}
       +H_f^0(\ce{NH3}) +H_f^0(\ce{H}) -H_f^0(\ce{NH4+})\nonumber\\
    &= E_0 +\frac{e^2}{a}
       +P_A(\ce{NH4+}) -13.6\rm\,eV  \, \label{eq:Hr}
\end{align}
where the proton affinity is $P_A(\ce{NH4+})=H_f^0(\ce{NH3}) +H_f^0(\ce{H+})-H_f^0(\ce{NH4+})$ and $H_f^0(\ce{H})-H_f^0(\ce{H+})=-13.6\rm\,eV$ is the ionization energy of hydrogen.
Since the term $e^2/a$ is negligible even for grains as small as $0.1\,\mu$m, Eq.~(\ref{eq:Hr}) allows us to state that a dissociative attachment reaction is exothermal when the work function minus the band gap plus the proton affinity of the respective molecule is smaller than 13.6\,eV, independent of $q$.

There is, however, a small correction term of order $e^2/a$ which complicates things. In our derivation, this term arises from the removal of one electron and one positive molecular ion from the electric field at the grain surface by the reaction, and this term is size-dependent. Similar correction terms appear in Eqs.\,(\ref{eq:EA}) and (\ref{eq:Emin}).  In the current code version, we use $E_0\!=\!5.89\,$eV, as explained in more detail in at the end of Section \ref{subsec:dustchargereac}.

\end{document}